\documentclass[%
 reprint,prb,
 amsmath,amssymb,
 aps,
floatfix,superscriptaddress
]{revtex4-2}
\usepackage{comment}
\usepackage{hyperref}
\usepackage{graphicx}
\usepackage{dcolumn}
\usepackage{bm}
\usepackage{color}
\usepackage{xcolor}
\usepackage[caption=false]{subfig}
\usepackage{appendix}
\usepackage{chngcntr}
\usepackage{natbib}

\hypersetup{
    colorlinks=true,   
    linkcolor=black,   
    citecolor=blue,    
    urlcolor=blue,     
}

\begin{document}

\preprint{APS/123-QED}

\title{Effect of dynamical electron correlations on the tunnelling magnetoresistance of Fe/MgO/Fe(001) junctions}

\author{Declan Nell}
\affiliation{School of Physics and CRANN Institute, Trinity College Dublin, The University of Dublin, Dublin 2, Ireland}
\author{Stefano Sanvito}
\affiliation{School of Physics and CRANN Institute, Trinity College Dublin, The University of Dublin, Dublin 2, Ireland}
\author{Ivan Rungger}
\affiliation{National Physical Laboratory, Hampton Road, Teddington TW11 0LW, United Kingdom}
\affiliation{Department of Computer Science, Royal Holloway, University of London, Egham, TW20 0EX, United Kingdom}
\author{Andrea Droghetti}
\email[]{andrea.droghetti@unive.it}
\affiliation{School of Physics and CRANN Institute, Trinity College Dublin, The University of Dublin, Dublin 2, Ireland}
\affiliation{Department of Molecular Sciences and Nanosystems, Ca’ Foscari University of Venice, Venice, Italy}

\begin{abstract}

We employ dynamical mean-field theory (DMFT) combined with density functional 
theory (DFT) and the non-equilibrium Green's function technique to investigate 
the steady-state transport properties of an Fe/MgO/Fe magnetic tunnel junction 
(MTJ), focusing on the impact of dynamical electron correlations on the Fe $3d$
orbitals. By applying the rigid shift approximation, we extend the calculations 
from zero- to finite-bias in a simple and computationally efficient manner, obtaining
the bias-dependent electronic structure and current-versus-voltage characteristic
curve in both the parallel and antiparallel configurations. In particular, we find
that dynamical electron correlation manifests as a reduction in the spin splitting 
of the Fe $3d_{z^2}$ state compared to DFT predictions and introduces a finite
relaxation time. The impact of these effects on the transport properties, however,
varies significantly between magnetic configurations. In the parallel configuration, the
characteristic curves obtained with DFT and DMFT are similar, as the transport is mostly
due to the coherent transmission of spin-up electrons through the MgO barrier. 
Conversely, in the antiparallel configuration, correlation effects become more 
significant, with DMFT predicting a sharp current increase due to bias-driven 
inelastic electron-electron scattering. As a consequence, DMFT gives a lower bias 
threshold for the suppression of the tunneling magnetoresistance ratio compared to 
DFT, matching experimental data more closely. 

\end{abstract}

\maketitle

\section{\label{sec:level1}Introduction}

Spintronics explores spin transport and dynamical phenomena in materials, with
applications in computing and data storage \cite{ts.zu.book}. One of the most
prominent phenomena in spintronics is the tunneling magnetoresistance (TMR), 
observed in magnetic tunnel junctions (MTJs) formed of two metallic ferromagnets
separated by a thin insulating barrier. The TMR measures a change in resistance
when the magnetization vectors of the ferromagnetic layers 
switch from an antiparallel (AP) to a parallel (P) configuration 
\cite{ju.75,mi.te.95,mo.ki.95,bo.cr.01}. The TMR ratio can then be defined as
\begin{equation}\label{eq: tmr}
    \textnormal{TMR} = \frac{I_{\textnormal{P}}-I_{\textnormal{AP}}}{I_{\textnormal{AP}}},
\end{equation}
where $I_{\textnormal{P}}$ ($I_{\textnormal{AP}}$) is the current at a bias 
voltage, $V$, for the P (AP) configuration. This is the so-called optimistic
definition, since in most cases $I_{\textnormal{AP}}<I_{\textnormal{P}}$.

In 2001, Butler {\it et al.} \cite{bu.zh.01} and Mathon and Umerski \cite{ma.um.01}
independently predicted a huge TMR ratio in epitaxial, all-crystalline Fe/MgO/Fe(001)
MJTs. These theoretical predictions were soon validated in Fe/MgO/Fe
\cite{Yu_Na_2004} and CoFe/MgO/CoFe \cite{Parkin2004,dj.ts.05} MTJs, reporting 
a very large TMR ratio, which now reaches up to about 600\% at room temperature 
and over 1000\% at a few Kelvin \cite{ik.ha.08}. 

The significant TMR values can be attributed to the spin polarization of the Fe 
density of states (DOS) and the specific details of the wave functions 
matching across the MgO barrier. In particular, states incoming from the Fe electrodes decay at different rates within the MgO barrier, depending on their symmetry. The most slowly decaying states are present at the Fermi energy for one spin channel (the majority channel) but not for the other, resulting in what is known as symmetry-enforced spin filtering \cite{Bu_2008,Mavropolous2000}. This means that the Fe/MgO system effectively behaves as a half-metal.

On the theoretical front, the charge transport in MTJs is primarily described 
in a phase-coherent and transverse momentum-conserving framework 
\cite{bu.zh.01,ma.um.01, sanvito}. First-principles studies are generally 
carried out by combining the Landauer-B\"uttiker approach to quantum transport
\cite{La.57,Bu.86,Bu.88} with Kohn-Sham (KS) density functional theory (DFT)
\cite{jo.gu.89,kohn.99,jone.15}, either within the local spin density approximation
(LSDA) \cite{ba.he.72,vo.wi.80} or the generalized gradient approximation (GGA)
\cite{pe.ch.92,pe.ch.93,pe.bu.96}. The conductance is determined by the transmission 
of the KS states from one electrode, through the barrier, and to the other electrode.
Various implementations of the approach exist (e.g., \cite{ku.dr.00, Kh.Br.05, Wo.Is.02, Wo.Is.02_2,Ma.Zh.99}), but the most convenient one is based on the Non-Equilibrium
Green's function (NEGF) technique, \cite{bookStefanucci,datta,book1}, and it is 
therefore known as DFT+NEGF \cite{Ta.Gu.01,Ba.Mo.02,ro.ga.06}.
In particular, a key advantage of DFT+NEGF, compared to other quantum transport
approaches, is that it extends calculations beyond the linear-response
limit of the original Landauer-B\"uttiker formalism. This enables the self-consistent
evaluation of charge redistribution between the electrodes at finite bias voltages and 
provides the current-versus-voltage ($I/V$) characteristic curve of a device. 

Over the years, DFT+NEGF has found extensive application in the study of the TMR, not only 
in Fe/MgO/Fe and related systems \cite{wa.ti.06, Ivan_Stefano_2009, PR_Ivan_Stefano_2008, Ivan_Rocha_2007, el.st.17}, but also in many other MTJs (see, for example, references \cite{ca.ar.11, sa.ka.12, do.na.14, st.st.21, Li.lu.19,pa.tu.19, Shao2021,pa.dl.22,do.li.22}). 
However, despite this widespread use, the approach has a number of conceptual
\cite{Kurth_2017} and practical issues.
In fact, it is based on the assumption that DFT can be used as an effective single-particle
theory and that the KS band structure provides a good approximation for the excitation
spectrum of a system. Crucially, beside the problems related to the mathematical 
consistency of DFT for transport \cite{Kurth_2017}, these assumptions often fail even 
at the practical level, particularly when comparing the calculated DOS to photoemission
experiments. In the ferromagnetic metals used in MTJs \cite{sa.fi.09,sp.pi.17}, (Fe, Ni, 
Co and related compounds), KS DFT drastically overestimates spin splitting of the $3d$ 
states, providing majority spin spectra that are too wide. In addition, it fails to 
account for the finite lifetime due to the electron-electron interaction that smears out the bands in energy and momentum \cite{mo.ma.02, Walter_2010}.  
Finally, the description of transport as phase-coherent, in terms of electron transmission,
may break down because of electron correlation effects. In fact, some experiments have 
reported features in the $I/V$ characteristic curves that have been interpreted as resulting
from electron-electron scattering  \cite{zh.le.97,ma.yu.05,li.ni.14,ba.dj.09}. 
These effects may be included into DFT+NEGF at a phenomenological level \cite{fa.22}, for example, through B\"uttiker probes \cite{bu.85}, but for a true {\it ab-initio} treatment, one has to go beyond DFT in the description of the electronic structure.

An improved picture of ferromagnetic $3d$ materials relevant for MTJs has been achieved 
by combining DFT with Dynamical Mean Field Theory (DMFT) \cite{me.vo.89,ge.ko.96,li.ka.98, li.ka.01,ko.vo.04,ko.sa.06}. This approach accounts for dynamical correlations, while
neglecting spatially non-local correlations. DMFT returns spectral properties, comprehensive 
of both quasi-particle and incoherent excitations, which generally agree well with 
experimental results \cite{br.mi.06, gr.ma.07, andrea_sigma_2, ja.dr.23}. Furthermore, 
DMFT can be implemented in a layer formulation \cite{po.no.99,ok.mi.04,free.04, ok.na.05,yu.mo.07,ze.fr.09,va.sa.10, va.sa.12, ok.07,ok.08}
to study heterostructures, and it can be integrated with quantum transport computational
frameworks \cite{andrea_Cu_co, liviu_Cu_Co_dmft} previously used for point contacts 
\cite{ja.ha.09,ja.ha.10,Ja.15} and molecular junctions \cite{ja.so.13,andrea_ivan_projection,ap.dr.18,ru.ba.19,bh.to.21,gandus2022strongly, ja.18}. 
However, applications for spin-dependent transport have thus far been limited to 
all-metallic heterostructures treated in the linear response regime \cite{andrea_Cu_co, liviu_Cu_Co_dmft,mo.ap.17,ha.ne.24}. To our knowledge, no study has addressed the effect of dynamical electron correlations on the (non-linear) $I/V$ 
characteristic curve and TMR of MTJs because calculations would require computationally efficient out-of-equilibrium multi-orbital DMFT solvers, which are not yet available. Our goal is to fill this knowledge gap, introducing simple approximations to practically bypass these technical challenges.

In this work, we apply DMFT combined with DFT+NEGF to compute the bias-dependent 
transport properties of an Fe/MgO/Fe MTJ beyond the single-particle 
framework and linear-response theory. In particular, we first treat dynamical electron
correlation effects on the Fe $3d$ orbitals to obtain an improved description of the 
system's electronic structure. Then, by introducing the so-called rigid shift approximation,
which allows for finite-bias calculations to be performed in a simple and computationally
efficient manner, we calculate the $I/V$ curve of the P and AP configurations, and 
consequently, the bias-dependent TMR.  

In the P configuration, the transport properties predicted by both DFT and DMFT are 
similar, as they are mostly related to the coherent transmission of spin-up electrons 
from the Fe layers through the MgO barrier. In contrast, in the AP configuration, correlation
effects become more significant, leading to a sharp increase of the current with bias. As 
a result, DMFT predicts a lower bias threshold for the suppression of the TMR ratio compared 
to DFT, improving the agreement with the experimental data available in the literature. 
Hence, we conclude that incorporating dynamic correlation effects is quite important 
to estimate the finite bias behavior of Fe/MgO/Fe MTJs.

The paper is organized as follows. In section~\ref{section: theory}, we present the setup 
used to represent an Fe/MgO/Fe junction and outline the theoretical methods, followed 
by the computational details in section~\ref{section: computational details}. In 
Section~\ref{sec.results}, we present our results, beginning with a description of the 
zero-bias electronic structure of the Fe/MgO/Fe junction, systematically comparing DFT 
with DMFT and highlighting the influence of dynamical electron correlations on spin-dependent
electron tunneling (subsection \ref{sec.zero_bias}). Next, we extend the discussion to 
finite-bias transport properties (subsection \ref{sec.finite_bias}), illustrating how 
the electronic structure evolves with bias and explaining how electron correlations 
impact the $I/V$ characteristics for both P and AP configurations, and consequently, 
the TMR. Finally, we compare our results with experimental data from the literature 
in section~\ref{sec.discussion} and conclude in section~\ref{sec.conclusion}.

\section{Method and implementation} \label{section: theory}
\subsection{\label{sec:system set}System setup}

The sensitive part of an MTJ can be represented as a two-terminal device, partitioned 
into three sections: a central region, along with a left-hand side (L) and a right-hand side 
(R) semi-infinite metallic leads \cite{book1}. The transport direction is assumed to be along the 
$z$ Cartesian axis. Periodic boundary conditions are applied in the transverse $x$-$y$ 
plane, and $\boldsymbol{k}=(k_x,k_y)$ indicates the wave-number in the transverse 
Brillouin zone. Mathematically, we achieve the partition into the leads and central region 
by employing a linear combination of atomic orbitals basis set so that each basis orbital 
can be attributed to one part of the device.

The central region, also called the scattering region, is the active section of the device. 
Following common practice in quantum transport \cite{haug1996quantum}, we assume that 
correlation effects are confined within this region, while the electrons in the 
leads are treated at the single-particle level. The leads, or electrodes, 
are effectively electronic baths characterized by their chemical potentials, $\mu_\mathrm{L}$
and $\mu_\mathrm{R}$, and their temperatures. The central region is in thermodynamic
equilibrium within the grand canonical ensemble, when the leads have the same 
chemical potential, $\mu_\mathrm{R}=\mu_\mathrm{L}\equiv E_\mathrm{F}$, and the same
temperature. The electrons are then distributed according to the Fermi-Dirac distribution 
with Fermi energy, $E_\mathrm{F}$. 

The effect of applying a finite bias voltage, $V$, is simulated by shifting the leads' chemical potentials such that $\mu_\mathrm{L}=E_\mathrm{F}+ eV/2$ and 
$\mu_\mathrm{R}=E_\mathrm{F}- eV/2$. In addition to the requirement of local charge neutrality due to the metallic nature of the leads, this results in a relative
displacement of the leads' entire band structure. Each lead remains in equilibrium and has a Fermi occupation function, 
$f_{\mathrm{L(R)}}(E,\mu_\mathrm{L(R)})=\big[1-e^{\beta(E-\mu_\mathrm{L(R)})}\big]^{-1}$.  However, the electrostatic potential drops across the central region, driving it out-of-equilibrium, resulting in a charge current that flows from one lead to the other and through the central region.

\begin{figure}[ht]
    \centering
    \subfloat[Central region]{%
      \includegraphics[clip,width=\columnwidth]{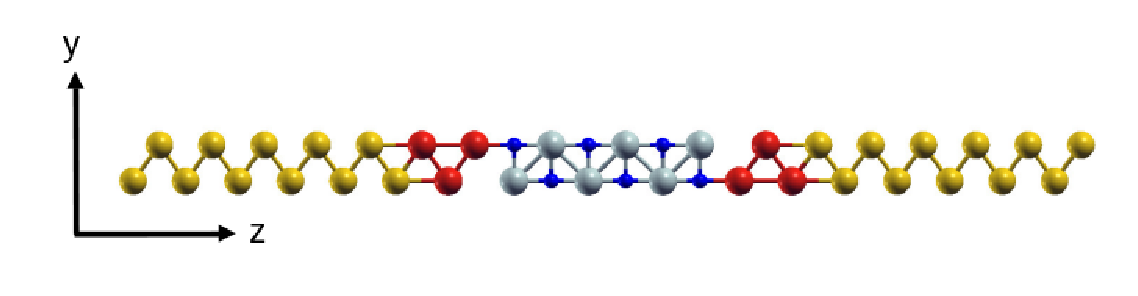}%
      \label{fig:central_region} 
    }

    \subfloat[Potential drop]{%
      \includegraphics[clip,width=\columnwidth]{potential_drop_smaller.eps}%
      \label{fig:potential_drop} 
    }

    \caption{
    Schematic representation of the studied MTJ. (a) Structure of the device, which comprises 
    three Fe layers on either side of the six-layer MgO barrier, and the 6$s$-only Au leads. Fe atoms 
    are in red, Mg atoms in grey, O atoms in blue, and Au atoms in yellow. (b) The electrostatic 
    potential, $V_\mathrm{e}(z)$, which drops linearly across the MgO barrier.
    } 
\end{figure}

The specific Fe/MgO/Fe(001) MTJ considered in this work is shown in Fig.~\ref{fig:central_region}.
The MgO barrier comprises six layers and the atomic structure of its interface with Fe is the same 
as in other works \cite{book1}. Since dynamical correlations are only included in the central 
region \cite{MW_current}, we can not employ the semi-infinite Fe leads that were considered 
in previous DFT+NEGF studies of Fe/MgO/Fe MTJs \cite{Ivan_Stefano_2009, PR_Ivan_Stefano_2008, Ivan_Rocha_2007}. 
Therefore, we place three Fe layers on either side of the MgO barrier in the central region, while the leads are
modelled by a {\it bcc} lattice of Au atoms with the same lattice parameters
as Fe. These Au electrodes are described using only $6s$ orbitals as a basis, a pragmatic 
choice that mimics the wide-band approximation often used in model calculations \cite{haug1996quantum}. 
Thus, the effect of the leads on the central region's electronic structure is merely to broaden and smooth the 
DOS of the Fe layers. Notably, three Fe atomic layers are sufficient to fully spin-polarize the charge current to the level of bulk Fe. A comparison of the zero-bias DFT+NEGF results for 
our simplified system and the corresponding system with realistic semi-infinite Fe leads, 
as used in previous studies \cite{Ivan_Stefano_2009,book1}, is provided in section S1 of the 
supplementary information (SI), showing that the differences are negligible.

\subsection{\label{sec:Extension beyond the single picture} NEGF equations}
The electronic structure of our Fe/MgO/Fe MTJ at zero- and finite-bias is obtained by solving 
the NEGF equations. Assuming that the system is in a stationary state, the NEGF equations read \cite{bookStefanucci,fe.ca.05,fe.ca.05_2}
\begin{equation}
\begin{split}
      G^{\sigma \; r}(E, \boldsymbol{k}) & = [(E+i0^+)S(\boldsymbol{k}) - H^{\sigma}(\boldsymbol{k})   - \Sigma^{\sigma \; r}_\mathrm{L}(E, \boldsymbol{k})\\
       & - \Sigma^{\sigma \; r}_\mathrm{R}(E, \boldsymbol{k}) - \Sigma_{\textnormal{MB}}^{\sigma \; r}(E, \boldsymbol{k})]^{-1},\\ \label{Eq: mb_retarded_gf}
\end{split}
\end{equation}
and
\begin{equation}
\begin{split}
    G^{\sigma \;  <} (E, \boldsymbol{k}) & = G^{\sigma \; r} (E, \boldsymbol{k}) \left[ \Sigma^{\sigma \; <}_{L}(E, \boldsymbol{k}) + \Sigma^{\sigma \; <}_{R}(E, \boldsymbol{k})\right.\\
    & \left. + \Sigma_{\textnormal{MB}}^{\sigma\; <}(E, \boldsymbol{k})\right] G^{\sigma \; a}(E, \boldsymbol{k}),\label{Eq: mb_lesser_gf}
\end{split}
\end{equation}
where $G^{\sigma \; r}(E, \boldsymbol{k})$, 
$G^{\sigma \; a}(E, \boldsymbol{k})=[G^{\sigma \; r}(E, \boldsymbol{k})]^\dagger$ 
and $G^{\sigma \; <}(E, \boldsymbol{k})$ are the retarded, advanced and lesser components of 
the Green's function of the central region respectively. $\sigma$ is the spin index and 
$H^{\sigma}(\boldsymbol{k})$ is the $\boldsymbol{k}$-dependent single-particle Hamiltonian 
of the central region, which also includes the electrostatic potential due to the voltage bias 
applied between the leads. Since, in general, the basis set is not orthogonal, $S(\boldsymbol{k})$ 
is the orbital overlap matrix, which differs from the identity. In the definition of the Green's
function, $\Sigma^{ \sigma \; r}_\mathrm{L(R)}(E,\boldsymbol{k})$ and 
$\Sigma^{ \sigma \; <}_\mathrm{L(R)}(E,\boldsymbol{k})$ are the retarded and lesser leads' self-energies, 
which describe the hybridization of the central region with the left-hand (right-hand) side lead, 
while $\Sigma_{\textnormal{MB}}^{\sigma \; r}(E, \boldsymbol{k})$ and 
$\Sigma_{\textnormal{MB}}^{\sigma \; <}(E, \boldsymbol{k})$ are the retarded and lesser many-body 
self-energies. These account for the effects due to the electron-electron interaction and therefore 
for the physics beyond the single-particle description. The Hamiltonian, the orbital overlap, the 
Green's functions, and the self-energies are square matrices of dimension equal to the number of 
basis orbitals in the central region. In principle, all of these matrices, except for 
$S(\boldsymbol{k})$, depend on the bias voltage, $V$, in addition to energy and spin. However, 
we have omitted the explicit $V$ dependence to keep the notation short.

Equation~(\ref{Eq: mb_retarded_gf}) is the Dyson equation. The retarded and advanced Green's 
functions, $G^{\sigma \; r}(E, \boldsymbol{k})$ and $G^{\sigma \; a}(E, \boldsymbol{k}) $, describe 
the energy levels in the central region, given by the spectral function, 
\begin{equation}
A(E, \boldsymbol{k})=i [G^{\sigma \; r}(E,\boldsymbol{k})-G^{\sigma \; a}(E, \boldsymbol{k})]\:.
\end{equation}
Then, the integral of $A(E, \boldsymbol{k})$ over the transverse wave-number, $\boldsymbol{k}$, gives the DOS. 

Equation~(\ref{Eq: mb_lesser_gf}) is sometimes called the Keldysh's equation. It describes the 
electron occupation of the central region. In particular, the integral of 
$G^{\sigma \; <}(E, \boldsymbol{k})$ over the energy and wave-number gives the density 
matrix,
\begin{equation}
 \rho^\sigma=\frac{1}{2\pi i}\frac{1}{\Omega_\mathbf{k}}\int d\boldsymbol{k}\int_{-\infty}^{\infty} dE\; G^{\sigma \;<}(E,\boldsymbol{k}),\label{eq.DFT_rho}
\end{equation}
where $\Omega_\mathbf{k}$ is the volume of the transverse Brillouin zone.

The retarded leads' self-energies can be calculated exactly, for example, through 
semi-analytical \cite{SS99,ru.sa.08} or iterative algorithms \cite{Sancho_1984}, since the electrons 
in the leads are assumed to be non-interacting. The self-energy, 
$\Sigma^{\sigma \; r}_\mathrm{L(R)}(E, \boldsymbol{k})$, is not Hermitian. Its anti-hermitian part,
\begin{equation}
\Gamma^{\sigma}_\mathrm{L(R)}(E, \boldsymbol{k}) = i[\Sigma^{\sigma \; r}_\mathrm{L(R)}(E, \boldsymbol{k}) - \Sigma^{\sigma \;r}_\mathrm{L(R)}(E, \boldsymbol{k})^\dagger],
\end{equation}
represents the electronic coupling of the central region to the left-hand (right-hand) side lead and
induces a broadening of the spectral function. The lesser leads' self-energies can be expressed 
according to the fluctuation-dissipation theorem \cite{bookStefanucci}, since the leads act as 
baths in local thermal equilibrium, as explained in section \ref{sec:system set}. We then write
\begin{equation}
\Sigma^{\sigma \; <}_{L(R)}(E, \boldsymbol{k})=if_\mathrm{L(R)}(E,\mu_\mathrm{L(R)})\Gamma^{\sigma}_\mathrm{L(R)}(E, \boldsymbol{k}),\label{eq.lead_sigma_lesser}
\end{equation}
for the left-hand (right-hand) side lead with Fermi function, $f_{L(R)}(E, \mu_\mathrm{L(R)})$. 

The shift of the leads' band structure in finite-bias calculations is achieved by setting 
\begin{equation}
\Sigma^{\sigma \; r}_\mathrm{L(R)}(E\pm eV/2,\boldsymbol{k},V)=\Sigma^{\sigma \; r}_\mathrm{L(R)}(E,\boldsymbol{k},V=0),\label{eq.SigmaLead_shift}
\end{equation}
where we have now made explicit the dependence on $V$ for clarity. With these boundary conditions, 
the electrostatic potential drop inside the central region is calculated by solving the Poisson equation 
for the charge density. In principle, Eqs. (\ref{Eq: mb_retarded_gf}), (\ref{Eq: mb_lesser_gf}) 
and (\ref{eq.DFT_rho}) need to be solved self-consistently, with the new electrostatic potential 
recalculated at each iteration \cite{sanvito}. However, in this paper, finite bias calculations 
are performed non-self-consistently by applying the rigid shift approximation 
(see Sec. \ref{sec.RSA}).

In contrast to the leads' self-energies, the lesser and retarded many-body self-energies, $\Sigma_{\textnormal{MB}}^{\sigma \; r}(E, \boldsymbol{k})$ and 
$\Sigma_{\textnormal{MB}}^{\sigma \; <}(E, \boldsymbol{k})$, are not easily calculated. Furthermore, 
in finite bias calculations, they do generally not satisfy a relation similar to 
Eq.~(\ref{eq.lead_sigma_lesser}), since the central region is out-of-equilibrium, and electrons 
are not distributed according to the Fermi-Dirac distribution. Therefore, suitable approximations 
are required. The simplest approach is to combine NEGF with DFT, known as the DFT+NEGF approach. 
A more advanced, and less explored, method is to further merge DFT+NEGF with DMFT.

\subsection{DFT+NEGF combined with DMFT}
In DFT+NEGF, the single-particle Hamiltonian $ H^{\sigma}(\boldsymbol{k})$ in 
Eq.~(\ref{Eq: mb_retarded_gf}) is replaced with the KS Hamiltonian, usually within 
the LSDA or the GGA, while the retarded and lesser many-body self-energies are neglected 
\cite{Ba.Mo.02,ro.ga.06,book1}. In this sense, DFT is used as an effective single-particle
theory with the exchange-correlation potential playing the role of a static (i.e. energy
independent) mean-field potential. To emphasis this approximation, in DFT+NEGF, the 
Green's functions are often called KS Green's functions. 

When merging DFT+NEGF with DMFT, the lesser and retarded self-energies are no longer 
neglected. However, the electron-electron interaction, beyond the effective KS potential, 
is included only inside a subspace, $\mathcal{C}$, of the central region 
\cite{ja.so.13, andrea_Cu_co}, where electron correlation is treated at the dynamical level, although its spatially non-local nature is neglected \cite{ko.sa.06,held.07}. In this work, such a correlated subspace comprises the five $3d$ 
orbitals of Fe with local average Coulomb interactions, $U$ and $J$. It therefore has a 
dimension of $2\times 5\times N$, where $N$ is the number of Fe atoms, for example, $N=6$ for the system 
of Fig. \ref{fig:central_region}, and $2$ accounts for the spin degree of freedom. 

The separation of $\mathcal{C}$ from the rest of the central region is achieved via 
a projection, or ``down-folding'', scheme which also orthogonalizes the orbitals
\cite{andrea_ivan_projection}. In practice, DFT+NEGF provides the material dependent
parameters, which describe the effective system within the subspace $\mathcal{C}$, while 
DMFT solves the many-body problem, obtaining the corresponding self-energies. In particular,
for quantum transport, one uses the layer formulation of DMFT described, for instance, 
in reference \cite{va.sa.12} and \cite{ok.07} for the zero- and finite-bias cases, 
respectively. 

Since DMFT assumes electron correlation to be local in space, the many-body retarded 
(lesser) self-energy matrix of the correlated subspace $\mathcal{C}$ is momentum 
independent and has a block diagonal form \cite{andrea_sigma_2, andrea_Cu_co}
\begin{equation}
 \tilde{\Sigma}^{\sigma \;  r(<)}_{\mathcal{C}}(E)= 
 \left( \begin{array}{cccc}
 \tilde{\Sigma}^{\sigma\;  r(<)}_1(E) &   0 & ... & 0          \\
 0 & \tilde{\Sigma}^{\sigma\;  r(<)}_2(E) & ... & 0           \\
 0 & 0 & ... & \tilde{\Sigma}^{\sigma\;  r(<)}_{N}(E)            \\
\end{array} \right),
\label{Sigma_local}
\end{equation}

where $\tilde{\Sigma}^{\sigma\;  r(<)}_i(E)$ is the $5 \times 5$ block for the $3d$ orbitals 
of the $i$-th Fe atom, and the ``0'' in this matrix represents a  corresponding $5 \times 5$ block of zeroes. In practice, $\tilde{\Sigma}^{\sigma\;  r(<)}_{\mathcal{C}}(E)$ is
computed by mapping the correlated subspace into a set of auxiliary impurity problems, one 
for each atom $i$, and imposing a self-consistent condition. In this work, each impurity
problem is solved as explained in section \ref{Sec.impurity}. 

The superscript `tilde' was added in Eq.~(\ref{Sigma_local}) to further stress that 
the self-energy $\tilde{\Sigma}^{\sigma\;  r(<)}_{\mathcal{C}}(E)$ is for the  subspace 
$\mathcal{C}$. In contrast, $\Sigma_{\textnormal{MB}}^{\sigma \; \;  r(<)}(E, \boldsymbol{k})$
in Eq.~(\ref{Eq: mb_retarded_gf}) and Eq.~(\ref{Eq: mb_lesser_gf}) is obtained by `upfolding' 
the correlated subspace back to the full central region space
\footnote{In our DFT+DMFT implementation, the upfolded DMFT self-energies acquire a 
$\boldsymbol{k}$-dependence because the transformation matrices are 
$\boldsymbol{k}$-dependent \cite{andrea_ivan_projection,  andrea_Cu_co,andrea_sigma_2}.}. 
The mathematical details of this operation are presented in 
references~\cite{andrea_ivan_projection,andrea_Cu_co}. After computing the self-energy, we can 
finally solve the NEGF equations, obtaining the charge density to be used to determine the 
potential drop and the DFT exchange-correlation potential in a self-consistent fashion, as 
outlined in the previous subsection. In this paper, however, we bypass this self-consistent 
procedure by applying the rigid shift approximation, as described in the next section.

\subsection{Rigid shift approximation for the bias voltage}\label{sec.RSA}

DMFT calculations at finite-bias are challenging due to the lack of efficient impurity 
solvers for out-of-equilibrium systems and the substantial computational overhead that 
would be required to determine the potential drop self-consistently. To our knowledge, 
finite-bias DMFT calculations have only been performed for one-orbital models, and not 
in charge self-consistent fashion \cite{ok.07,ok.08}. However, it is important to note 
that DMFT calculations can be significantly simplified in the case of Fe/MgO/Fe MTJs. 
Several works \cite{zhang12,xie2016spin} have shown that the electrostatic potential under 
an applied bias is trivial (see also Sec. S2 in the SI). More specifically, it remains 
flat in the metals due to near complete screening, and it varies linearly across the MgO 
insulating barrier, since charging effects are minimal. As a consequence, instead of computing 
the potential self-consistently, we are justified to use the KS Hamiltonian of the central 
region calculated at zero bias and apply a rigid energy shift to the leads' band structure 
and chemical potentials, bridged by an ideal linear drop inside the MgO barrier, as 
illustrated in Fig.~\ref{fig:potential_drop}. In practice, we add a ramp-like
electrostatic potential $V_\mathrm{e}(z)=-eVz/l+eV/2$ in the central region, where
$z=0$ $(z=l)$ is the position of left-hand (right-hand) side Fe/MgO interface, and $l$ 
is the thickness of the MgO barrier, as shown in Fig. ~\ref{fig:potential_drop}.
This approach implements the rigid shift approximation 
\cite{PhysRevB.72.180406,Rudnev_Sci_Adv2017}. 

Within the rigid shift approximation, the Fe layers on either side of the junction are in
equilibrium with the relevant lead. Thus, the DMFT self-energies in Eq.~(\ref{Sigma_local})
can be computed by performing `equilibrium' DMFT calculations and by applying
fluctuation-dissipation theorem. In mathematical terms, each block of the retarded and 
lesser correlated subspace's self-energies in Eq.~(\ref{Sigma_local}) is obtained through
relations similar to Eq.~(\ref{eq.SigmaLead_shift}) and Eq.~(\ref{eq.lead_sigma_lesser}), 
namely 
\begin{eqnarray}
&\tilde{\Sigma}^{\sigma \; r}_{i}(E\pm eV/2,V)=\tilde{\Sigma}^{\sigma \; r}_{i}(E,V=0)\:,\label{eq.SigmaDMFT_shift}\\ 
&\tilde{\Sigma}^{\sigma \; <}_{i}(E,V)=if_\mathrm{L(R)}(E,\mu_\mathrm{L(R)})\tilde{\Gamma}^{\sigma}_{i}(E,V)\:,\label{eq.sigmaDMFT_lesser}
\end{eqnarray}
where 
\begin{equation}
\tilde{\Gamma}^{\sigma}_{i}(E, V) = i[\tilde{\Sigma}^{\sigma \; r}_{i}(E, V) - \tilde{\Sigma}^{\sigma \;r}_{i}(E, V)^\dagger]\:,
\end{equation}
gives the effective broadening of the states in $\mathcal{C}$ due to the electron-electron
interaction. The plus (minus) sign in Eq.~(\ref{eq.SigmaDMFT_shift}) and the Fermi functions
$f_\mathrm{L(R)}$ in Eq.~(\ref{eq.sigmaDMFT_lesser}) are used when the index $i$ corresponds 
to a Fe atom on the left-hand (right-hand) side of the MgO barrier. By using 
Eqs.~(\ref{eq.SigmaDMFT_shift}) and (\ref{eq.sigmaDMFT_lesser}), we only need to compute 
$\tilde{\Sigma}^{\sigma \; r}_{i}(E,V=0)$ at zero-bias to obtain the lesser and retarded 
self-energies at any finite-bias voltage, avoiding challenging out-of-equilibrium many-body
calculations. Notably, a similar approach was also proposed to simulate transport through 
correlated molecules in a scanning tunneling microscope setup \cite{ja.ku.18,ja.18}.

In practice, we find that a straightforward application of Eq.~(\ref{eq.SigmaDMFT_shift}) 
leads to very poor current conservation (see the next subsection), partly because it misses 
an important electronic structure effect. In fact, when the linear potential is applied between 
the Fe layers, the energy level alignment across the junction changes, thereby modifying the
hybridization between the Fe and MgO orbitals, as well as the Fe orbitals on either side 
of the junction. In order to account for this electronic effect, we find it convenient to 
calculate $\tilde{\Sigma}^{\sigma \; r}_{i}(E,V)$ in an equilibrium calculation, but with 
the ramp potential $V_\mathrm{e}$ added to the zero-bias KS Hamiltonian of the central 
region. After obtaining $\tilde{\Sigma}^{\sigma \; r}_{i}(E,V)$ in this way, we use 
Eq.~(\ref{eq.sigmaDMFT_lesser}) to obtain the corresponding lesser DMFT self-energy. The 
improvement arising from this approach is verified by noting that it significantly enhances 
current conservation compared to the application of Eq.~(\ref{eq.SigmaDMFT_shift}), although 
some issues, which we believe are mostly numerical, remain (see the following subsection 
and, moreover, Section S6 in the SI).

\subsection{Charge current}\label{sec.theory.current}
An expression for the charge current in terms of the NEGF was derived by Meir and Wingreen 
in a seminal paper \cite{MW_current}. Specifically, they demonstrated that the current between 
the central region and the left-hand (right-hand) side lead reads
\begin{widetext}
\begin{equation}
    I_\mathrm{L(R)} = 
    \frac{ie}{h}\frac{1}{\Omega_{\boldsymbol{k}}}
   \sum_\sigma \int  d\boldsymbol{k} \int_{-\infty}^{\infty} 
    dE\,\textnormal{Tr}\Big[f_\mathrm{L(R)}(E, \mu_{L(R)})\Gamma^\sigma_\mathrm{L(R)}(E,\boldsymbol{k}) 
    A^\sigma(E,\boldsymbol{k}) 
   + i \Gamma^\sigma_\mathrm{L(R)}(E,\boldsymbol{k})G^{\sigma\,<}(E,\boldsymbol{k})\Big],\label{eq.IL_MW}
\end{equation}
\end{widetext}
%
and it is therefore determined by the coupling between the conductor and the lead, 
described by $\Gamma^\sigma_\mathrm{L(R)}(E,\boldsymbol{k})$, the energy levels in the 
conductor, given by the spectral function  $A^\sigma(E,\boldsymbol{k})$, and the occupations 
of energy levels in the lead and in the conductor, $f_\mathrm{L(R)}(E, \mu_{L(R)})$ and
$G^{\sigma\,<}(E,\boldsymbol{k})$.

Since, in the steady-state, the current is conserved, we expect that $I_\mathrm{L}=-I_\mathrm{R}$,
ensuring charge conservation. The current through the central region is then usually expressed 
in the symmetrized form, $I=(I_\mathrm{L}-I_\mathrm{R})/2$, referred to as the Meir-Wingreen formula. 
In this paper, we will also adopt that convention. However, we note that some care is needed in 
the calculations. Although DMFT is a conserving many-body theory in the Baym-Kadanoff sense \cite{ko.sa.06}, ensuring charge conservation is in practice a difficult task, as we have anticipated
above and further discuss in section S6 of the SI. This difficulty stems from two mains issues. 
The first is fundamental: we cannot guarantee that our computational implementation remains conserving after introducing the necessary approximations for treating our Fe/MgO/Fe MTJs. The second is instead computational: the current in the Meir-Wingreen formula arises from a very small difference (on the order of $10^{-9}$ to $10^{-12}$ A) between two large terms. For accurate results, each quantity involved in these terms must be computed with extremely high numerical precision, to a level that is practically unreachable.

The Meir-Wingreen formula is general and applies to both the interacting and non-interacting
case. However, in the latter situation, it can be greatly simplified. By setting the many-body 
self-energies equal to zero in Eqs.~(\ref{Eq: mb_retarded_gf}) and (\ref{Eq: mb_lesser_gf}), 
it is easy to show \cite{MW_current} that the current can be rewritten as
\begin{equation}
\begin{split}
    & I \equiv I_\mathrm{c} = \frac{e}{h} \sum_\sigma \int_{-\infty}^{\infty}\! \!dE \;
 T^\sigma(E) \big[f_\mathrm{L}(E,\mu_{L})-f_\mathrm{R}(E, \mu_{R})\big],\label{eq.Landauer}
\end{split}
\end{equation}
where
\begin{equation}\begin{split}
&T^\sigma(E)=\\ &\frac{1}{\Omega_\mathbf{k}}\int d\boldsymbol{k}\,  \textnormal{Tr}[\Gamma^{\sigma}_{L}(E, \boldsymbol{k})  G^{\sigma\; r}(E, \boldsymbol{k})\Gamma^{\sigma}_{R}(E, \boldsymbol{k})G^{\sigma\; a}(E, \boldsymbol{k})],
\end{split}\label{eq.TRC}
\end{equation}
is the transmission coefficient. Equation~(\ref{eq.Landauer}) is the Laudauer-B\"uttiker 
formula generalized to finite bias (see, for example, reference \cite{datta}) and commonly applied 
in DFT+NEGF calculations (e.g. \cite{sanvito, ro.ga.06,rungger_2009}). The current is due 
to the coherent transmission of electrons through the central region and is determined by 
the occupation of the incoming and out-coming states, $f_\mathrm{L}$ and $f_\mathrm{R}$. 
The integrand in Eq.~(\ref{eq.Landauer}) is significantly different from zero only within 
an energy range approximately comprised between $E_\mathrm{F}-eV/2$ and $E+eV/2$, which is 
called the bias window. Only electrons occupying states within the bias window and propagating
across the central region contribute to the transport. In practical terms, the coherent current 
can be quickly estimated by looking at the area under the $T(E)$ curve within the bias window.

In the case of interacting systems, the Meir-Wingreen formula no longer reduces to 
Eq.~(\ref{eq.Landauer}). In fact, the current is no longer carried solely by electrons 
coherently transmitted from one lead to the other. The difference between the current 
obtained from the Meir-Wingreen formula and that computed with the Landauer-B\"uttiker formula 
defines the incoherent contribution to the current \cite{andrea_ivan_projection,ne.da.10} 
due to inelastic electron-electron scattering. This non-coherent contribution to transport 
is often neglected in studies of real material systems. In this paper, we move beyond this
limitation by applying the Meir-Wingreen formula to fully account for the both the coherent 
and non-coherent contributions to the current, comparing their relative importance 
(see sections \ref{sec.coherent_current} and \ref{sec.incoherent_current}).

\subsection{Solution of the impurity problem}\label{Sec.impurity}
The solution of the auxiliary impurity problem in DMFT for transport is a difficult task, 
even assuming the rigid shift approximation, which effectively allows us to use equilibrium 
solvers in finite-bias calculations. 

The most established (equilibrium) impurity solvers, such as continuous-time quantum 
Monte Carlo \cite{gu.mi.11} and the spin-polarized $T$-matrix fluctuating exchange 
approximation \cite{li.ka.98,ka.li.99,ka.li.02,po.ka.06}, are formulated on the imaginary 
frequency axis. The spectral functions are then obtained via analytical continuation 
\cite{ja.gu.96,sa.98,mi.pr.20,fu.pr.10}. However, analytical continuation of discrete 
numerical data is an ambiguous mathematical operation, making practical calculations 
problematic. Peaks in the spectral functions are often artificiality smeared out over 
too large energy intervals so that fine details can not be resolved. In the context of 
transport, this makes an accurate evaluation of the charge current, via Eq.~(\ref{eq.IL_MW}),
impossible. For this reason, here we consider the second-order perturbative treatment 
proposed in reference~\cite{andrea_sigma_2}, which was implemented to provide the self-energy 
directly on the real energy axis. In spite of its simplicity, it was shown  
\cite{andrea_sigma_2, andrea_FeO} that the second-order self-energy already accounts 
for all characteristic spectroscopic features due to electron correlation in ferromagnetic 
$3d$ metals.

\section{Computational Details}\label{section: computational details}
The calculations are performed by using the {\sc Smeagol} code 
\cite{ro.ga.06, book1}, which interfaces the NEGF scheme and DMFT 
\cite{dr.ru.17,andrea_sigma_2,andrea_Cu_co} with an in-house development version 
of the {\sc Siesta} DFT package \cite{so.ar.02}.

The DFT+NEGF calculations assume the local spin density approximation (LSDA)
\cite{ba.he.72,vo.wi.80} for the exchange-correlation functional. The leads' self-energies 
are calculated using the semi-analytic algorithm from reference~\cite{ru.sa.08}. 
Norm-conserving Troullier-Martins pseudopotentials \cite{Tr.Ma.91} are employed to treat 
the core electrons. Valence states are expanded using a numerical atomic orbital basis set, 
which includes multiple-$\zeta$ and polarized functions \cite{so.ar.02}. The electronic 
temperature is set to 300 K, and the real space mesh is determined by an equivalent energy 
cutoff of 300 Ry. 

The equilibrium DFT density matrix, Eq.~(\ref{eq.DFT_rho}), is computed self consistently 
using a $20 \times 20$ transverse $\boldsymbol{k}$-point mesh. The integration over energy 
is performed using a semi-circular contour in the complex energy plane \cite{ro.ga.06}. In 
this approach, 16 poles are used to represent the Fermi distribution, and 16 energy points 
are sampled along both the semi-circular arc and the imaginary line that form the contour. 

The converged density matrix is read as an input for a non-self-consistent DFT calculation 
using a $200 \times 200$ transverse $\boldsymbol{k}$-mesh to obtain the zero-bias DFT DOS 
and transmission coefficient. All energy values are shifted to align the Fermi level at 0 eV.

Finite-bias calculations are performed non-self-consistently, using the Hamiltonian of the 
central region calculated at zero-bias with an added ramp potential \cite{Rudnev_Sci_Adv2017} 
that reproduces the linear voltage drop inside the MgO barrier, as described in 
section~\ref{sec.RSA}. Similar to the equilibrium case, a $200 \times 200$ transverse 
$\boldsymbol{k}$-mesh is used to obtain the finite bias DFT DOS and transmission coefficient.

DMFT calculations are performed assuming general orbital-dependent screened Coulomb 
interaction parameters for the $3d$ orbitals within each Fe atom. These parameters are 
expressed in terms of Slater integrals $F^0$, $F^2$ and $F^4$ \cite{im.fu.98}. 
The ratio $F^4/F^2$ is assumed to correspond to the atomic value $\approx 0.625$ (see
\cite{an.gu.91}. The average $U$ and $J$ interaction parameters are given by 
the relations $U=F^0$ and $J=(F^2+F^4)/14$.

The local Green's function in the DMFT loop \cite{andrea_Cu_co} is calculated by summing 
the retarded Green's function over $16 \times 16$ $\boldsymbol{k}$ points. This $\boldsymbol{k}$-mesh is then used to plot the DMFT DOS and transmission coefficient. 

In the perturbative impurity solver, we split the calculation of the first- and second-order
contributions to the self-energy as explained in reference~\cite{andrea_sigma_2}. The first-order
contribution accounts for static (energy-independent) mean-field corrections to the LSDA KS
Hamiltonian and is approximated with the $U$-potential by Dudarev {\it et al.} \cite{dudarev}. 
The second-order contribution introduces dynamic correlation. It is computed on an energy grid
comprising $3800$ points and extending from $-18$~eV to $8$~eV at low bias. As the bias increases,
the energy grid is progressively expanded to a maximum of $5525$ points, covering the range 
[-26, 12]~eV to account for the shifting of the Fe states towards higher and lower energies 
in the left-hand side and right-hand side leads, respectively.  

\section{Results}\label{sec.results}
\begin{figure}[b]
\centering\includegraphics[width=0.48\textwidth]{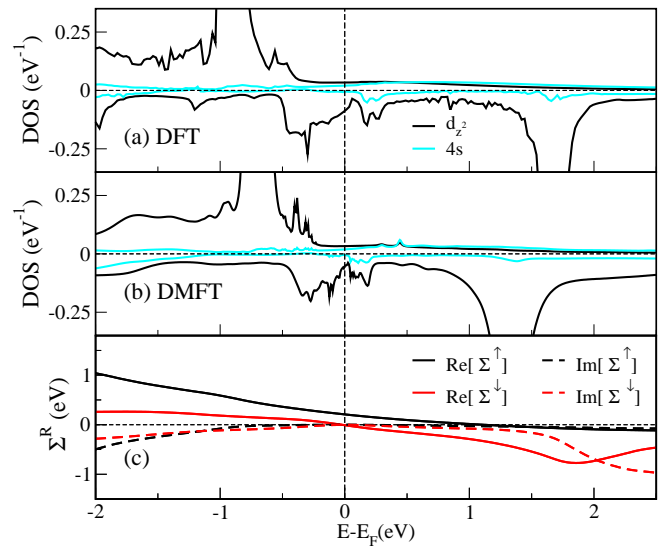}
\caption{Electronic structure of the Fe atom placed at the left-hand side Fe/MgO interface. 
(a) DFT- and (b) DMFT-calculated DOS projected onto the $3d_{z^2}$ and $4s$ 
Fe orbitals. Positive (negative) values are for spin-up (down) states. (c) Spin-up and spin-down 
retarded self energy for the $3d_{z^2}$ Fe orbital.}
\label{fig:self_energy_dos}
\end{figure}

\subsection{Zero-bias transport}\label{sec.zero_bias}

The tunneling through an Fe/MgO/Fe junction is dominated by spin-split evanescent states with 
$\Delta_1$ symmetry, which have the slowlest decay rate inside the MgO barrier \cite{Bu_Zh+2001, Bu_2008}.
Within our reference frame, where the transport direction is aligned along the $z$ axis 
[see Fig.~\ref{fig:central_region}], and for a cubic space group, the $\Delta_1$ states are 
those that transform as a linear combination of $s$, $p_{z}$, and $d_{z^2}$ atomic orbitals. Therefore, 
the spin-filtering properties of the junction can be understood through an analysis focusing 
primarily on these orbitals, while neglecting all the others \cite{Bu_2008}, as done in the following. 

\subsubsection{Fe electronic structure}

\begin{figure}[b]
\centering\includegraphics[width=0.48\textwidth]{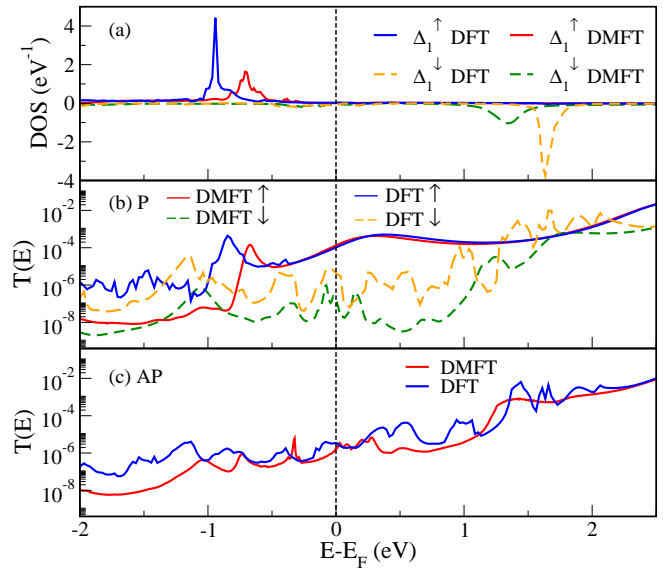}
\caption{Zero-bias transmission properties of the Fe/MgO/Fe MTJ. (a) $\Delta_{1}^\sigma$-PDOS of 
the Fe atom at the left-hand side Fe/MgO interface, calculated by DFT and DMFT. (b) Spin-up and 
spin-down transmission coefficient for the P configuration. (c) Spin-up and spin-down transmission 
coefficient for the AP configuration. Note that in panels (a) and (b), the transmission is on a
logarithmic scale.} 
\label{fig:trc_dmft_Fe_MgO_equilibirum}
\end{figure}

The spin-polarized DOS projected onto the $4s$ and $3d_{z^2}$ orbitals of the Fe atom at the 
left-hand side Fe/MgO interface, as calculated by DFT and DMFT, are presented in panels (a) and (b) 
of Fig.~\ref{fig:self_energy_dos}, respectively. The same results, namely the sum of the $4s$ and
$3d_{z^2}$-PDOS, collectively called $\Delta_1^\sigma$-PDOS (with $\sigma$ denoting the spin, 
$\sigma=\uparrow, \downarrow$), are also presented in Fig.~\ref{fig:trc_dmft_Fe_MgO_equilibirum}(a) 
on a different scale. This offers a clearer display of the height of the main peaks, while forgoing 
the fine details. The Fe $4p_z$-PDOS is neglected in our analysis as this state is significantly 
higher in energy than the Fe-$4s$ and $3d$ ones. In the junction's P configuration, the PDOS of the 
right interface mirrors that of the left interface due to the junction's symmetry. In contrast, in 
the AP configuration, the spin-up and spin-down PDOS of the right interface are swapped. All DFT 
calculations are carried out within the LSDA, while the DMFT ones are performed using the average screened 
interaction parameters of $U =2.0$ eV and $J=0.5$ eV, which were found appropriate to describe Fe 
nanostructures in previous studies using our implementation of DFT+DMFT \cite{andrea_sigma_2}. Nonetheless, for a more complete analysis, 
we further discuss the dependence of our results on varying the interaction parameters in section S3 
of the SI.

DFT [see Fig.~\ref{fig:self_energy_dos}(a)] returns us a picture where the spin-up and spin-down 
$3d_{z^2}$ Fe states are split in accordance to the Stoner picture of ferromagnetism. Their PDOS 
are characterized by sharp peaks located at $E-E_\mathrm{F}\sim -0.9$ eV and $\sim 1.6 $ eV, 
respectively. In addition, in the spin-down $3d_{z^2}$-PDOS, there are some extra features around 
the Fermi energy, which are associated to interface states, as already reported in other DFT studies 
for Fe/MgO/Fe junctions \cite{Ivan_Stefano_2009, FeMgO_GW}. These interface states were found to 
significantly impact the tunneling current for thin barriers (less than four MgO layers)
\cite{Ivan_Stefano_2009}. However, for the six-layer barrier considered here, they become less 
important for transport.  

The $4s$ orbital is strongly hybridized with the $3d_{z^2}$ one and it is then spin split accordingly. 
As seen in Fig. \ref{fig:self_energy_dos}(a), the spin-up and spin-down $4s$-PDOS begin at the same 
energy positions as the spin-up and spin-down $3d_{z^2}$-states and have maxima at 
$E-E_\mathrm{F}\sim 0.2$ eV and $\sim 2.8 $ eV, respectively. These states are very broad, extending 
towards high energies above $E_\mathrm{F}$.

Compared to DFT, DMFT gives a significant broadening and shift in energy of the $3d_{z^2}$-PDOS for 
both spin channels, resulting in a reduction of the spin splitting, as shown in Figs.~\ref{fig:self_energy_dos} 
and \ref{fig:trc_dmft_Fe_MgO_equilibirum}. These features arising from dynamical electron correlation in transition 
metal ferromagnets have already been reported in numerous studies 
\cite{br.mi.06, gr.ma.07}, showing that DFT+DMFT provides an accurate description 
of electronic spectra, closely matching experimental results \cite{andrea_sigma_2, ja.dr.23}. In this regard, it is important to note that including only static correlation, for example, by means of the so-called DFT+$U$ scheme \cite{an.za.91,li.an.95,du.bo.98,co.gi.05}, would have the opposite effect, namely, it would increase the spin splitting between the $3d$ states \cite{co.gi.05,andrea_Cu_co,andrea_sigma_2}, thus worsening the agreement between the calculated Fe electronic structure and experiments \cite{andrea_sigma_2}. As such, DFT+$U$ is not appropriate for $3d$ ferromagnets, and we will not consider it in this work.

The energy shift of the states is caused by the DMFT retarded self-energy, specifically by its real part, presented 
in Fig.~\ref{fig:self_energy_dos}(c). On the one hand, since $\textnormal{Re}[\Sigma^{\uparrow}(E)]$ is 
positive for $E<E_\mathrm{F}$, the spin-up $3d_{z^2}$ state is pushed towards higher energies. On the 
other hand, since $\textnormal{Re}[\Sigma^{\downarrow}(E)]$ is negative for $E > E_F$, the spin-down 
$3d_{z^2}$ state is pulled down in energy towards the Fermi level. As a result, DMFT places the
$3d_{z^2}$-PDOS peaks for spin-up and spin-down at around $E-E_\mathrm{F}\sim -0.7$~eV and 
$\sim 1.3 $~eV, respectively. 

The DMFT broadening and smoothing of the $3d_{z^2}$-PDOS is due to the imaginary part of the DMFT 
self-energy, plotted as the dashed line in Fig. \ref{fig:self_energy_dos}(c). This imaginary part 
is typical of a Fermi-liquid, namely $\textnormal{Im}[\Sigma^{\sigma}(E)]\propto -(E-E_\mathrm{F})^2$ 
for both spin channels, as expected for $3d$ ferromagnetic transition metals \cite{gr.ma.07}. It accounts for the fact that, within DMFT, the Fe electronic excitations 
have quasi-particles character with a finite lifetime, $\tau\propto-\hbar/\textnormal{Im}[\Sigma^{\sigma}(E)]$, which is reduced with increasing energies both below and above $E_\mathrm{F}$. 

\subsubsection{Spin-dependent tunneling}\label{sec.tunneling}

The tunneling through an Fe/MgO/Fe junction is quantitatively described by the spin-polarized
transmission coefficient, as defined in Eq.~(\ref{eq.TRC}). Specifically, one finds that the 
transmission coefficient is enhanced at energies where the $\Delta_1^\sigma$ states of the 
left-hand side Fe layers are resonant with the corresponding states at right-hand side, as 
incoming electrons can find channels to tunnel into \cite{ju.75}.

In the P configuration, the resonant condition is exactly satisfied. The transmission coefficient, 
$T_\mathrm{P}^\sigma(E)$, which is shown in  \ref{fig:trc_dmft_Fe_MgO_equilibirum}(b), reflects 
the Fe $\Delta^\sigma_1$-PDOS. Within DFT, in the spin-up channel, $T_\mathrm{P}^\uparrow$ shows 
a sharp increase of about two orders of magnitude (note the logarithmic scale), with a peak
located at $E-E_\mathrm{F}\sim -0.9$ eV, corresponding to the maximum spin-up $3d_{z^2}$-PDOS. Such 
peak marks the onset of a high-transmission region, where $T^\uparrow\sim 10^{-4}$. This region 
appears as a wide plateau, which spans across $E_\mathrm{F}$ and extends over several eV. It is associated to the $\Delta_1^\uparrow$ states around $\boldsymbol{k}=(0,0)$ \cite{Bu_Zh+2001, Bu_2008, Ivan_Stefano_2009}, with an $s$ character which increases with the energy at the expense of the $3d_{z^2}$ one, as evident from the PDOS (see also section S4 of the SI). 

Conversely, in the spin-down channel, $T_\mathrm{P}^\downarrow$ 
shows a less steep increase, it reaches a maximum followed by a plateau well above $E_\mathrm{F}$, 
namely at $E-E_\mathrm{F}\sim 1.6$ eV. This corresponds to the energy of the spin-down $3d_{z^2}$ 
state. Consequently at $E_\mathrm{F}$, the spin-down transmission coefficient is low and negligible 
compared to the spin-up counterpart.
As is well-established, the unique spin-filtering behavior of Fe/MgO/Fe junctions arises from the $\Delta_1^\sigma$ states, which are only present at $E_\mathrm{F}$ in the the spin-up 
channel \cite{bu.zh.95}.

The DFT+NEGF results for our system, which consists of a finite number of Fe layers attached to model
electrodes, are qualitatively and even quantitatively similar to those of junctions with semi-infinite 
Fe electrodes studied in the literature \cite{Ivan_Stefano_2009}. The main difference is that in the 
semi-infinite case, the spin-down transmission coefficient is somewhat smoother due to the absence of 
quantum confinement effects, as shown in Section S1 of the SI. 

When including dynamical correlations, the peaks in both $T_\mathrm{P}^\uparrow$ and 
$T_\mathrm{P}^\downarrow$ are shifted in energy compared to DFT, reflecting the shift of the 
$3d_{z^2}$-PDOS induced by the real part of the DMFT self-energy. More in detail, we find that 
the transmission maximum in the spin-up channel is now at $E-E_\mathrm{F}\sim -0.7$. Nonetheless, beyond 
$E - E_\mathrm{F} \sim -0.4$ eV, the high-transmission plateau remains identical to that found in
DFT. This is because the DMFT self-energy is small near $E_\mathrm{F}$ (effectively vanishing owing to the Fermi liquid character of the Fe layers), and as one moves to higher energies, the $4s$ component of the Fe $\Delta_1^\uparrow$ state becomes more prominent.

In the spin-down channel, DMFT leads to a smoothing of $T^\downarrow_\mathrm{P}$, reflecting 
the broadening of the PDOS. We observe the onset of the high-transmission plateau at 
$E - E_\mathrm{F} \sim 1.5$ eV, which becomes more distinct than in DFT. Similarly, we can also 
better recognize the peaks associated to the interface states in the energy region between 
$E-E_\mathrm{F}\sim -0.5$ and $\sim 0.2$ eV. 

Notably, both $T^\uparrow_\mathrm{P}$ and $T^\downarrow_\mathrm{P}$ are massively reduced, 
up to two orders magnitude, when compared to their DFT counterparts, for energies below the 
corresponding $\Delta_1^\uparrow$ and $\Delta_1^\downarrow$ band-edges. These are, respectively, 
at $E-E_\mathrm{F}< -1 $ eV and $< 1.5 $ eV. At these energies, one finds that the transmission 
is due to evanescent states with the main contribution from the Fe $d_{xz}$ and $d_{yz}$ orbitals, 
neglected so far in our description. These states are correlated and because of the imaginary 
part of the DMFT self-energy, they acquire an additional finite relaxation time, which always reduces the 
transmission coefficient \cite{andrea_Cu_co}.

\begin{figure*}
\centering\includegraphics[width=0.75\textwidth]{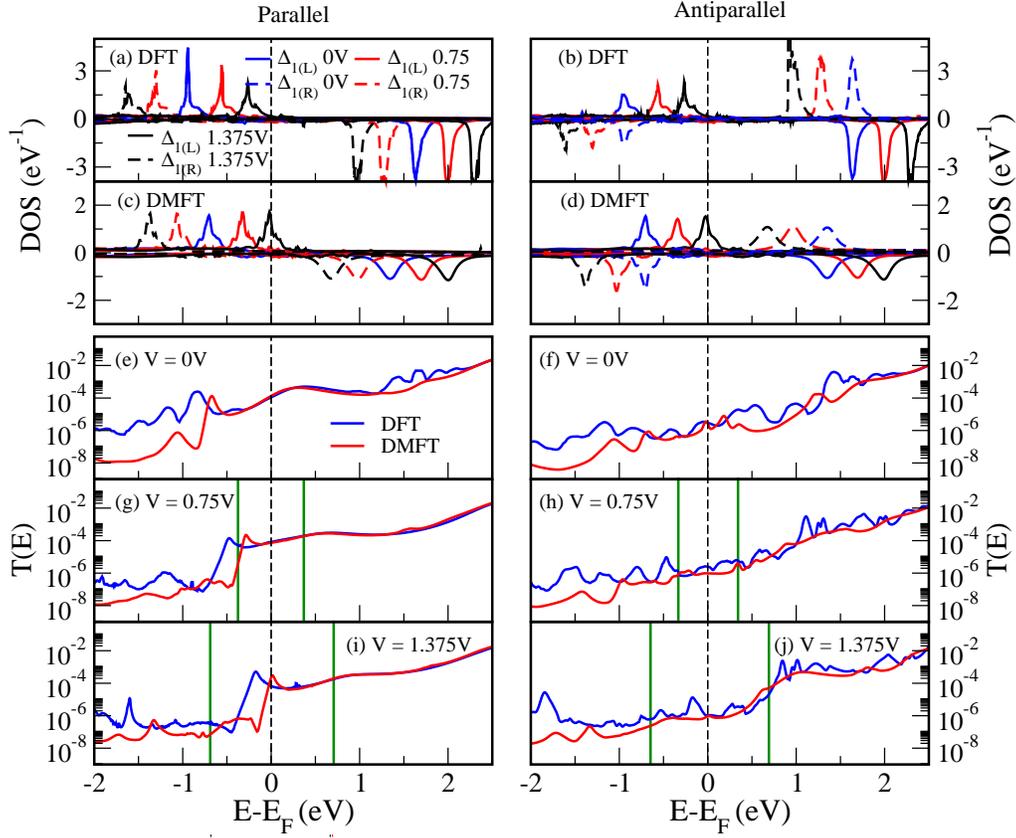}
\caption{Transport properties at finite bias for the Fe/MgO/Fe junction investigated. 
$\Delta^\sigma_{1(L/R)}$ states at the left-hand and right-hand side Fe/MgO interface 
in the P and AP configurations at $V=0$, $0.75$ and $1.375$ V, calculated by DFT [panels 
(a) and (b)] and DMFT [panels (c) and (d)]. Total DFT and DMFT transmission coefficient 
for the P and AP configurations for a bias of $V=0$ V [panels (e) and (f)], $V =0.75$ V 
[panels (g) and (h)] and  $V =1.375$ V [panels (i) and (j)]. The Green vertical bars in 
panels (g), (h), (i) and (j) delimit the bias window. } 
\label{fig: dos_trc_bias}
\end{figure*}

When we switch from the P to the AP configuration, the spin-up and spin-down states of the 
right-hand side Fe layers are exchanged with respect to those of the left-hand side 
(see Fig. \ref{fig: dos_trc_bias}). In first approximation, the AP transmission coefficient, shown in Fig.\ref{fig:trc_dmft_Fe_MgO_equilibirum}(c), 
can be written as a convolution of the spin-up and spin down P transmission coefficients, 
$T^\sigma_\mathrm{AP}\approx\sqrt{T^\uparrow_\mathrm{P}T^\downarrow_\mathrm{P}}$. Hence, $T^\sigma_\mathrm{AP}$ 
is much lower than $T^\uparrow_\mathrm{P}$ over a large energy range, and in particular at 
$E_\mathrm{F}$. In DFT, we find that the AP transmission coefficient at $E_\mathrm{F}$ is 
$\sim10^{-6}$, namely it is two orders of magnitude smaller than that of the P configuration. 
The onset of a high-transmission region is observed only at energies $E-E_\mathrm{F}\sim 1.4$~eV, 
where both $T^\uparrow_\mathrm{P}$ and $T^\downarrow_\mathrm{P}$ display the plateau. 

The difference between the DFT and the DMFT results for the AP configuration can be  
understood in terms of the differences between the DFT and DMFT $T^\sigma_\mathrm{P}$ owing to the approximate relation $T^\sigma_\mathrm{AP}\approx\sqrt{T^\uparrow_\mathrm{P}T^\downarrow_\mathrm{P}}$. In particular, 
we note that the DMFT $T^\sigma_\mathrm{AP}$ is smoother than the DFT one, due to the additional broadening 
of the PDOS, and that there is a plateau region beyond $E-E_\mathrm{F}\sim 1.4$~eV, whose onset 
and shape reflects that seen in $T^\downarrow_\mathrm{P}$. 

In summary, the main effect of dynamical electron correlation in Fe/MgO/Fe is to reduce the spin 
splitting of the $3d_{z^2}$ Fe states and to lower the  transmission through the other $3d$ states 
because of the real and imaginary part of the self-energy, respectively. Despite this, the zero-bias 
spin-dependent tunnelling through the junction is similar for electrons at $E_\mathrm{F}$.

\subsection{Finite bias transport}\label{sec.finite_bias}
The study of our Fe/MgO/Fe junction is here extended from zero- to finite-bias, allowing 
for the calculation of the $I/V$ characteristics for the P and AP configurations, and therefore, 
of the bias-dependent TMR.
As explained in Section \ref{sec.theory.current}, the current generally consists of two 
contributions: a coherent and a non-coherent current. The coherent current is expressed by 
the Landauer-B\"uttiker formula of Eq.~(\ref{eq.Landauer}). In contrast, the incoherent part 
is related to inelastic electron-electron scattering. This is absent in single-particle KS 
DFT, while in DMFT it is mathematically accounted for by the lesser DMFT self-energy \cite{dr.ru.17}. 
In the following, we start by analysing the junction's bias-dependent electronic structure 
and then proceed to estimate its coherent and non-coherent currents separately, assessing 
their relative importance for the overall transport properties of the system. 

\subsubsection{Electronic structure and transmission coefficient}

The finite-bias behavior of the junction is determined by the energy shift of the 
spin-polarized Fe $\Delta_1^\sigma$ state at one side of the MgO barrier relative to that 
at the other side, as a function of $V$ ~\cite{Ivan_Stefano_2009}. As mentioned in 
section \ref{sec.RSA}, the Fe state at the left-hand side Fe/MgO interface, $\Delta_{1(L)}^\sigma$, 
shifts upwards in energy by $\frac{eV}{2}$, while that at the right-hand side interface, 
$\Delta_{1(R)}^\sigma$, shifts downwards by $-\frac{eV}{2}$ compared to the zero-bias case. 
This is clearly observed in the PDOS calculated at $V=0$, $0.75$, and $1.375$ V presented in 
Fig.~\ref{fig: dos_trc_bias}(a) and Fig. \ref{fig: dos_trc_bias}(b) for the P and AP configurations,
respectively. The resulting change in energy level alignment across the junction causes the Fe 
states on either side of the barrier to move out of resonance, thereby suppressing tunnelling 
at energies where the left and right $\Delta_{1(L)}^\sigma$- and $\Delta_{1(R)}^\sigma$-PDOS no 
longer overlap. In practice, this implies that electrons spin-filtered by the left-hand side 
interface can not find states at the same energy on the right interface to tunnel into. This 
physical picture holds both within a single-particle framework and in presence of dynamical 
correlation, and it can be quantitatively analyzed in terms of the changes of the transmission 
coefficient with $V$, extending the discussion of section \ref{sec.tunneling} to finite-bias.

The total transmission coefficient, $T_\mathrm{P(AP)}=T_\mathrm{P(AP)}^\uparrow+T_\mathrm{P(AP)}^\downarrow$,
at $V=0$, $0.75$, and $1.375$ V is presented in Fig. \ref{fig: dos_trc_bias} panels (e) through (i) 
[(f) through (j)] for the P (AP) configuration, calculated using both DFT and DMFT. In the P 
configuration, the DFT main peak, which marks the onset of the high-transmission plateau, follows 
the spin-up Fe $d_{z^2}$ PDOS of the left-hand interface. Therefore, it 
shifts from $E-E_F\approx -0.8$ eV at $V=0$ V to $-0.45$ eV and $-0.15$ eV at $V=0.75$ V and 
$V=1.375$ V, respectively. In contrast, in the AP configuration, the peak follows the spin-up Fe $d_{z^2}$-PDOS of the right interface, shifting from $E-E_F\approx 1.4$ eV 
at $V=0$ V to $1.0$ eV and $0.8$ eV at $V=0.75$ V and $V=1.375$ V respectively.

In the case of DMFT, the bias-induced changes in the electronic structure are similar to those 
observed with DFT. Quantitatively, the differences between the DMFT and DFT transmission coefficients 
reflect the shift of the Fe $3d_{z^2}$ states induced by the real part of the self-energy. This 
is clearly seen by comparing Fig.~\ref{fig: dos_trc_bias}(a) with Fig.~\ref{fig: dos_trc_bias}(c) 
and Fig.~\ref{fig: dos_trc_bias}(b) with Fig.~\ref{fig: dos_trc_bias}(d). For the P configuration, 
the $\Delta_{1(L)}^\downarrow$ state is approximately 0.2~eV higher in energy (i.e., closer to 
$E_\mathrm{F}$) in DMFT compared to DFT, at any $V$. This results in the transmission peak to be 
accordingly moved up in energy by 0.2~eV relative to the DFT position. At the same time, for the AP 
configuration, the $\Delta^\uparrow_{1(R)}$ state, which lies above $E_\mathrm{F}$, is shifted 
down in energy by approximately 0.3 eV when compared to DFT, leading to a corresponding downward 
shift of AP transmission peak, at any $V$. The bias-induced changes in the electronic structure 
and transmission coefficients are reflected in the coherent current, since this is approximately 
equal to the area under the transmission coefficient-vs-energy curve inside the bias window. 

\begin{figure}[h]
\centering\includegraphics[width=0.48\textwidth]{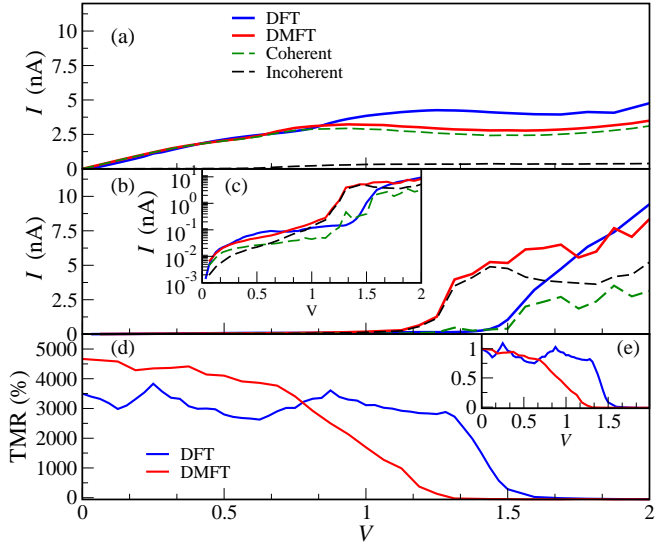}
\caption{$I$-$V$ curves and TMR for the Fe/MgO/Fe junction investigated. (a) DFT and DMFT current 
for the P configuration. (b) DFT and DMFT current for the AP configuration. (c) DFT and DMFT current 
for the AP configuration on a logarithmic scale. (d) Comparison of the DFT and DMFT TMR.
(e) Comparison of the normalised DFT and DMFT TMR.} 
\label{fig: current_Fe_MgO}
\end{figure}

\subsubsection{Coherent current}\label{sec.coherent_current}

The coherent current as a function of $V$ is plotted in Fig.~\ref{fig: current_Fe_MgO} [panels (a) and 
(b)] for the P and AP configurations. The DFT (DMFT) curve is represented by the blue continuous (green 
dashed) line. In the P configuration, at low bias, the coherent current increases linearly with $V$ as 
the bias window opens over the nearly featureless high-transmission plateau of $T_\mathrm{P}$ [see 
Fig. \ref{fig: dos_trc_bias}(g) where the bias window is delimited by the green vertical bars]. However, 
as the bias increases further, the $\Delta^\uparrow_{1(L)}$ edge, followed by the low-transmission 
region, eventually enters the bias window [see Fig.~\ref{fig: dos_trc_bias}(i)]. At this point, the 
coherent current tends to saturate.

The $I/V$ curves of the P configuration calculated by DFT and DMFT are similar at a qualitative level, and 
they resemble the curves reported in previous DFT+NEGF studies \cite{Ivan_Stefano_2009,PhysRevB.78.024430}.
However, quantitative differences emerge between the DFT and DMFT results. Firstly, the crossover 
between the linear and nearly flat regions of the $I/V$ occurs at $V\approx 1.2$ V in DFT and 
$V\approx 0.9$ V in DMFT. This is because, as discussed above, the DMFT $\Delta^\uparrow_{1(L)}$ edge, 
which determines the position of the transmission peak, is about 0.2 eV higher in energy than the DFT
one, and therefore enters the bias window at a lower $V$ [see, for instance, Fig. \ref{fig: dos_trc_bias}(g) 
where the $\Delta^\uparrow_{1(L)}$ edge is outside and inside the bias windows for DFT and DMFT, 
respectively]. Secondly, in the nearly flat $I/V$ region, the DMFT coherent current is reduced relative 
to the DFT one. This reduction is due to the transmission coefficient being smaller for energies below 
the $\Delta_{1(L)}^\uparrow$ edge in DMFT ($T_\mathrm{P}\sim 10^{-7}$) than in DFT 
($T_\mathrm{P}\sim 10^{-6}$) as the transport occurs through the Fe $d_{xz}$ and $d_{yz}$ states, 
which acquire a finite relaxation time because of the imaginary part of the many-body retarded self-energy.

In the AP configuration, the coherent current at low bias is three orders of magnitude smaller 
than the corresponding current in the P configuration. This is because the $T_\mathrm{AP}$ 
high-transmission plateau is located at high energies, outside the bias window, rather than 
around $E_\mathrm{F}$. The $I/V$ curve of the AP configuration remains linear up to relatively 
large biases. Then, the current starts rising sharply as the high-transmission plateau begins 
to enter the bias window [see Fig. \ref{fig: dos_trc_bias}(j)]. The crossover between linear 
and sharp rise of the $I/V$ curve occurs at different biases, $V\sim 1.4$ eV and $V\sim 1.1$ V, in DFT and DMFT, respectively, reflecting the different energy position of 
the $\Delta^\uparrow_{1(R)}$ state. Finally, when the bias increases further ($\gtrsim 1.7$ V), 
the $I/V$ curve returns linear as the high-transmission plateau is well inside the bias window.

In summary, the DFT and DMFT results for the coherent current are rather similar, with the main 
differences that can be traced back to the shift and broadening in energy of the Fe states induced 
by the dynamical correlation.

\subsubsection{Incoherent current}\label{sec.incoherent_current}
The DMFT incoherent current as a function of $V$ is plotted as a black dashed line in 
Fig.~\ref{fig: current_Fe_MgO}(a) and Fig.~\ref{fig: current_Fe_MgO}(b) for the P and AP 
configurations, respectively. 
In the P configuration, the incoherent current is negligible at low bias. The incoherent current starts increasing 
at the critical bias where the coherent current displays the crossover between the linear and 
constant behaviour, namely as the $d_{z^2}$ peak of the $\Delta^\uparrow_{1(L)}$-PDOS enters the bias window, followed by the Fe $d_{xz}$ and $d_{yz}$ states. However, 
the incoherent current remains a small fraction of the total current.

In contrast, in the AP configuration, the incoherent current is comparable to the coherent current 
already at low bias. Moreover, as the bias increases, a crossover occurs at $V\sim 0.6$ V, where the incoherent 
current surpasses the coherent one. Beyond this point, the incoherent current increases sharply 
as the peak of the $\Delta^\uparrow_{1(L)}$ state, with $d_{z^2}$ character, enters the bias window [see Fig. \ref{fig: dos_trc_bias}(d)]. In this situation, transport is through this correlated state and electron-electron scattering opens new channels for transport. 
Finally, at higher biases ($V\gtrsim 1$ V), the incoherent current remains nearly constant, as 
no additional correlated orbitals enter the bias window. The bias window instead opens to include the $s$ 
states, leading to a rapid increase of the coherent rather than incoherent current, as already 
discussed above.

The scattering processes leading to a sharp current increase can be understood as inelastic excitations, 
where electrons scatter off other electrons. This is an energy-losing process, which leads to the 
creation of particle-hole pairs with different spins, such as magnons and Stoner excitations. These 
processes are already captured by DMFT with our perturbative impurity solver \cite{andrea_sigma_2}
that, despite the locality approximation inherent in DMFT, effectively extends the many-body approaches
introduced to include electron-magnon interactions in itinerant ferromagnets \cite{Hertz_1973, Edwards_1973}.
Similar current-driven excitations have been studied in magnetic atoms and molecules (see for example \cite{ja.ma.2000, he.gu.04,hi.lu.06, ba.sc.09,lo.co.10, kh.lo.11,Rossier2009, hu.ba.11, Ternes_2015,ja.or.21,ga.de.17}), where they are measured in inelastic tunneling spectroscopy as 
steps in the differential conductance. However, to the best of our knowledge, these phenomena 
have not yet been extensively explored in solid-state MTJs so that our results could provide valuable 
insights for future studies.

\subsubsection{Bias-dependent TMR} \label{sec.TMR}

By using the P and AP charge currents, we can finally compute the TMR ratio as defined in 
Eq.~(\ref{eq: tmr}) at any given bias. The DFT and DMFT results are presented in 
Fig.~\ref{fig: current_Fe_MgO}(d) (blue and red lines, respectively).
DFT gives a nearly constant TMR ratio of approximately $3000\%$ up to $V \sim1.3$ V. Beyond this 
voltage, the TMR ratio is sharply suppressed and eventually vanishes for $V\sim 1.6$ V, as the 
AP current rises, while the P current saturates. 

DMFT exhibits a quantitatively different behavior compared to DFT, underscoring the 
significant impact of dynamical electron correlations on spin transport. Around zero-bias, 
the maximum TMR ratio is equal to approximately $4600\%$, which is larger than the DFT value 
because of the slightly larger P current predicted by DMFT. As the bias increases, the TMR is 
smoothly reduced to about $4000\%$ at $V \sim 0.7$ V, coinciding with the saturation of the P 
current. Finally, at even larger bias voltages, the TMR ratio eventually drops to zero driven 
by the rise of the AP incoherent, rather than coherent, current. Eventually, the TMR ratio vanishes at $V \sim 1.3$ V in DMFT.

In summary, the differences in the shape of the TMR ratio-versus-$V$ curve calculated by DFT 
and DMFT can be attributed to two main electron correlation effects discussed above. Firstly, 
the P current saturates at a lower bias in DMFT compared to DFT due to the correlation-induced 
shift of the Fe states. Secondly, the AP current increases rapidly above a certain bias due to 
inelastic scattering.

\section{Discussion}\label{sec.discussion}
Our predicted TMR ratio can be compared to values in the literature, and in particular, to 
experimental data, to assess whether DMFT provides an improved description of the electronic 
structure and transport properties of Fe/MgO/Fe junctions. 
At zero-bias, our DFT results are in good agreement with those in the DFT+NEGF literature 
\cite{bu.zh.01,rungger_2009, ru.sa.08}. However, the calculated TMR ratio is approximately three 
times higher than the maximum reported experimental values, which are around $900\%$ for epitaxial 
Fe/MgO/Fe junctions \cite{Scheike2021} and $1100\%$ for CoFe/MgO/CoFe junctions \cite{Scheike2023} 
at 3 K. In order to explain the thoretical overestimation, it has been speculated \cite{book1} that 
many-body effects may contribute to reduce the TMR. However, our DMFT calculations clearly show that 
this is not the case. In fact, dynamical electron correlations even enhance the low bias TMR in our 
system. Hence, other effects such as disorder \cite{disorder_1, disorder_2}, partial oxidation \cite{FeO_oxidation} of the interface Fe layer, or minor magnetization tilts in the P configuration,
may all be responsible for the reduction of the TMR, as already argued in early works. 
However, these effects remain challenging to incorporate into first-principles studies because they would require large supercells and averaging over various disorder configurations, making the computations highly demanding.

At finite-bias, the behavior of the junctions can be analyzed by considering the normalized TMR ratio, 
displayed in the inset of Fig.~\ref{fig: current_Fe_MgO}(e), and by computing the bias voltage, $V_{1/2}$, at which 
this normalized TMR reaches half of its zero-bias value. This analysis reveals that DMFT generally improves the agreement between theoretical results and experimental data.
Specifically DFT and 
DMFT yield $V_{1/2}= 1.375$ V and $V_{1/2}=0.9$ V, respectively, while, on the experimental side, reference \cite{Yu_Na_2004} 
reported $V_{1/2}$ equal to approximately $0.5$ V for negative bias and $1.2$ V for positive bias, with 
the different values due to the asymmetry of the junction. Similarly, reference~\cite{dj.ts.05}, which considered
amorphous CoFeB electrodes, reported $V_{1/2}= 0.68$ V for negative bias and $V_{1/2}=0.590$ V for 
positive bias. Comparable values for Fe/MgO/Fe were also found in reference \cite{cm.sm_06}. Despite 
the large variation in experimental results, it is evident that DFT consistently overestimates $V_{1/2}$, 
whereas DMFT reduces this overestimation. This suggest that incorporating dynamic correlation effects appears 
quite important to estimate the finite-bias behavior of Fe/MgO/Fe MTJs. Nonetheless, the agreement 
between the experimental and the calculate normalized TMR is not yet perfect. In particular, 
the experimental curves show a much faster decay of the TMR at low bias \cite{M.Bowen_A.Fert_01,Yu_Na_2004, dj.ts.05, cm.sm_06}, whereas in DMFT, the decay appears 
significantly smoother. This discrepancy may be attributed to having neglected disorder effects, 
which have been shown to sharpen the TMR curve, even when incorporated through a simple homogeneous 
broadening of the states \cite{rungger_2009}.

\section{Summary and conclusion}\label{sec.conclusion}
We have combined DMFT with DFT and the NEGF technique to analyse the effect of dynamical correlation 
on the electronic structure and steady-state transport properties of an Fe/MgO/Fe MTJ. In particular, 
by introducing the rigid shift approximation, we have been able to extend our approach beyond the 
linear response limit in a straightforward and computationally efficient manner, enabling the calculation 
of the junction's bias-dependent electronic structure and $I/V$ characteristic curve.

The main effect of dynamical electron correlation on the electronic structure is to reduce the spin 
splitting of the Fe $3d_{z^2}$ state and to introduce a finite relaxation time, which depends on the 
energy, following Fermi liquid theory. Despite these effects, at zero-bias, electron tunneling through 
the junction remains primarily governed by the symmetry-enforced spin-filtering mechanism proposed by 
early DFT calculations \cite{bu.zh.01,ma.um.01}. 

At finite bias, in the P configuration, the spin-dependent transport through the junction is similar 
in both DFT and DMFT, as it is mostly due to the coherent transmission of spin-up electrons. At low bias, the current increases linearly with 
$V$, and as the bias increases, it reaches a near-constant value when the Fe $3d_{z^2}$ states enter 
the bias window. The crossover between the linear and flat regions of the $I/V$ curve differs between 
DFT and DMFT, reflecting the correlation-induced shift of the Fe $3d_{z^2}$ state. 

In the AP configuration, correlation effects become very pronounced, as electrons tunnel through the Fe 
$3d$ states, in particular the $3d_{z^2}$. Then, we 
predicted that DMFT gives a sharp increase of the current at $V\gtrsim 1$ V due to inelastic 
electron-electron scattering. These phenomenon has so far not been extensively explored in MTJs, 
and our results could provide valuable insights for future studies.

From the current for the P and AP configurations, we could estimate the TMR ratio. The maximum TMR 
is found at zero-bias, where DFT yields a value of about $3500\%$, which is further enhanced by DMFT. 
However, the dependence of the TMR on the bias is different between DFT and DMFT. Specifically, we 
found that the TMR ratio is suppressed at a significantly lower bias in DMFT compared to DFT because 
of the correlation-driven sharp increase in AP current. Overall, the DMFT results better reproduce 
the experimental data from the literature, demonstrating the importance of incorporating dynamical 
correlation effects in quantum transport when dealing with devices incorporating ferromagnetic transition metals.

\section*{Supplementary Information}

\setcounter{section}{0}
\renewcommand{\thefigure}{S\arabic{figure}}
\setcounter{figure}{0}

\section*{Acknowledgements}
DN was supported by the Irish 
Research Council (Grant No. GOIPG/2021/1468 ). AD acknowledges funding by Science Foundation Ireland 
(SFI) and the Royal Society through the University Research Fellowship URF/R1/191769 during the initial 
stage of the project, when he was employed at Trinity College Dublin. SS acknowledges Science Foundation Ireland (AMBER Center grant 12/RC/2278-P2). IR acknowledges the support of the UK government Department for Science, Innovation and Technology through the UK National Quantum Technologies Programme. Computational resources were provided by Trinity College Dublin Research IT.

\section{Model versus semi-infinite Fe electrodes} 

\begin{figure}[ht]
    \centering
    \subfloat[Device 1: Fe leads]{%
      \includegraphics[clip,width=\columnwidth]{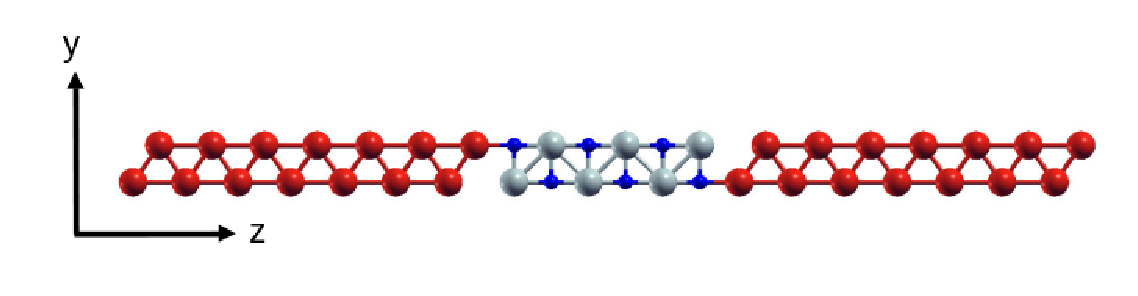}%
      \label{fig:Fe_lead}
    }

    \subfloat[Device 2: Au Leads]{%
      \includegraphics[clip,width=\columnwidth]{Au_lead.eps}%
      \label{fig:Au_lead}
    }
    \caption{Atomic structures of device 1 with  semi-infinite Fe leads (a), and device 2 with $6s$-only Au leads (b), which was described in the paper. Fe atoms are represented in red, Mg atoms in grey, O atoms in blue, and the Au atoms in yellow. }
    \label{fig:atomic_structure_si}
\end{figure}

In this section, we compare the results for the Fe/MgO/Fe MTJ with semi-infinite Fe leads [device 1 in Fig. \ref{fig:Fe_lead}] with those for the MTJ studied in the paper [device 2 in Fig. \ref{fig:Au_lead}], where the central region consists of three Fe layers on each side of the MgO barrier, connected to bcc $6s$-only Au leads. This comparison allows us to assess the impact of different leads on the spin-dependent transport properties.

\begin{figure}[t]
\centering\includegraphics[width=0.45\textwidth]{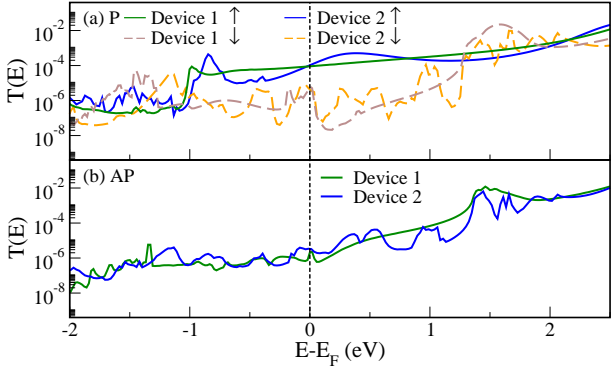}
\caption{Spin-up and spin-down transmission coefficient calculated by DFT for the device 1 and 2 in the P (a) and AP (b) configurations.  } 
\label{fig: trc_dos_Fe_MgO_equilibirum}
\end{figure}

Fig. \ref{fig: trc_dos_Fe_MgO_equilibirum} shows the spin-dependent transmission coefficients for both devices. In the P configuration [Fig. \ref{fig: trc_dos_Fe_MgO_equilibirum}(a)], the peak marking the onset of the high-transmission plateau in the spin-up channel is less pronounced in device 1 compared to device 2, with the plateau itself being flatter in device 1. However, these differences are overall minor. In the spin-down channel, both devices exhibit features related to Fe/MgO interface states between $E - E_\mathrm{F} \sim -0.4$ and $\sim 0.1$ eV. Beyond this energy range, the transmission coefficient for device 2 is relatively smooth, whereas the transmission coefficient for device 1 shows multiple sharp peaks, reflecting the formation of $3d$ minibands due to the finite number of Fe layers.

\begin{figure}[b]
\centering\includegraphics[width=0.48\textwidth]{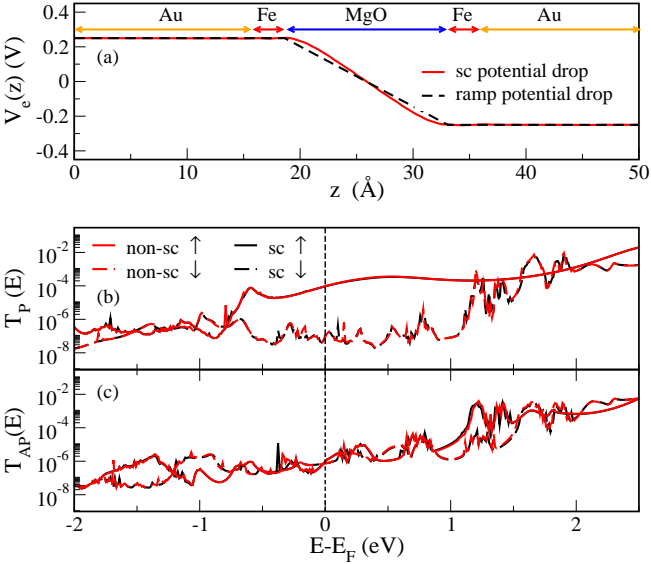}
\caption{
(a) Self-consistent (sc) potential $\Delta V_{e}(z)$ versus model ramp potential at a finite bias, $V=0.5$ V. 
(b) Transmission coefficients for the P configuration calculated self-consistently (black curve) and non-self-consistent (non-sc), applying the ramp potential (red curve). (c) Transmission coefficients for the AP configuration calculated self-consistently (black curve) and non-self-consistent, applying the ramp potential (red curve).} 
\label{fig: self_consistent bias}
\end{figure}

Similarly, in the AP configuration [Fig. \ref{fig: trc_dos_Fe_MgO_equilibirum}(b)], the transmission coefficient for device 1 is smoother than for device 2.  Despite this, the overall shapes and magnitudes of the transmission coefficients are similar across the energy range considered.

In conclusion, we see that the use of bcc $6s$-only Au electrodes in device 2 is well justified for the scope of our work, as the differences in spin-dependent transport transmission coefficients between the two devices are negligible.

\begin{figure}[b]
\centering\includegraphics[width=0.48\textwidth]{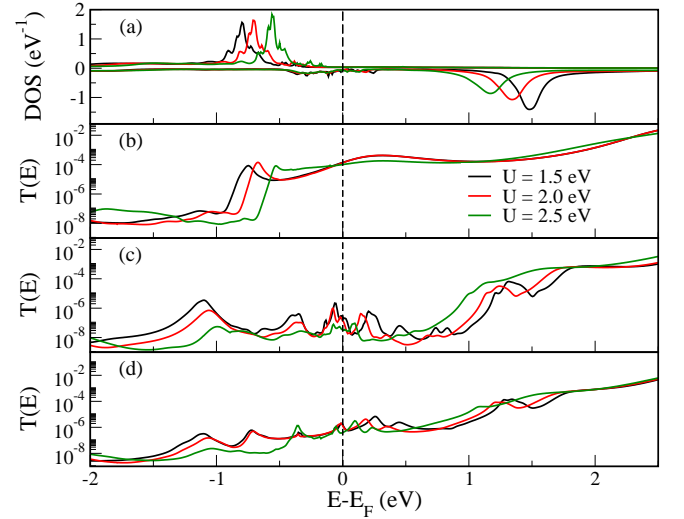}
\caption{DMFT zero-bias results for several values of the local Coulomb interaction, namely $U$ = 1.5 eV, $U$ = 2.0 eV, and $U$ = 2.5 eV ($J$ is fixed at 0.5 eV). (a) $\Delta_1^\sigma$-PDOS of the Fe at the left Fe/MgO interface. (b) Spin-up transmission coefficient for the P configuration. (c) Spin-down transmission coefficient for the P configuration. (d) Total transmission coefficient for the AP configuration. } 
\label{fig: trc_u}
\end{figure}

\section{Electrostatic potential drop across the insulating barrier}
 
To assess the validity of the rigid shift approximation, we have performed DFT+NEGF calculations at finite bias with full charge self-consistency. The results of these self-consistent calculations are presented in Fig. \ref{fig: self_consistent bias} and compared to those from the paper, carried out non-self-consistently with a ramp potential added to the Hamiltonian.

Fig. \ref{fig: self_consistent bias}(a) shows the self-consistent potential drop across the junction (red line), estimated as the difference between the planar average of the self-consistent Hartree potential at finite bias ($V=0.5$ V) and at zero bias, denoted as $\Delta V_\mathrm{H}(z)$, along the transport direction $z$. $\Delta V_\mathrm{H}(z)$ exhibits an almost linear decay across the MgO barrier and remains flat within the left and right electrodes, where it equals $V/2$ and $-V/2$, respectively. The overall shape of $\Delta V_\mathrm{H}(z)$ is very similar to the ramp potential  (black dashed line). Similar results are observed for all bias voltages. 

Fig. \ref{fig: self_consistent bias}(b) and (c) compare the DFT transmission coefficients for the P and AP configurations calculated self-consistently (black curve) and within the rigid shift approximation (red curve). The self-consistent and non-self-consistent results are nearly indistinguishable, further validating the rigid shift approximation.

 \section{Dependence of the DMFT results on the
$U$ effective interaction parameter}

In this section, we analyse the zero-bias DMFT transport properties of our Fe/MgO/Fe MTJ for several values of the local Coulomb interaction, namely $U = 1.5$ eV, $U = 2.0$ eV, and $U = 2.5$ eV (with $J$ fixed at 0.5 eV). 

Fig. \ref{fig: trc_u}(a) shows the DOS projected over the $\Delta_{1}$ state of the Fe atom at the left interface with the MgO barrier. In the spin-up channel, the $\Delta_{1}^\uparrow$ peak shifts upward in energy towards $E_\mathrm{F}$ as $U$ increases. In contrast, in the spin-down channel, it shifts downward in energy by nearly the same amount. Thus, the spin splitting is reduced with $U$. This is reflected in the transmission coefficient, which is plotted in Fig. \ref{fig: trc_u} (b), (c) and (d) for the P configuration (spin-up channel), P configuration (spin-down channel), and AP configuration, respectively.

\begin{figure}[t]
\centering\includegraphics[width=0.45\textwidth]{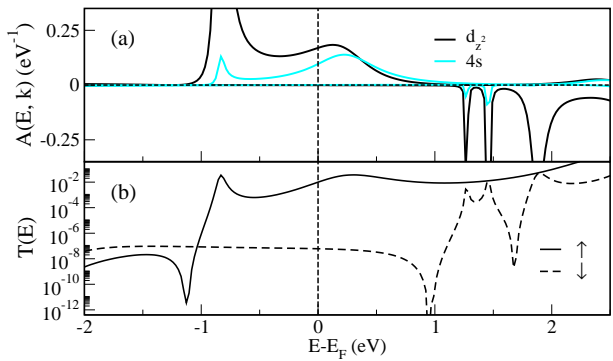}
\caption{(a) DFT spectral function for $\boldsymbol{k}=(0,0)$, projected over the $3d_{z^2}$ and $4s$ orbitals of the left-hand side Fe/MgO interface in the P configuration. (b) DFT $\boldsymbol{k}=(0,0)$-resolved transmission coefficient for the P configuration.} 
\label{fig: dos_gamma}
\end{figure}

  \begin{figure*}[!t]
\includegraphics[width=0.75\textwidth]{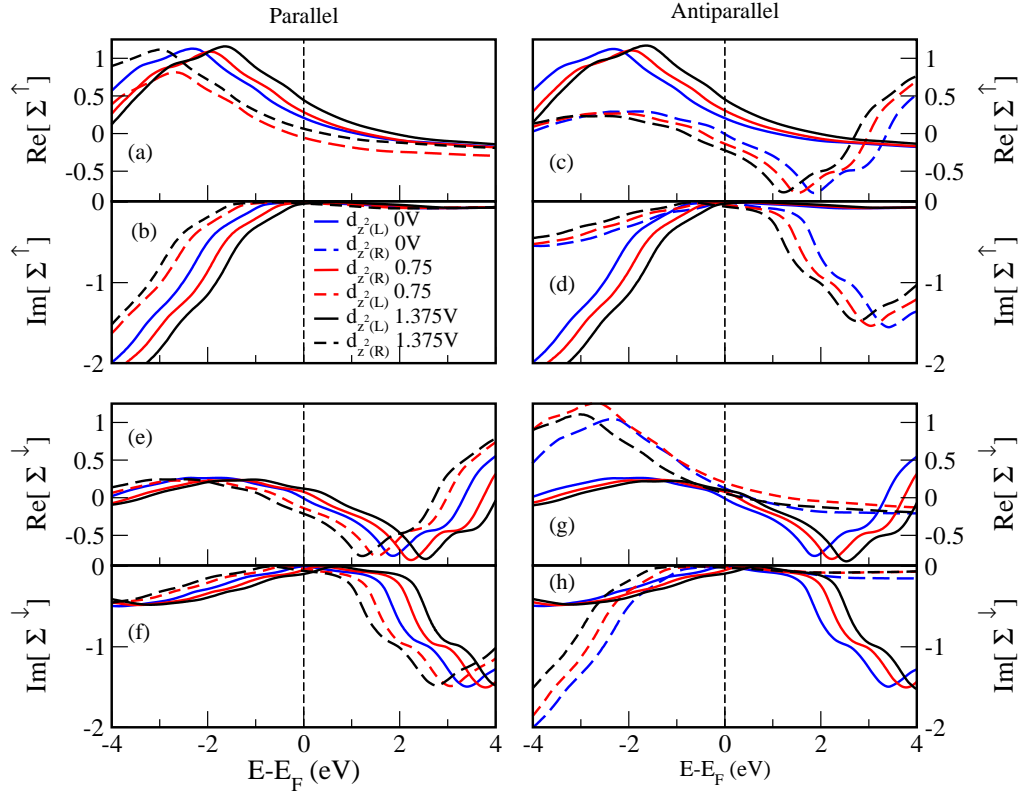}
\caption{(a) and (b): Real and imaginary parts of $\tilde{\Sigma}^{\uparrow \; r}_{d_{z^2},\mathrm{L(R)}}(E,V)$ for the P configuration. (c) and (d): Real and imaginary parts of $\tilde{\Sigma}^{\uparrow \; r}_{d_{z^2},\mathrm{L(R)}}(E,V)$ for the AP configuration. (e) and (f): Real and imaginary parts of $\tilde{\Sigma}^{\downarrow \; r}_{d_{z^2},\mathrm{R}}(E,V)$ for the P configuration. (g) and (h): Real and imaginary parts of $\tilde{\Sigma}^{\downarrow \; r}_{d_{z^2},\mathrm{L(R)}}(E,V)$ for the AP configuration. }
\label{fig: self_energy_bias}
\end{figure*}

In the P configuration, the onset of the high-transmission plateau in the spin-up (down) channel moves upward (downward) in energy towards $E_\mathrm{F}$ with $U$, following the shift of the spin-up (down) $\Delta_{1}^{\uparrow(\downarrow)}$ state, as shown in Fig. \ref{fig: trc_u}(b) [(c)]. 
Additionally, in the spin-down channel, at energies below this plateau, particularly below $E_\mathrm{F}$, the transmission coefficient decreases with increasing $U$ due to the reduced lifetimes of the $3d$ states, as the imaginary part of the DMFT self-energy increases with $U$.

In the AP configuration, the transmission coefficient, which is shown in Fig. \ref{fig: trc_u}(d), can be expressed as $T_\mathrm{AP}\sim\sqrt{T^\uparrow_\mathrm{P}T^\downarrow_\mathrm{P}}$. Therefore, the changes with $U$ merely follow those for the P transmission coefficient. We observe that the onset of the high-transmission region above $E_\mathrm{F}$ shifts down in energy, while the transmission coefficient is drastically reduced below $E_\mathrm{F}$ with increasing $U$.

The dependence of the transmission coefficient on $U$ will translate in modifications of the $I/V$ curve. Specifically, in the P configuration, increasing (decreasing) $U$ is expected to slightly lower (raise) the bias at which the crossover between the linear and constant current regions occurs. In contrast, for the AP configuration, changes in the spin splitting of the $\Delta_1$ states will lead to a shift in the bias where the inelastic current exhibits a sharp jump. 

\section{Zero-bias results at the $\Gamma$ point}

Fig. \ref{fig: dos_gamma}(a) and (b) display the spectral function, $A(E, \mathbf{k})$, projected over the orbitals of Fe atoms at the left-hand side Fe/MgO interface, along with the transmission coefficient, both calculated at the $\Gamma$ point, i.e., $\boldsymbol{k}=(0,0)$. The spectral function has a perfect half metallic character, with the spin-up component extending across $E_\mathrm{F}$, while the spin-down component begins at $E-E\mathrm{F}\approx 1.2$ eV. The $\Gamma$-point-resolved transmission coefficient closely mirrors this spectral function and accounts for most of the total transmission in both spin channels. This indicates that tunneling is dominated by half-metallic states with zero transverse wave-vector \cite{xie2016spin},
 resulting in the system nearly perfect spin-filtering behavior \cite{Bu_Zh+2001, Bu_2008, Ivan_Stefano_2009}. Furthermore, we observe that the high-transmission plateau begins with the $3d_{z^2}$ state, but, as energy increases, it becomes primarily due to $4s$ Fe states, explaining why this plateau is featureless.

\section{Bias-dependence of the DMFT self-energy}
Fig. \ref{fig: self_energy_bias} presents the DMFT retarded self-energy of the Fe $d_{z^2}$ orbital on either side of the MgO barrier, $\tilde{\Sigma}^{\sigma \; r}_{d_{z^2},\mathrm{L(R)}}(E,V)$, for three different bias voltages $V$ and for both the P and AP configurations. 

We observe that $\tilde{\Sigma}^{\sigma \; r}_{d_{z^2},\mathrm{L(R)}}(E,V)$ shifts nearly exactly in energy with bias, following the relation:  
\begin{equation}
\tilde{\Sigma}^{\sigma \; r}_{d_{z^2},\mathrm{L(R)}}(E+(-) eV/2,V)\approx\tilde{\Sigma}^{\sigma \; r}_{d_{z^2}\mathrm{L(R)}}(E,V=0),\label{eq.approxSigmaDMFT_shift}\\ 
\end{equation}
as implied by the the rigid shift approximation.
Small deviations from this equality are due to changes in the energy level alignment across the junction when the ramp potential is applied in the central region, as explained at the end of section II-D of the paper.

\section{Current conservation}\label{appendix: current conservation}
\begin{figure}[hbt]
\centering\includegraphics[width=0.48\textwidth]{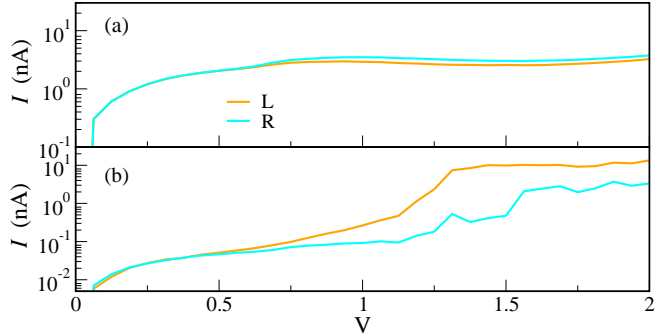}
\caption{(a) Absolute value of $I_\mathrm{L}$ and $I_\mathrm{R}$ calculated by DMFT for the (a) P and (b) AP configuration. Note the logarithmic scale.  } 
\label{fig: current_left_right_Fe_MgO}
\end{figure}

Charge conservation in steady-state quantum transport  dictates that the current flowing from the left lead into the central region must be equal to the current flowing from the central region into the right lead, $I_\mathrm{L} = -I_\mathrm{R}$, where $I_\mathrm{L}$ and $ I_\mathrm{R}$ are defined in Eq. (13) of the paper. This condition is always satisfied with a great accuracy in DFT+NEGF calculations. In contrast, this may not be the case in DMFT calculations because of two main issues.

The first issue is fundamental. Although DMFT is conserving in the Kadanoff-Baym sense \cite{ko.sa.06}, we cannot guarantee that our computational implementation remains conserving after introducing the necessary approximations for treating real systems.

The second issue is computational instead. The current in the Meir-Wingreen formula arises from a very small difference (on the order of $10^{-9}$ to $10^{-12}$ A) between two large terms. To achieve accurate results, every quantity involved in these terms must be computed with extreme numerical precision—something that is practically unreachable. For example, in model calculations with a simple central region consisting of a few sites with single orbitals and wide-band-limit electrodes, current conservation can be achieved, but only if the many-body self-energy is evaluated on an energy grid with hundreds of thousands of points and extending from $\sim -100 W$ to $\sim 100 W$ with $W$ the band width.

For real material systems, this level of precision is unattainable due to the computational cost, which restricts the energy grid to a few thousand points. Therefore, current conservation must be carefully checked to assess the margin of error in the results. This is done in Fig. \ref{fig: current_left_right_Fe_MgO} (a) and (b), which present the DMFT left and right currents as functions of bias for the P and AP configurations, respectively.

In the P configuration, where transport is largely coherent, the left and right currents are nearly identical, and their difference is small enough to be negligible for the scope of our work. In contrast, in the AP configuration, we observe that $I_\mathrm{L}$ remains equal to $I_\mathrm{R}$ for small bias ($V\lesssim 0.7$ V), but as the bias increases, they begin to diverge, with the non-coherent contribution to transport becoming dominant. 
Despite this divergence, both currents display a jump at around $1$ V although $I_R$ rises somewhat more smoothly than $I_L$. Consequently, while the quantitative prediction of the current at high biases may not be entirely accurate, the overall phenomenology associated with electron-electron scattering opening new channels for transport can still be reliably captured.

\bibliography{FeMgO}

\begin{thebibliography}{154}
\expandafter\ifx\csname natexlab\endcsname\relax\def\natexlab#1{#1}\fi
\expandafter\ifx\csname bibnamefont\endcsname\relax
  \def\bibnamefont#1{#1}\fi
\expandafter\ifx\csname bibfnamefont\endcsname\relax
  \def\bibfnamefont#1{#1}\fi
\expandafter\ifx\csname citenamefont\endcsname\relax
  \def\citenamefont#1{#1}\fi
\expandafter\ifx\csname url\endcsname\relax
  \def\url#1{\texttt{#1}}\fi
\expandafter\ifx\csname urlprefix\endcsname\relax\def\urlprefix{URL }\fi
\providecommand{\bibinfo}[2]{#2}
\providecommand{\eprint}[2][]{\url{#2}}

\bibitem[{\citenamefont{Tsymbal and Žutić}(2019)}]{ts.zu.book}
\bibinfo{editor}{\bibfnamefont{E.}~\bibnamefont{Tsymbal}} \bibnamefont{and} \bibinfo{editor}{\bibfnamefont{I.}~\bibnamefont{Žutić}}, eds., \emph{\bibinfo{title}{Spintronics Handbook: Spin Transport and Magnetism (2nd ed.)}} (\bibinfo{publisher}{CRC Press}, \bibinfo{year}{2019}).

\bibitem[{\citenamefont{Julliere}(1975)}]{ju.75}
\bibinfo{author}{\bibfnamefont{M.}~\bibnamefont{Julliere}}, \bibinfo{journal}{Physics Letters A} \textbf{\bibinfo{volume}{54}}, \bibinfo{pages}{225} (\bibinfo{year}{1975}), ISSN \bibinfo{issn}{0375-9601}, \bibinfo{note}{\href{http://www.sciencedirect.com/science/article/pii/0375960175901747}{URL}}.

\bibitem[{\citenamefont{Miyazaki and Tezuka}(1995)}]{mi.te.95}
\bibinfo{author}{\bibfnamefont{T.}~\bibnamefont{Miyazaki}} \bibnamefont{and} \bibinfo{author}{\bibfnamefont{N.}~\bibnamefont{Tezuka}}, \bibinfo{journal}{Journal of Magnetism and Magnetic Materials} \textbf{\bibinfo{volume}{139}}, \bibinfo{pages}{L231} (\bibinfo{year}{1995}), ISSN \bibinfo{issn}{0304-8853}, \bibinfo{note}{\href{http://www.sciencedirect.com/science/article/pii/0304885395900012}{URL}}.

\bibitem[{\citenamefont{Moodera et~al.}(1995)\citenamefont{Moodera, Kinder, Wong, and Meservey}}]{mo.ki.95}
\bibinfo{author}{\bibfnamefont{J.~S.} \bibnamefont{Moodera}}, \bibinfo{author}{\bibfnamefont{L.~R.} \bibnamefont{Kinder}}, \bibinfo{author}{\bibfnamefont{T.~M.} \bibnamefont{Wong}}, \bibnamefont{and} \bibinfo{author}{\bibfnamefont{R.}~\bibnamefont{Meservey}}, \bibinfo{journal}{Phys. Rev. Lett.} \textbf{\bibinfo{volume}{74}}, \bibinfo{pages}{3273} (\bibinfo{year}{1995}), \bibinfo{note}{\href{http://link.aps.org/doi/10.1103/PhysRevLett.74.3273}{URL}}.

\bibitem[{\citenamefont{Bowen et~al.}(2001{\natexlab{a}})\citenamefont{Bowen, Cros, Petroff, Fert, Martinez~Boubeta, Costa-Kramer, Anguita, Cebollada, Briones, de~Teresa et~al.}}]{bo.cr.01}
\bibinfo{author}{\bibfnamefont{M.}~\bibnamefont{Bowen}}, \bibinfo{author}{\bibfnamefont{V.}~\bibnamefont{Cros}}, \bibinfo{author}{\bibfnamefont{F.}~\bibnamefont{Petroff}}, \bibinfo{author}{\bibfnamefont{A.}~\bibnamefont{Fert}}, \bibinfo{author}{\bibfnamefont{C.}~\bibnamefont{Martinez~Boubeta}}, \bibinfo{author}{\bibfnamefont{J.~L.} \bibnamefont{Costa-Kramer}}, \bibinfo{author}{\bibfnamefont{J.~V.} \bibnamefont{Anguita}}, \bibinfo{author}{\bibfnamefont{A.}~\bibnamefont{Cebollada}}, \bibinfo{author}{\bibfnamefont{F.}~\bibnamefont{Briones}}, \bibinfo{author}{\bibfnamefont{J.~M.} \bibnamefont{de~Teresa}}, \bibnamefont{et~al.}, \bibinfo{journal}{Applied Physics Letters} \textbf{\bibinfo{volume}{79}}, \bibinfo{pages}{1655} (\bibinfo{year}{2001}{\natexlab{a}}), \bibinfo{note}{\href{https://doi.org/10.1063/1.1404125}{URL}}.

\bibitem[{\citenamefont{Butler et~al.}(2001{\natexlab{a}})\citenamefont{Butler, Zhang, Schulthess, and MacLaren}}]{bu.zh.01}
\bibinfo{author}{\bibfnamefont{W.~H.} \bibnamefont{Butler}}, \bibinfo{author}{\bibfnamefont{X.-G.} \bibnamefont{Zhang}}, \bibinfo{author}{\bibfnamefont{T.~C.} \bibnamefont{Schulthess}}, \bibnamefont{and} \bibinfo{author}{\bibfnamefont{J.~M.} \bibnamefont{MacLaren}}, \bibinfo{journal}{Phys. Rev. B} \textbf{\bibinfo{volume}{63}}, \bibinfo{pages}{054416} (\bibinfo{year}{2001}{\natexlab{a}}), \bibinfo{note}{\href{http://link.aps.org/doi/10.1103/PhysRevB.63.054416}{URL}}.

\bibitem[{\citenamefont{Mathon and Umerski}(2001)}]{ma.um.01}
\bibinfo{author}{\bibfnamefont{J.}~\bibnamefont{Mathon}} \bibnamefont{and} \bibinfo{author}{\bibfnamefont{A.}~\bibnamefont{Umerski}}, \bibinfo{journal}{Phys. Rev. B} \textbf{\bibinfo{volume}{63}}, \bibinfo{pages}{220403} (\bibinfo{year}{2001}), \bibinfo{note}{\href{http://link.aps.org/doi/10.1103/PhysRevB.63.220403}{URL}}.

\bibitem[{\citenamefont{Yuasa et~al.}(2004)\citenamefont{Yuasa, Nagahama, Fukushima, Suzuki, and Ando}}]{Yu_Na_2004}
\bibinfo{author}{\bibfnamefont{S.}~\bibnamefont{Yuasa}}, \bibinfo{author}{\bibfnamefont{T.}~\bibnamefont{Nagahama}}, \bibinfo{author}{\bibfnamefont{A.}~\bibnamefont{Fukushima}}, \bibinfo{author}{\bibfnamefont{Y.}~\bibnamefont{Suzuki}}, \bibnamefont{and} \bibinfo{author}{\bibfnamefont{K.}~\bibnamefont{Ando}}, \bibinfo{journal}{Nature Materials} \textbf{\bibinfo{volume}{3}}, \bibinfo{pages}{868} (\bibinfo{year}{2004}), ISSN \bibinfo{issn}{1476-4660}, \bibinfo{note}{\href{ttps://doi.org/10.1038/nmat1257}{URL}}.

\bibitem[{\citenamefont{Parkin et~al.}(2004)\citenamefont{Parkin, Kaiser, Panchula, Rice, Hughes, Samant, and Yang}}]{Parkin2004}
\bibinfo{author}{\bibfnamefont{S.~S.~P.} \bibnamefont{Parkin}}, \bibinfo{author}{\bibfnamefont{C.}~\bibnamefont{Kaiser}}, \bibinfo{author}{\bibfnamefont{A.}~\bibnamefont{Panchula}}, \bibinfo{author}{\bibfnamefont{P.~M.} \bibnamefont{Rice}}, \bibinfo{author}{\bibfnamefont{B.}~\bibnamefont{Hughes}}, \bibinfo{author}{\bibfnamefont{M.}~\bibnamefont{Samant}}, \bibnamefont{and} \bibinfo{author}{\bibfnamefont{S.-H.} \bibnamefont{Yang}}, \bibinfo{journal}{Nature Materials} \textbf{\bibinfo{volume}{3}}, \bibinfo{pages}{862} (\bibinfo{year}{2004}), ISSN \bibinfo{issn}{1476-4660}, \bibinfo{note}{\href{ttps://doi.org/10.1038/nmat1256}{URL}}.

\bibitem[{\citenamefont{Djayaprawira et~al.}(2005)\citenamefont{Djayaprawira, Tsunekawa, Nagai, Maehara, Yamagata, Watanabe, Yuasa, Suzuki, and Ando}}]{dj.ts.05}
\bibinfo{author}{\bibfnamefont{D.~D.} \bibnamefont{Djayaprawira}}, \bibinfo{author}{\bibfnamefont{K.}~\bibnamefont{Tsunekawa}}, \bibinfo{author}{\bibfnamefont{M.}~\bibnamefont{Nagai}}, \bibinfo{author}{\bibfnamefont{H.}~\bibnamefont{Maehara}}, \bibinfo{author}{\bibfnamefont{S.}~\bibnamefont{Yamagata}}, \bibinfo{author}{\bibfnamefont{N.}~\bibnamefont{Watanabe}}, \bibinfo{author}{\bibfnamefont{S.}~\bibnamefont{Yuasa}}, \bibinfo{author}{\bibfnamefont{Y.}~\bibnamefont{Suzuki}}, \bibnamefont{and} \bibinfo{author}{\bibfnamefont{K.}~\bibnamefont{Ando}}, \bibinfo{journal}{Applied Physics Letters} \textbf{\bibinfo{volume}{86}}, \bibinfo{pages}{092502} (\bibinfo{year}{2005}), \bibinfo{note}{\href{https://doi.org/10.1063/1.1871344}{URL}}.

\bibitem[{\citenamefont{Ikeda et~al.}(2008)\citenamefont{Ikeda, Hayakawa, Ashizawa, Lee, Miura, Hasegawa, Tsunoda, Matsukura, and Ohno}}]{ik.ha.08}
\bibinfo{author}{\bibfnamefont{S.}~\bibnamefont{Ikeda}}, \bibinfo{author}{\bibfnamefont{J.}~\bibnamefont{Hayakawa}}, \bibinfo{author}{\bibfnamefont{Y.}~\bibnamefont{Ashizawa}}, \bibinfo{author}{\bibfnamefont{Y.~M.} \bibnamefont{Lee}}, \bibinfo{author}{\bibfnamefont{K.}~\bibnamefont{Miura}}, \bibinfo{author}{\bibfnamefont{H.}~\bibnamefont{Hasegawa}}, \bibinfo{author}{\bibfnamefont{M.}~\bibnamefont{Tsunoda}}, \bibinfo{author}{\bibfnamefont{F.}~\bibnamefont{Matsukura}}, \bibnamefont{and} \bibinfo{author}{\bibfnamefont{H.}~\bibnamefont{Ohno}}, \bibinfo{journal}{Applied Physics Letters} \textbf{\bibinfo{volume}{93}}, \bibinfo{pages}{082508} (\bibinfo{year}{2008}), \bibinfo{note}{\href{https://doi.org/10.1063/1.2976435}{URL}}.

\bibitem[{\citenamefont{Butler}(2008)}]{Bu_2008}
\bibinfo{author}{\bibfnamefont{W.~H.} \bibnamefont{Butler}}, \bibinfo{journal}{Science and Technology of Advanced Materials} \textbf{\bibinfo{volume}{9}}, \bibinfo{pages}{014106} (\bibinfo{year}{2008}), \bibinfo{note}{\href{https://doi.org/10.1088/1468-6996/9/1/014106} {URL}}.

\bibitem[{\citenamefont{Mavropoulos et~al.}(2000)\citenamefont{Mavropoulos, Papanikolaou, and Dederichs}}]{Mavropolous2000}
\bibinfo{author}{\bibfnamefont{P.}~\bibnamefont{Mavropoulos}}, \bibinfo{author}{\bibfnamefont{N.}~\bibnamefont{Papanikolaou}}, \bibnamefont{and} \bibinfo{author}{\bibfnamefont{P.~H.} \bibnamefont{Dederichs}}, \bibinfo{journal}{Phys. Rev. Lett.} \textbf{\bibinfo{volume}{85}}, \bibinfo{pages}{1088} (\bibinfo{year}{2000}), \bibinfo{note}{\href{http://link.aps.org/doi/10.1103/PhysRevLett.85.1088}{URL}}.

\bibitem[{\citenamefont{Sanvito}(2005)}]{sanvito}
\bibinfo{author}{\bibfnamefont{S.}~\bibnamefont{Sanvito}} (\bibinfo{publisher}{American Scientific Publishers}, \bibinfo{year}{2005}), \bibinfo{note}{\href{http://doi.org/10.48550/arXiv.cond-mat/0503445}{URL}}.

\bibitem[{\citenamefont{Landauer}(1957)}]{La.57}
\bibinfo{author}{\bibfnamefont{R.}~\bibnamefont{Landauer}}, \bibinfo{journal}{IBM Journal of Research and Development} \textbf{\bibinfo{volume}{1}}, \bibinfo{pages}{223} (\bibinfo{year}{1957}).

\bibitem[{\citenamefont{B\"uttiker}(1986)}]{Bu.86}
\bibinfo{author}{\bibfnamefont{M.}~\bibnamefont{B\"uttiker}}, \bibinfo{journal}{Phys. Rev. Lett.} \textbf{\bibinfo{volume}{57}}, \bibinfo{pages}{1761} (\bibinfo{year}{1986}), \bibinfo{note}{\href{http://link.aps.org/doi/10.1103/PhysRevLett.57.1761}{URL}}.

\bibitem[{\citenamefont{Buttiker}(1988)}]{Bu.88}
\bibinfo{author}{\bibfnamefont{M.}~\bibnamefont{Buttiker}}, \bibinfo{journal}{IBM Journal of Research and Development} \textbf{\bibinfo{volume}{32}}, \bibinfo{pages}{317} (\bibinfo{year}{1988}).

\bibitem[{\citenamefont{Jones and Gunnarsson}(1989)}]{jo.gu.89}
\bibinfo{author}{\bibfnamefont{R.~O.} \bibnamefont{Jones}} \bibnamefont{and} \bibinfo{author}{\bibfnamefont{O.}~\bibnamefont{Gunnarsson}}, \bibinfo{journal}{Rev. Mod. Phys.} \textbf{\bibinfo{volume}{61}}, \bibinfo{pages}{689} (\bibinfo{year}{1989}), \bibinfo{note}{\href{http://link.aps.org/doi/10.1103/RevModPhys.61.689}{URL}}.

\bibitem[{\citenamefont{Kohn}(1999)}]{kohn.99}
\bibinfo{author}{\bibfnamefont{W.}~\bibnamefont{Kohn}}, \bibinfo{journal}{Rev. Mod. Phys.} \textbf{\bibinfo{volume}{71}}, \bibinfo{pages}{1253} (\bibinfo{year}{1999}).

\bibitem[{\citenamefont{Jones}(2015)}]{jone.15}
\bibinfo{author}{\bibfnamefont{R.~O.} \bibnamefont{Jones}}, \bibinfo{journal}{Rev. Mod. Phys.} \textbf{\bibinfo{volume}{87}}, \bibinfo{pages}{897} (\bibinfo{year}{2015}), \bibinfo{note}{\href{http://link.aps.org/doi/10.1103/RevModPhys.87.897}{URL}}.

\bibitem[{\citenamefont{von Barth and Hedin}(1972)}]{ba.he.72}
\bibinfo{author}{\bibfnamefont{U.}~\bibnamefont{von Barth}} \bibnamefont{and} \bibinfo{author}{\bibfnamefont{L.}~\bibnamefont{Hedin}}, \bibinfo{journal}{J. Phys. C} \textbf{\bibinfo{volume}{5}}, \bibinfo{pages}{1629} (\bibinfo{year}{1972}).

\bibitem[{\citenamefont{Vosko et~al.}(1980)\citenamefont{Vosko, Wilk, and Nusair}}]{vo.wi.80}
\bibinfo{author}{\bibfnamefont{S.~H.} \bibnamefont{Vosko}}, \bibinfo{author}{\bibfnamefont{L.}~\bibnamefont{Wilk}}, \bibnamefont{and} \bibinfo{author}{\bibfnamefont{M.}~\bibnamefont{Nusair}}, \bibinfo{journal}{Can. J. Phys.} \textbf{\bibinfo{volume}{58}}, \bibinfo{pages}{1200} (\bibinfo{year}{1980}).

\bibitem[{\citenamefont{Perdew et~al.}(1992)\citenamefont{Perdew, Chevary, Vosko, Jackson, Pederson, Singh, and Fiolhais}}]{pe.ch.92}
\bibinfo{author}{\bibfnamefont{J.~P.} \bibnamefont{Perdew}}, \bibinfo{author}{\bibfnamefont{J.~A.} \bibnamefont{Chevary}}, \bibinfo{author}{\bibfnamefont{S.~H.} \bibnamefont{Vosko}}, \bibinfo{author}{\bibfnamefont{K.~A.} \bibnamefont{Jackson}}, \bibinfo{author}{\bibfnamefont{M.~R.} \bibnamefont{Pederson}}, \bibinfo{author}{\bibfnamefont{D.~J.} \bibnamefont{Singh}}, \bibnamefont{and} \bibinfo{author}{\bibfnamefont{C.}~\bibnamefont{Fiolhais}}, \bibinfo{journal}{Phys. Rev. B} \textbf{\bibinfo{volume}{46}}, \bibinfo{pages}{6671} (\bibinfo{year}{1992}), \bibinfo{note}{\href{http://link.aps.org/doi/10.1103/PhysRevB.46.6671}{URL}}.

\bibitem[{\citenamefont{Perdew et~al.}(1993)\citenamefont{Perdew, Chevary, Vosko, Jackson, Pederson, Singh, and Fiolhais}}]{pe.ch.93}
\bibinfo{author}{\bibfnamefont{J.~P.} \bibnamefont{Perdew}}, \bibinfo{author}{\bibfnamefont{J.~A.} \bibnamefont{Chevary}}, \bibinfo{author}{\bibfnamefont{S.~H.} \bibnamefont{Vosko}}, \bibinfo{author}{\bibfnamefont{K.~A.} \bibnamefont{Jackson}}, \bibinfo{author}{\bibfnamefont{M.~R.} \bibnamefont{Pederson}}, \bibinfo{author}{\bibfnamefont{D.~J.} \bibnamefont{Singh}}, \bibnamefont{and} \bibinfo{author}{\bibfnamefont{C.}~\bibnamefont{Fiolhais}}, \bibinfo{journal}{Phys. Rev. B} \textbf{\bibinfo{volume}{48}}, \bibinfo{pages}{4978} (\bibinfo{year}{1993}), \bibinfo{note}{\href{http://link.aps.org/doi/10.1103/PhysRevB.48.4978.2}{URL}}.

\bibitem[{\citenamefont{Perdew et~al.}(1996)\citenamefont{Perdew, Burke, and Ernzerhof}}]{pe.bu.96}
\bibinfo{author}{\bibfnamefont{J.~P.} \bibnamefont{Perdew}}, \bibinfo{author}{\bibfnamefont{K.}~\bibnamefont{Burke}}, \bibnamefont{and} \bibinfo{author}{\bibfnamefont{M.}~\bibnamefont{Ernzerhof}}, \bibinfo{journal}{Phys. Rev. Lett.} \textbf{\bibinfo{volume}{77}}, \bibinfo{pages}{3865} (\bibinfo{year}{1996}), \bibinfo{note}{\href{http://link.aps.org/doi/10.1103/PhysRevLett.77.3865}{URL}}.

\bibitem[{\citenamefont{Kudrnovsk\'y et~al.}(2000)\citenamefont{Kudrnovsk\'y, Drchal, Blaas, Weinberger, Turek, and Bruno}}]{ku.dr.00}
\bibinfo{author}{\bibfnamefont{J.}~\bibnamefont{Kudrnovsk\'y}}, \bibinfo{author}{\bibfnamefont{V.}~\bibnamefont{Drchal}}, \bibinfo{author}{\bibfnamefont{C.}~\bibnamefont{Blaas}}, \bibinfo{author}{\bibfnamefont{P.}~\bibnamefont{Weinberger}}, \bibinfo{author}{\bibfnamefont{I.}~\bibnamefont{Turek}}, \bibnamefont{and} \bibinfo{author}{\bibfnamefont{P.}~\bibnamefont{Bruno}}, \bibinfo{journal}{Phys. Rev. B} \textbf{\bibinfo{volume}{62}}, \bibinfo{pages}{15084} (\bibinfo{year}{2000}), \bibinfo{note}{\href{http://link.aps.org/doi/10.1103/PhysRevB.62.15084}{URL}}.

\bibitem[{\citenamefont{Khomyakov et~al.}(2005)\citenamefont{Khomyakov, Brocks, Karpan, Zwierzycki, and Kelly}}]{Kh.Br.05}
\bibinfo{author}{\bibfnamefont{P.~A.} \bibnamefont{Khomyakov}}, \bibinfo{author}{\bibfnamefont{G.}~\bibnamefont{Brocks}}, \bibinfo{author}{\bibfnamefont{V.}~\bibnamefont{Karpan}}, \bibinfo{author}{\bibfnamefont{M.}~\bibnamefont{Zwierzycki}}, \bibnamefont{and} \bibinfo{author}{\bibfnamefont{P.~J.} \bibnamefont{Kelly}}, \bibinfo{journal}{Phys. Rev. B} \textbf{\bibinfo{volume}{72}}, \bibinfo{pages}{035450} (\bibinfo{year}{2005}), \bibinfo{note}{\href{http://link.aps.org/doi/10.1103/PhysRevB.72.035450}{URL}}.

\bibitem[{\citenamefont{Wortmann et~al.}(2002{\natexlab{a}})\citenamefont{Wortmann, Ishida, and Bl\"ugel}}]{Wo.Is.02}
\bibinfo{author}{\bibfnamefont{D.}~\bibnamefont{Wortmann}}, \bibinfo{author}{\bibfnamefont{H.}~\bibnamefont{Ishida}}, \bibnamefont{and} \bibinfo{author}{\bibfnamefont{S.}~\bibnamefont{Bl\"ugel}}, \bibinfo{journal}{Phys. Rev. B} \textbf{\bibinfo{volume}{65}}, \bibinfo{pages}{165103} (\bibinfo{year}{2002}{\natexlab{a}}), \bibinfo{note}{\href{http://link.aps.org/doi/10.1103/PhysRevB.65.165103}{URL}}.

\bibitem[{\citenamefont{Wortmann et~al.}(2002{\natexlab{b}})\citenamefont{Wortmann, Ishida, and Bl\"ugel}}]{Wo.Is.02_2}
\bibinfo{author}{\bibfnamefont{D.}~\bibnamefont{Wortmann}}, \bibinfo{author}{\bibfnamefont{H.}~\bibnamefont{Ishida}}, \bibnamefont{and} \bibinfo{author}{\bibfnamefont{S.}~\bibnamefont{Bl\"ugel}}, \bibinfo{journal}{Phys. Rev. B} \textbf{\bibinfo{volume}{66}}, \bibinfo{pages}{075113} (\bibinfo{year}{2002}{\natexlab{b}}), \bibinfo{note}{\href{http://link.aps.org/doi/10.1103/PhysRevB.66.075113}{URL}}.

\bibitem[{\citenamefont{MacLaren et~al.}(1999)\citenamefont{MacLaren, Zhang, Butler, and Wang}}]{Ma.Zh.99}
\bibinfo{author}{\bibfnamefont{J.~M.} \bibnamefont{MacLaren}}, \bibinfo{author}{\bibfnamefont{X.-G.} \bibnamefont{Zhang}}, \bibinfo{author}{\bibfnamefont{W.~H.} \bibnamefont{Butler}}, \bibnamefont{and} \bibinfo{author}{\bibfnamefont{X.}~\bibnamefont{Wang}}, \bibinfo{journal}{Phys. Rev. B} \textbf{\bibinfo{volume}{59}}, \bibinfo{pages}{5470} (\bibinfo{year}{1999}), \bibinfo{note}{\href{http://link.aps.org/doi/10.1103/PhysRevB.59.5470}{URL}}.

\bibitem[{\citenamefont{Stefanucci and van Leeuwen}(2013)}]{bookStefanucci}
\bibinfo{author}{\bibfnamefont{G.}~\bibnamefont{Stefanucci}} \bibnamefont{and} \bibinfo{author}{\bibfnamefont{R.}~\bibnamefont{van Leeuwen}}, \emph{\bibinfo{title}{Nonequilibrium Many-Body Theory of Quantum Systems: A Modern Introduction}} (\bibinfo{publisher}{Cambridge University Press}, \bibinfo{year}{2013}).

\bibitem[{\citenamefont{Datta}(1995)}]{datta}
\bibinfo{author}{\bibfnamefont{S.}~\bibnamefont{Datta}}, \emph{\bibinfo{title}{Electronic Transport in Mesoscopic Systems}} (\bibinfo{publisher}{Cambridge University Press}, \bibinfo{address}{Cambridge, UK}, \bibinfo{year}{1995}).

\bibitem[{\citenamefont{Rungger et~al.}(2019)\citenamefont{Rungger, Droghetti, and Stamenova}}]{book1}
\bibinfo{author}{\bibfnamefont{I.}~\bibnamefont{Rungger}}, \bibinfo{author}{\bibfnamefont{A.}~\bibnamefont{Droghetti}}, \bibnamefont{and} \bibinfo{author}{\bibfnamefont{M.}~\bibnamefont{Stamenova}}, in \emph{\bibinfo{booktitle}{Handbook of Materials Modeling. Vol. 1 Methods: Theory and Modeling}}, edited by \bibinfo{editor}{\bibfnamefont{S.}~\bibnamefont{Yip}} \bibnamefont{and} \bibinfo{editor}{\bibfnamefont{W.}~\bibnamefont{W.~Andreoni}} (\bibinfo{publisher}{Springer International Publishing}, \bibinfo{year}{2019}).

\bibitem[{\citenamefont{Taylor et~al.}(2001)\citenamefont{Taylor, Guo, and Wang}}]{Ta.Gu.01}
\bibinfo{author}{\bibfnamefont{J.}~\bibnamefont{Taylor}}, \bibinfo{author}{\bibfnamefont{H.}~\bibnamefont{Guo}}, \bibnamefont{and} \bibinfo{author}{\bibfnamefont{J.}~\bibnamefont{Wang}}, \bibinfo{journal}{Phys. Rev. B} \textbf{\bibinfo{volume}{63}}, \bibinfo{pages}{245407} (\bibinfo{year}{2001}), \bibinfo{note}{\href{http://link.aps.org/doi/10.1103/PhysRevB.63.245407}{URL}}.

\bibitem[{\citenamefont{Brandbyge et~al.}(2002)\citenamefont{Brandbyge, Mozos, Ordej\'on, Taylor, and Stokbro}}]{Ba.Mo.02}
\bibinfo{author}{\bibfnamefont{M.}~\bibnamefont{Brandbyge}}, \bibinfo{author}{\bibfnamefont{J.-L.} \bibnamefont{Mozos}}, \bibinfo{author}{\bibfnamefont{P.}~\bibnamefont{Ordej\'on}}, \bibinfo{author}{\bibfnamefont{J.}~\bibnamefont{Taylor}}, \bibnamefont{and} \bibinfo{author}{\bibfnamefont{K.}~\bibnamefont{Stokbro}}, \bibinfo{journal}{Phys. Rev. B} \textbf{\bibinfo{volume}{65}}, \bibinfo{pages}{165401} (\bibinfo{year}{2002}), \bibinfo{note}{\href{http://link.aps.org/doi/10.1103/PhysRevB.65.165401}{URL}}.

\bibitem[{\citenamefont{Rocha et~al.}(2006)\citenamefont{Rocha, Garc\'{\i}a-Su\'arez, Bailey, Lambert, Ferrer, and Sanvito}}]{ro.ga.06}
\bibinfo{author}{\bibfnamefont{A.~R.} \bibnamefont{Rocha}}, \bibinfo{author}{\bibfnamefont{V.~M.} \bibnamefont{Garc\'{\i}a-Su\'arez}}, \bibinfo{author}{\bibfnamefont{S.}~\bibnamefont{Bailey}}, \bibinfo{author}{\bibfnamefont{C.}~\bibnamefont{Lambert}}, \bibinfo{author}{\bibfnamefont{J.}~\bibnamefont{Ferrer}}, \bibnamefont{and} \bibinfo{author}{\bibfnamefont{S.}~\bibnamefont{Sanvito}}, \bibinfo{journal}{Phys. Rev. B} \textbf{\bibinfo{volume}{73}}, \bibinfo{pages}{085414} (\bibinfo{year}{2006}), \bibinfo{note}{\href{http://link.aps.org/doi/10.1103/PhysRevB.73.085414}{URL}}.

\bibitem[{\citenamefont{Waldron et~al.}(2006)\citenamefont{Waldron, Timoshevskii, Hu, Xia, and Guo}}]{wa.ti.06}
\bibinfo{author}{\bibfnamefont{D.}~\bibnamefont{Waldron}}, \bibinfo{author}{\bibfnamefont{V.}~\bibnamefont{Timoshevskii}}, \bibinfo{author}{\bibfnamefont{Y.}~\bibnamefont{Hu}}, \bibinfo{author}{\bibfnamefont{K.}~\bibnamefont{Xia}}, \bibnamefont{and} \bibinfo{author}{\bibfnamefont{H.}~\bibnamefont{Guo}}, \bibinfo{journal}{Phys. Rev. Lett.} \textbf{\bibinfo{volume}{97}}, \bibinfo{pages}{226802} (\bibinfo{year}{2006}), \bibinfo{note}{\href{http://link.aps.org/doi/10.1103/PhysRevLett.97.226802}{URL}}.

\bibitem[{\citenamefont{Rungger et~al.}(2009)\citenamefont{Rungger, Mryasov, and Sanvito}}]{Ivan_Stefano_2009}
\bibinfo{author}{\bibfnamefont{I.}~\bibnamefont{Rungger}}, \bibinfo{author}{\bibfnamefont{O.}~\bibnamefont{Mryasov}}, \bibnamefont{and} \bibinfo{author}{\bibfnamefont{S.}~\bibnamefont{Sanvito}}, \bibinfo{journal}{Phys. Rev. B} \textbf{\bibinfo{volume}{79}}, \bibinfo{pages}{094414} (\bibinfo{year}{2009}), \bibinfo{note}{\href{http://link.aps.org/doi/10.1103/PhysRevB.79.094414}{URL}}.

\bibitem[{\citenamefont{Peralta-Ramos et~al.}(2008{\natexlab{a}})\citenamefont{Peralta-Ramos, Llois, Rungger, and Sanvito}}]{PR_Ivan_Stefano_2008}
\bibinfo{author}{\bibfnamefont{J.}~\bibnamefont{Peralta-Ramos}}, \bibinfo{author}{\bibfnamefont{A.~M.} \bibnamefont{Llois}}, \bibinfo{author}{\bibfnamefont{I.}~\bibnamefont{Rungger}}, \bibnamefont{and} \bibinfo{author}{\bibfnamefont{S.}~\bibnamefont{Sanvito}}, \bibinfo{journal}{Phys. Rev. B} \textbf{\bibinfo{volume}{78}}, \bibinfo{pages}{024430} (\bibinfo{year}{2008}{\natexlab{a}}), \bibinfo{note}{\href{https://link.aps.org/doi/10.1103/PhysRevB.78.024430}{URL}}.

\bibitem[{\citenamefont{Rungger et~al.}(2007)\citenamefont{Rungger, {Reily Rocha}, Mryasov, Heinonen, and Sanvito}}]{Ivan_Rocha_2007}
\bibinfo{author}{\bibfnamefont{I.}~\bibnamefont{Rungger}}, \bibinfo{author}{\bibfnamefont{A.}~\bibnamefont{{Reily Rocha}}}, \bibinfo{author}{\bibfnamefont{O.}~\bibnamefont{Mryasov}}, \bibinfo{author}{\bibfnamefont{O.}~\bibnamefont{Heinonen}}, \bibnamefont{and} \bibinfo{author}{\bibfnamefont{S.}~\bibnamefont{Sanvito}}, \bibinfo{journal}{Journal of Magnetism and Magnetic Materials} \textbf{\bibinfo{volume}{316}}, \bibinfo{pages}{481} (\bibinfo{year}{2007}), ISSN \bibinfo{issn}{0304-8853}, \bibinfo{note}{\href{http://www.sciencedirect.com/science/article/pii/S0304885307004994}{URL}}.

\bibitem[{\citenamefont{Ellis et~al.}(2017)\citenamefont{Ellis, Stamenova, and Sanvito}}]{el.st.17}
\bibinfo{author}{\bibfnamefont{M.~O.~A.} \bibnamefont{Ellis}}, \bibinfo{author}{\bibfnamefont{M.}~\bibnamefont{Stamenova}}, \bibnamefont{and} \bibinfo{author}{\bibfnamefont{S.}~\bibnamefont{Sanvito}}, \bibinfo{journal}{Phys. Rev. B} \textbf{\bibinfo{volume}{96}}, \bibinfo{pages}{224410} (\bibinfo{year}{2017}), \bibinfo{note}{\href{http://link.aps.org/doi/10.1103/PhysRevB.96.224410}{URL}}.

\bibitem[{\citenamefont{Caffrey et~al.}(2011)\citenamefont{Caffrey, Archer, Rungger, and Sanvito}}]{ca.ar.11}
\bibinfo{author}{\bibfnamefont{N.~M.} \bibnamefont{Caffrey}}, \bibinfo{author}{\bibfnamefont{T.}~\bibnamefont{Archer}}, \bibinfo{author}{\bibfnamefont{I.}~\bibnamefont{Rungger}}, \bibnamefont{and} \bibinfo{author}{\bibfnamefont{S.}~\bibnamefont{Sanvito}}, \bibinfo{journal}{Phys. Rev. B} \textbf{\bibinfo{volume}{83}}, \bibinfo{pages}{125409} (\bibinfo{year}{2011}), \bibinfo{note}{\href{http://link.aps.org/doi/10.1103/PhysRevB.83.125409}{URL}}.

\bibitem[{\citenamefont{Saha et~al.}(2012)\citenamefont{Saha, Blom, Thygesen, and Nikoli\ifmmode~\acute{c}\else \'{c}\fi{}}}]{sa.ka.12}
\bibinfo{author}{\bibfnamefont{K.~K.} \bibnamefont{Saha}}, \bibinfo{author}{\bibfnamefont{A.}~\bibnamefont{Blom}}, \bibinfo{author}{\bibfnamefont{K.~S.} \bibnamefont{Thygesen}}, \bibnamefont{and} \bibinfo{author}{\bibfnamefont{B.~K.} \bibnamefont{Nikoli\ifmmode~\acute{c}\else \'{c}\fi{}}}, \bibinfo{journal}{Phys. Rev. B} \textbf{\bibinfo{volume}{85}}, \bibinfo{pages}{184426} (\bibinfo{year}{2012}), \bibinfo{note}{\href{http://link.aps.org/doi/10.1103/PhysRevB.85.184426}{URL}}.

\bibitem[{\citenamefont{Dolui et~al.}(2014)\citenamefont{Dolui, Narayan, Rungger, and Sanvito}}]{do.na.14}
\bibinfo{author}{\bibfnamefont{K.}~\bibnamefont{Dolui}}, \bibinfo{author}{\bibfnamefont{A.}~\bibnamefont{Narayan}}, \bibinfo{author}{\bibfnamefont{I.}~\bibnamefont{Rungger}}, \bibnamefont{and} \bibinfo{author}{\bibfnamefont{S.}~\bibnamefont{Sanvito}}, \bibinfo{journal}{Phys. Rev. B} \textbf{\bibinfo{volume}{90}}, \bibinfo{pages}{041401} (\bibinfo{year}{2014}), \bibinfo{note}{\href{http://link.aps.org/doi/10.1103/PhysRevB.90.041401}{URL}}.

\bibitem[{\citenamefont{Stamenova et~al.}(2021)\citenamefont{Stamenova, Stamenov, Mahfouzi, Sun, Kioussis, and Sanvito}}]{st.st.21}
\bibinfo{author}{\bibfnamefont{M.}~\bibnamefont{Stamenova}}, \bibinfo{author}{\bibfnamefont{P.}~\bibnamefont{Stamenov}}, \bibinfo{author}{\bibfnamefont{F.}~\bibnamefont{Mahfouzi}}, \bibinfo{author}{\bibfnamefont{Q.}~\bibnamefont{Sun}}, \bibinfo{author}{\bibfnamefont{N.}~\bibnamefont{Kioussis}}, \bibnamefont{and} \bibinfo{author}{\bibfnamefont{S.}~\bibnamefont{Sanvito}}, \bibinfo{journal}{Phys. Rev. B} \textbf{\bibinfo{volume}{103}}, \bibinfo{pages}{094403} (\bibinfo{year}{2021}), \bibinfo{note}{\href{http://link.aps.org/doi/10.1103/PhysRevB.103.094403}{URL}}.

\bibitem[{\citenamefont{Li et~al.}(2019)\citenamefont{Li, Lü, Zhang, You, Su, and Tsymbal}}]{Li.lu.19}
\bibinfo{author}{\bibfnamefont{X.}~\bibnamefont{Li}}, \bibinfo{author}{\bibfnamefont{J.-T.} \bibnamefont{Lü}}, \bibinfo{author}{\bibfnamefont{J.}~\bibnamefont{Zhang}}, \bibinfo{author}{\bibfnamefont{L.}~\bibnamefont{You}}, \bibinfo{author}{\bibfnamefont{Y.}~\bibnamefont{Su}}, \bibnamefont{and} \bibinfo{author}{\bibfnamefont{E.~Y.} \bibnamefont{Tsymbal}}, \bibinfo{journal}{Nano Letters} \textbf{\bibinfo{volume}{19}}, \bibinfo{pages}{5133} (\bibinfo{year}{2019}), \bibinfo{note}{\href{ https://doi.org/10.1021/acs.nanolett.9b01506}{URL}}.

\bibitem[{\citenamefont{Paudel and Tsymbal}(2019)}]{pa.tu.19}
\bibinfo{author}{\bibfnamefont{T.~R.} \bibnamefont{Paudel}} \bibnamefont{and} \bibinfo{author}{\bibfnamefont{E.~Y.} \bibnamefont{Tsymbal}}, \bibinfo{journal}{ACS Applied Materials \& Interfaces} \textbf{\bibinfo{volume}{11}}, \bibinfo{pages}{15781} (\bibinfo{year}{2019}), \bibinfo{note}{\href{ https://doi.org/10.1021/acsami.9b01942}{URL}}.

\bibitem[{\citenamefont{Shao et~al.}(2021)\citenamefont{Shao, Zhang, Li, Eom, and Tsymbal}}]{Shao2021}
\bibinfo{author}{\bibfnamefont{D.-F.} \bibnamefont{Shao}}, \bibinfo{author}{\bibfnamefont{S.-H.} \bibnamefont{Zhang}}, \bibinfo{author}{\bibfnamefont{M.}~\bibnamefont{Li}}, \bibinfo{author}{\bibfnamefont{C.-B.} \bibnamefont{Eom}}, \bibnamefont{and} \bibinfo{author}{\bibfnamefont{E.~Y.} \bibnamefont{Tsymbal}}, \bibinfo{journal}{Nature Communications} \textbf{\bibinfo{volume}{12}}, \bibinfo{pages}{7061} (\bibinfo{year}{2021}), ISSN \bibinfo{issn}{2041-1723}, \bibinfo{note}{\href{ttps://doi.org/10.1038/s41467-021-26915-3}{URL}}.

\bibitem[{\citenamefont{Martin et~al.}(2022)\citenamefont{Martin, Dlubak, Mattana, Seneor, Martin, Henner, Godel, Sander, Collin, Chen et~al.}}]{pa.dl.22}
\bibinfo{author}{\bibfnamefont{P.}~\bibnamefont{Martin}}, \bibinfo{author}{\bibfnamefont{B.}~\bibnamefont{Dlubak}}, \bibinfo{author}{\bibfnamefont{R.}~\bibnamefont{Mattana}}, \bibinfo{author}{\bibfnamefont{P.}~\bibnamefont{Seneor}}, \bibinfo{author}{\bibfnamefont{M.-B.} \bibnamefont{Martin}}, \bibinfo{author}{\bibfnamefont{T.}~\bibnamefont{Henner}}, \bibinfo{author}{\bibfnamefont{F.}~\bibnamefont{Godel}}, \bibinfo{author}{\bibfnamefont{A.}~\bibnamefont{Sander}}, \bibinfo{author}{\bibfnamefont{S.}~\bibnamefont{Collin}}, \bibinfo{author}{\bibfnamefont{L.}~\bibnamefont{Chen}}, \bibnamefont{et~al.}, \bibinfo{journal}{Nanoscale} \textbf{\bibinfo{volume}{14}}, \bibinfo{pages}{12692} (\bibinfo{year}{2022}), \bibinfo{note}{\href{http://dx.doi.org/10.1039/D2NR01917E}{URL}}.

\bibitem[{\citenamefont{Dong et~al.}(2022)\citenamefont{Dong, Li, Gurung, Zhu, Zhang, Zheng, Tsymbal, and Zhang}}]{do.li.22}
\bibinfo{author}{\bibfnamefont{J.}~\bibnamefont{Dong}}, \bibinfo{author}{\bibfnamefont{X.}~\bibnamefont{Li}}, \bibinfo{author}{\bibfnamefont{G.}~\bibnamefont{Gurung}}, \bibinfo{author}{\bibfnamefont{M.}~\bibnamefont{Zhu}}, \bibinfo{author}{\bibfnamefont{P.}~\bibnamefont{Zhang}}, \bibinfo{author}{\bibfnamefont{F.}~\bibnamefont{Zheng}}, \bibinfo{author}{\bibfnamefont{E.~Y.} \bibnamefont{Tsymbal}}, \bibnamefont{and} \bibinfo{author}{\bibfnamefont{J.}~\bibnamefont{Zhang}}, \bibinfo{journal}{Phys. Rev. Lett.} \textbf{\bibinfo{volume}{128}}, \bibinfo{pages}{197201} (\bibinfo{year}{2022}), \bibinfo{note}{\href{http://link.aps.org/doi/10.1103/PhysRevLett.128.197201}{URL}}.

\bibitem[{\citenamefont{Kurth and Stefanucci}(2017)}]{Kurth_2017}
\bibinfo{author}{\bibfnamefont{S.}~\bibnamefont{Kurth}} \bibnamefont{and} \bibinfo{author}{\bibfnamefont{G.}~\bibnamefont{Stefanucci}}, \bibinfo{journal}{Journal of Physics: Condensed Matter} \textbf{\bibinfo{volume}{29}}, \bibinfo{pages}{413002} (\bibinfo{year}{2017}), \bibinfo{note}{\href{http://dx.doi.org/10.1088/1361-648X/aa7e36}{URL}}.

\bibitem[{\citenamefont{S\'anchez-Barriga et~al.}(2009)\citenamefont{S\'anchez-Barriga, Fink, Boni, Di~Marco, Braun, Min\'ar, Varykhalov, Rader, Bellini, Manghi et~al.}}]{sa.fi.09}
\bibinfo{author}{\bibfnamefont{J.}~\bibnamefont{S\'anchez-Barriga}}, \bibinfo{author}{\bibfnamefont{J.}~\bibnamefont{Fink}}, \bibinfo{author}{\bibfnamefont{V.}~\bibnamefont{Boni}}, \bibinfo{author}{\bibfnamefont{I.}~\bibnamefont{Di~Marco}}, \bibinfo{author}{\bibfnamefont{J.}~\bibnamefont{Braun}}, \bibinfo{author}{\bibfnamefont{J.}~\bibnamefont{Min\'ar}}, \bibinfo{author}{\bibfnamefont{A.}~\bibnamefont{Varykhalov}}, \bibinfo{author}{\bibfnamefont{O.}~\bibnamefont{Rader}}, \bibinfo{author}{\bibfnamefont{V.}~\bibnamefont{Bellini}}, \bibinfo{author}{\bibfnamefont{F.}~\bibnamefont{Manghi}}, \bibnamefont{et~al.}, \bibinfo{journal}{Phys. Rev. Lett.} \textbf{\bibinfo{volume}{103}}, \bibinfo{pages}{267203} (\bibinfo{year}{2009}), \bibinfo{note}{\href{http://link.aps.org/doi/10.1103/PhysRevLett.103.267203}{URL}}.

\bibitem[{\citenamefont{Sponza et~al.}(2017)\citenamefont{Sponza, Pisanti, Vishina, Pashov, Weber, van Schilfgaarde, Acharya, Vidal, and Kotliar}}]{sp.pi.17}
\bibinfo{author}{\bibfnamefont{L.}~\bibnamefont{Sponza}}, \bibinfo{author}{\bibfnamefont{P.}~\bibnamefont{Pisanti}}, \bibinfo{author}{\bibfnamefont{A.}~\bibnamefont{Vishina}}, \bibinfo{author}{\bibfnamefont{D.}~\bibnamefont{Pashov}}, \bibinfo{author}{\bibfnamefont{C.}~\bibnamefont{Weber}}, \bibinfo{author}{\bibfnamefont{M.}~\bibnamefont{van Schilfgaarde}}, \bibinfo{author}{\bibfnamefont{S.}~\bibnamefont{Acharya}}, \bibinfo{author}{\bibfnamefont{J.}~\bibnamefont{Vidal}}, \bibnamefont{and} \bibinfo{author}{\bibfnamefont{G.}~\bibnamefont{Kotliar}}, \bibinfo{journal}{Phys. Rev. B} \textbf{\bibinfo{volume}{95}}, \bibinfo{pages}{041112} (\bibinfo{year}{2017}), \bibinfo{note}{\href{http://link.aps.org/doi/10.1103/PhysRevB.95.041112}{URL}}.

\bibitem[{\citenamefont{Monastra et~al.}(2002)\citenamefont{Monastra, Manghi, Rozzi, Arcangeli, Wetli, Neff, Greber, and Osterwalder}}]{mo.ma.02}
\bibinfo{author}{\bibfnamefont{S.}~\bibnamefont{Monastra}}, \bibinfo{author}{\bibfnamefont{F.}~\bibnamefont{Manghi}}, \bibinfo{author}{\bibfnamefont{C.~A.} \bibnamefont{Rozzi}}, \bibinfo{author}{\bibfnamefont{C.}~\bibnamefont{Arcangeli}}, \bibinfo{author}{\bibfnamefont{E.}~\bibnamefont{Wetli}}, \bibinfo{author}{\bibfnamefont{H.-J.} \bibnamefont{Neff}}, \bibinfo{author}{\bibfnamefont{T.}~\bibnamefont{Greber}}, \bibnamefont{and} \bibinfo{author}{\bibfnamefont{J.}~\bibnamefont{Osterwalder}}, \bibinfo{journal}{Phys. Rev. Lett.} \textbf{\bibinfo{volume}{88}}, \bibinfo{pages}{236402} (\bibinfo{year}{2002}), \bibinfo{note}{\href{http://link.aps.org/doi/10.1103/PhysRevLett.88.236402}{URL}}.

\bibitem[{\citenamefont{Walter et~al.}(2010)\citenamefont{Walter, Riley, and Rader}}]{Walter_2010}
\bibinfo{author}{\bibfnamefont{A.~L.} \bibnamefont{Walter}}, \bibinfo{author}{\bibfnamefont{J.~D.} \bibnamefont{Riley}}, \bibnamefont{and} \bibinfo{author}{\bibfnamefont{O.}~\bibnamefont{Rader}}, \bibinfo{journal}{New Journal of Physics} \textbf{\bibinfo{volume}{12}}, \bibinfo{pages}{013007} (\bibinfo{year}{2010}), \bibinfo{note}{\href{http://dx.doi.org/10.1088/1367-2630/12/1/013007}{URL}}.

\bibitem[{\citenamefont{Zhang et~al.}(1997)\citenamefont{Zhang, Levy, Marley, and Parkin}}]{zh.le.97}
\bibinfo{author}{\bibfnamefont{S.}~\bibnamefont{Zhang}}, \bibinfo{author}{\bibfnamefont{P.~M.} \bibnamefont{Levy}}, \bibinfo{author}{\bibfnamefont{A.~C.} \bibnamefont{Marley}}, \bibnamefont{and} \bibinfo{author}{\bibfnamefont{S.~S.~P.} \bibnamefont{Parkin}}, \bibinfo{journal}{Phys. Rev. Lett.} \textbf{\bibinfo{volume}{79}}, \bibinfo{pages}{3744} (\bibinfo{year}{1997}), \bibinfo{note}{\href{http://link.aps.org/doi/10.1103/PhysRevLett.79.3744}{URL}}.

\bibitem[{\citenamefont{Matsumoto et~al.}(2005)\citenamefont{Matsumoto, Hamada, Mizuguchi, Shiraishi, Maehara, Tsunekawa, Djayaprawira, Watanabe, Kurosaki, Nagahama et~al.}}]{ma.yu.05}
\bibinfo{author}{\bibfnamefont{R.}~\bibnamefont{Matsumoto}}, \bibinfo{author}{\bibfnamefont{Y.}~\bibnamefont{Hamada}}, \bibinfo{author}{\bibfnamefont{M.}~\bibnamefont{Mizuguchi}}, \bibinfo{author}{\bibfnamefont{M.}~\bibnamefont{Shiraishi}}, \bibinfo{author}{\bibfnamefont{H.}~\bibnamefont{Maehara}}, \bibinfo{author}{\bibfnamefont{K.}~\bibnamefont{Tsunekawa}}, \bibinfo{author}{\bibfnamefont{D.~D.} \bibnamefont{Djayaprawira}}, \bibinfo{author}{\bibfnamefont{N.}~\bibnamefont{Watanabe}}, \bibinfo{author}{\bibfnamefont{Y.}~\bibnamefont{Kurosaki}}, \bibinfo{author}{\bibfnamefont{T.}~\bibnamefont{Nagahama}}, \bibnamefont{et~al.}, \bibinfo{journal}{Solid State Communications} \textbf{\bibinfo{volume}{136}}, \bibinfo{pages}{611} (\bibinfo{year}{2005}), ISSN \bibinfo{issn}{0038-1098}, \bibinfo{note}{\href{http://www.sciencedirect.com/science/article/pii/S0038109805008070}{URL}}.

\bibitem[{\citenamefont{Liu et~al.}(2014)\citenamefont{Liu, Niu, Xiang, Wei, Li, Feng, Han, Zhang, and Coey}}]{li.ni.14}
\bibinfo{author}{\bibfnamefont{L.}~\bibnamefont{Liu}}, \bibinfo{author}{\bibfnamefont{J.}~\bibnamefont{Niu}}, \bibinfo{author}{\bibfnamefont{L.}~\bibnamefont{Xiang}}, \bibinfo{author}{\bibfnamefont{J.}~\bibnamefont{Wei}}, \bibinfo{author}{\bibfnamefont{D.-L.} \bibnamefont{Li}}, \bibinfo{author}{\bibfnamefont{J.-F.} \bibnamefont{Feng}}, \bibinfo{author}{\bibfnamefont{X.-F.} \bibnamefont{Han}}, \bibinfo{author}{\bibfnamefont{X.-G.} \bibnamefont{Zhang}}, \bibnamefont{and} \bibinfo{author}{\bibfnamefont{J.~M.~D.} \bibnamefont{Coey}}, \bibinfo{journal}{Phys. Rev. B} \textbf{\bibinfo{volume}{90}}, \bibinfo{pages}{195132} (\bibinfo{year}{2014}), \bibinfo{note}{\href{http://link.aps.org/doi/10.1103/PhysRevB.90.195132}{URL}}.

\bibitem[{\citenamefont{Bang et~al.}(2009)\citenamefont{Bang, Nozaki, Djayaprawira, Shiraishi, Suzuki, Fukushima, Kubota, Nagahama, Yuasa, Maehara et~al.}}]{ba.dj.09}
\bibinfo{author}{\bibfnamefont{D.}~\bibnamefont{Bang}}, \bibinfo{author}{\bibfnamefont{T.}~\bibnamefont{Nozaki}}, \bibinfo{author}{\bibfnamefont{D.~D.} \bibnamefont{Djayaprawira}}, \bibinfo{author}{\bibfnamefont{M.}~\bibnamefont{Shiraishi}}, \bibinfo{author}{\bibfnamefont{Y.}~\bibnamefont{Suzuki}}, \bibinfo{author}{\bibfnamefont{A.}~\bibnamefont{Fukushima}}, \bibinfo{author}{\bibfnamefont{H.}~\bibnamefont{Kubota}}, \bibinfo{author}{\bibfnamefont{T.}~\bibnamefont{Nagahama}}, \bibinfo{author}{\bibfnamefont{S.}~\bibnamefont{Yuasa}}, \bibinfo{author}{\bibfnamefont{H.}~\bibnamefont{Maehara}}, \bibnamefont{et~al.}, \bibinfo{journal}{Journal of Applied Physics} \textbf{\bibinfo{volume}{105}}, \bibinfo{pages}{07C924} (\bibinfo{year}{2009}), ISSN \bibinfo{issn}{0021-8979}, \bibinfo{note}{\href{http://doi.org/10.1063/1.3063674}{URL}}.

\bibitem[{\citenamefont{Fabian et~al.}(2022)\citenamefont{Fabian, Gradhand, Czerner, and Heiliger}}]{fa.22}
\bibinfo{author}{\bibfnamefont{A.}~\bibnamefont{Fabian}}, \bibinfo{author}{\bibfnamefont{M.}~\bibnamefont{Gradhand}}, \bibinfo{author}{\bibfnamefont{M.}~\bibnamefont{Czerner}}, \bibnamefont{and} \bibinfo{author}{\bibfnamefont{C.}~\bibnamefont{Heiliger}}, \bibinfo{journal}{Phys. Rev. B} \textbf{\bibinfo{volume}{105}}, \bibinfo{pages}{165106} (\bibinfo{year}{2022}), \bibinfo{note}{\href{http://link.aps.org/doi/10.1103/PhysRevB.105.165106}{URL}}.

\bibitem[{\citenamefont{B\"uttiker}(1985)}]{bu.85}
\bibinfo{author}{\bibfnamefont{M.}~\bibnamefont{B\"uttiker}}, \bibinfo{journal}{Phys. Rev. B} \textbf{\bibinfo{volume}{32}}, \bibinfo{pages}{1846} (\bibinfo{year}{1985}), \bibinfo{note}{\href{http://link.aps.org/doi/10.1103/PhysRevB.32.1846}{URL}}.

\bibitem[{\citenamefont{Metzner and Vollhardt}(1989)}]{me.vo.89}
\bibinfo{author}{\bibfnamefont{W.}~\bibnamefont{Metzner}} \bibnamefont{and} \bibinfo{author}{\bibfnamefont{D.}~\bibnamefont{Vollhardt}}, \bibinfo{journal}{Phys. Rev. Lett.} \textbf{\bibinfo{volume}{62}}, \bibinfo{pages}{324} (\bibinfo{year}{1989}).

\bibitem[{\citenamefont{Georges et~al.}(1996)\citenamefont{Georges, Kotliar, Krauth, and Rozenberg}}]{ge.ko.96}
\bibinfo{author}{\bibfnamefont{A.}~\bibnamefont{Georges}}, \bibinfo{author}{\bibfnamefont{G.}~\bibnamefont{Kotliar}}, \bibinfo{author}{\bibfnamefont{W.}~\bibnamefont{Krauth}}, \bibnamefont{and} \bibinfo{author}{\bibfnamefont{M.~J.} \bibnamefont{Rozenberg}}, \bibinfo{journal}{Rev. Mod. Phys.} \textbf{\bibinfo{volume}{68}}, \bibinfo{pages}{13} (\bibinfo{year}{1996}).

\bibitem[{\citenamefont{Lichtenstein and Katsnelson}(1998)}]{li.ka.98}
\bibinfo{author}{\bibfnamefont{A.~I.} \bibnamefont{Lichtenstein}} \bibnamefont{and} \bibinfo{author}{\bibfnamefont{M.~I.} \bibnamefont{Katsnelson}}, \bibinfo{journal}{Phys. Rev. B} \textbf{\bibinfo{volume}{57}}, \bibinfo{pages}{6884} (\bibinfo{year}{1998}), \bibinfo{note}{\href{http://link.aps.org/doi/10.1103/PhysRevB.57.6884}{URL}}.

\bibitem[{\citenamefont{Lichtenstein et~al.}(2001)\citenamefont{Lichtenstein, Katsnelson, and Kotliar}}]{li.ka.01}
\bibinfo{author}{\bibfnamefont{A.~I.} \bibnamefont{Lichtenstein}}, \bibinfo{author}{\bibfnamefont{M.~I.} \bibnamefont{Katsnelson}}, \bibnamefont{and} \bibinfo{author}{\bibfnamefont{G.}~\bibnamefont{Kotliar}}, \bibinfo{journal}{Phys. Rev. Lett.} \textbf{\bibinfo{volume}{87}}, \bibinfo{pages}{067205} (\bibinfo{year}{2001}).

\bibitem[{\citenamefont{Kotliar and Vollhardt}(2004)}]{ko.vo.04}
\bibinfo{author}{\bibfnamefont{G.}~\bibnamefont{Kotliar}} \bibnamefont{and} \bibinfo{author}{\bibfnamefont{D.}~\bibnamefont{Vollhardt}}, \bibinfo{journal}{Phys. Today} \textbf{\bibinfo{volume}{57}}, \bibinfo{pages}{53} (\bibinfo{year}{2004}).

\bibitem[{\citenamefont{Kotliar et~al.}(2006)\citenamefont{Kotliar, Savrasov, Haule, Oudovenko, Parcollet, and Marianetti}}]{ko.sa.06}
\bibinfo{author}{\bibfnamefont{G.}~\bibnamefont{Kotliar}}, \bibinfo{author}{\bibfnamefont{S.~Y.} \bibnamefont{Savrasov}}, \bibinfo{author}{\bibfnamefont{K.}~\bibnamefont{Haule}}, \bibinfo{author}{\bibfnamefont{V.~S.} \bibnamefont{Oudovenko}}, \bibinfo{author}{\bibfnamefont{O.}~\bibnamefont{Parcollet}}, \bibnamefont{and} \bibinfo{author}{\bibfnamefont{C.~A.} \bibnamefont{Marianetti}}, \bibinfo{journal}{Rev. Mod. Phys.} \textbf{\bibinfo{volume}{78}}, \bibinfo{pages}{865} (\bibinfo{year}{2006}).

\bibitem[{\citenamefont{Braun et~al.}(2006)\citenamefont{Braun, Min\'ar, Ebert, Katsnelson, and Lichtenstein}}]{br.mi.06}
\bibinfo{author}{\bibfnamefont{J.}~\bibnamefont{Braun}}, \bibinfo{author}{\bibfnamefont{J.}~\bibnamefont{Min\'ar}}, \bibinfo{author}{\bibfnamefont{H.}~\bibnamefont{Ebert}}, \bibinfo{author}{\bibfnamefont{M.~I.} \bibnamefont{Katsnelson}}, \bibnamefont{and} \bibinfo{author}{\bibfnamefont{A.~I.} \bibnamefont{Lichtenstein}}, \bibinfo{journal}{Phys. Rev. Lett.} \textbf{\bibinfo{volume}{97}}, \bibinfo{pages}{227601} (\bibinfo{year}{2006}), \bibinfo{note}{\href{http://link.aps.org/doi/10.1103/PhysRevLett.97.227601}{URL}}.

\bibitem[{\citenamefont{Grechnev et~al.}(2007)\citenamefont{Grechnev, Di~Marco, Katsnelson, Lichtenstein, Wills, and Eriksson}}]{gr.ma.07}
\bibinfo{author}{\bibfnamefont{A.}~\bibnamefont{Grechnev}}, \bibinfo{author}{\bibfnamefont{I.}~\bibnamefont{Di~Marco}}, \bibinfo{author}{\bibfnamefont{M.~I.} \bibnamefont{Katsnelson}}, \bibinfo{author}{\bibfnamefont{A.~I.} \bibnamefont{Lichtenstein}}, \bibinfo{author}{\bibfnamefont{J.~M.} \bibnamefont{Wills}}, \bibnamefont{and} \bibinfo{author}{\bibfnamefont{O.}~\bibnamefont{Eriksson}}, \bibinfo{journal}{Phys. Rev. B} \textbf{\bibinfo{volume}{76}}, \bibinfo{pages}{035107} (\bibinfo{year}{2007}), \bibinfo{note}{\href{http://link.aps.org/doi/10.1103/PhysRevB.76.035107}{URL}}.

\bibitem[{\citenamefont{Droghetti et~al.}(2022{\natexlab{a}})\citenamefont{Droghetti, Radonji\ifmmode~\acute{c}\else \'{c}\fi{}, Halder, Rungger, and Chioncel}}]{andrea_sigma_2}
\bibinfo{author}{\bibfnamefont{A.}~\bibnamefont{Droghetti}}, \bibinfo{author}{\bibfnamefont{M.~M.} \bibnamefont{Radonji\ifmmode~\acute{c}\else \'{c}\fi{}}}, \bibinfo{author}{\bibfnamefont{A.}~\bibnamefont{Halder}}, \bibinfo{author}{\bibfnamefont{I.}~\bibnamefont{Rungger}}, \bibnamefont{and} \bibinfo{author}{\bibfnamefont{L.}~\bibnamefont{Chioncel}}, \bibinfo{journal}{Phys. Rev. B} \textbf{\bibinfo{volume}{105}}, \bibinfo{pages}{115129} (\bibinfo{year}{2022}{\natexlab{a}}), \bibinfo{note}{\href{https://link.aps.org/doi/10.1103/PhysRevB.105.115129}{URL}}.

\bibitem[{\citenamefont{Janas et~al.}(2023)\citenamefont{Janas, Droghetti, Ponzoni, Cojocariu, Jugovac, Feyer, Radonjić, Rungger, Chioncel, Zamborlini et~al.}}]{ja.dr.23}
\bibinfo{author}{\bibfnamefont{D.~M.} \bibnamefont{Janas}}, \bibinfo{author}{\bibfnamefont{A.}~\bibnamefont{Droghetti}}, \bibinfo{author}{\bibfnamefont{S.}~\bibnamefont{Ponzoni}}, \bibinfo{author}{\bibfnamefont{I.}~\bibnamefont{Cojocariu}}, \bibinfo{author}{\bibfnamefont{M.}~\bibnamefont{Jugovac}}, \bibinfo{author}{\bibfnamefont{V.}~\bibnamefont{Feyer}}, \bibinfo{author}{\bibfnamefont{M.~M.} \bibnamefont{Radonjić}}, \bibinfo{author}{\bibfnamefont{I.}~\bibnamefont{Rungger}}, \bibinfo{author}{\bibfnamefont{L.}~\bibnamefont{Chioncel}}, \bibinfo{author}{\bibfnamefont{G.}~\bibnamefont{Zamborlini}}, \bibnamefont{et~al.}, \bibinfo{journal}{Advanced Materials} \textbf{\bibinfo{volume}{35}}, \bibinfo{pages}{2205698} (\bibinfo{year}{2023}), \bibinfo{note}{\href{http://onlinelibrary.wiley.com/doi/abs/10.1002/adma.202205698}{URL}}.

\bibitem[{\citenamefont{Potthoff and Nolting}(1999)}]{po.no.99}
\bibinfo{author}{\bibfnamefont{M.}~\bibnamefont{Potthoff}} \bibnamefont{and} \bibinfo{author}{\bibfnamefont{W.}~\bibnamefont{Nolting}}, \bibinfo{journal}{Phys. Rev. B} \textbf{\bibinfo{volume}{59}}, \bibinfo{pages}{2549} (\bibinfo{year}{1999}), \bibinfo{note}{\href{http://link.aps.org/doi/10.1103/PhysRevB.59.2549}{URL}}.

\bibitem[{\citenamefont{Okamoto and Millis}(2004)}]{ok.mi.04}
\bibinfo{author}{\bibfnamefont{S.}~\bibnamefont{Okamoto}} \bibnamefont{and} \bibinfo{author}{\bibfnamefont{A.~J.} \bibnamefont{Millis}}, \bibinfo{journal}{Phys. Rev. B} \textbf{\bibinfo{volume}{70}}, \bibinfo{pages}{241104} (\bibinfo{year}{2004}), \bibinfo{note}{\href{http://link.aps.org/doi/10.1103/PhysRevB.70.241104}{URL}}.

\bibitem[{\citenamefont{Freericks}(2004)}]{free.04}
\bibinfo{author}{\bibfnamefont{J.~K.} \bibnamefont{Freericks}}, \bibinfo{journal}{Phys. Rev. B} \textbf{\bibinfo{volume}{70}}, \bibinfo{pages}{195342} (\bibinfo{year}{2004}), \bibinfo{note}{\href{http://link.aps.org/doi/10.1103/PhysRevB.70.195342}{URL}}.

\bibitem[{\citenamefont{Oka and Nagaosa}(2005)}]{ok.na.05}
\bibinfo{author}{\bibfnamefont{T.}~\bibnamefont{Oka}} \bibnamefont{and} \bibinfo{author}{\bibfnamefont{N.}~\bibnamefont{Nagaosa}}, \bibinfo{journal}{Phys. Rev. Lett.} \textbf{\bibinfo{volume}{95}}, \bibinfo{pages}{266403} (\bibinfo{year}{2005}), \bibinfo{note}{\href{http://link.aps.org/doi/10.1103/PhysRevLett.95.266403}{URL}}.

\bibitem[{\citenamefont{Yunoki et~al.}(2007)\citenamefont{Yunoki, Moreo, Dagotto, Okamoto, Kancharla, and Fujimori}}]{yu.mo.07}
\bibinfo{author}{\bibfnamefont{S.}~\bibnamefont{Yunoki}}, \bibinfo{author}{\bibfnamefont{A.}~\bibnamefont{Moreo}}, \bibinfo{author}{\bibfnamefont{E.}~\bibnamefont{Dagotto}}, \bibinfo{author}{\bibfnamefont{S.}~\bibnamefont{Okamoto}}, \bibinfo{author}{\bibfnamefont{S.~S.} \bibnamefont{Kancharla}}, \bibnamefont{and} \bibinfo{author}{\bibfnamefont{A.}~\bibnamefont{Fujimori}}, \bibinfo{journal}{Phys. Rev. B} \textbf{\bibinfo{volume}{76}}, \bibinfo{pages}{064532} (\bibinfo{year}{2007}), \bibinfo{note}{\href{http://link.aps.org/doi/10.1103/PhysRevB.76.064532}{URL}}.

\bibitem[{\citenamefont{Zenia et~al.}(2009)\citenamefont{Zenia, Freericks, Krishnamurthy, and Pruschke}}]{ze.fr.09}
\bibinfo{author}{\bibfnamefont{H.}~\bibnamefont{Zenia}}, \bibinfo{author}{\bibfnamefont{J.~K.} \bibnamefont{Freericks}}, \bibinfo{author}{\bibfnamefont{H.~R.} \bibnamefont{Krishnamurthy}}, \bibnamefont{and} \bibinfo{author}{\bibfnamefont{T.}~\bibnamefont{Pruschke}}, \bibinfo{journal}{Phys. Rev. Lett.} \textbf{\bibinfo{volume}{103}}, \bibinfo{pages}{116402} (\bibinfo{year}{2009}), \bibinfo{note}{\href{http://link.aps.org/doi/10.1103/PhysRevLett.103.116402}{URL}}.

\bibitem[{\citenamefont{Valli et~al.}(2010)\citenamefont{Valli, Sangiovanni, Gunnarsson, Toschi, and Held}}]{va.sa.10}
\bibinfo{author}{\bibfnamefont{A.}~\bibnamefont{Valli}}, \bibinfo{author}{\bibfnamefont{G.}~\bibnamefont{Sangiovanni}}, \bibinfo{author}{\bibfnamefont{O.}~\bibnamefont{Gunnarsson}}, \bibinfo{author}{\bibfnamefont{A.}~\bibnamefont{Toschi}}, \bibnamefont{and} \bibinfo{author}{\bibfnamefont{K.}~\bibnamefont{Held}}, \bibinfo{journal}{Phys. Rev. Lett.} \textbf{\bibinfo{volume}{104}}, \bibinfo{pages}{246402} (\bibinfo{year}{2010}), \bibinfo{note}{\href{http://link.aps.org/doi/10.1103/PhysRevLett.104.246402}{URL}}.

\bibitem[{\citenamefont{Valli et~al.}(2012)\citenamefont{Valli, Sangiovanni, Toschi, and Held}}]{va.sa.12}
\bibinfo{author}{\bibfnamefont{A.}~\bibnamefont{Valli}}, \bibinfo{author}{\bibfnamefont{G.}~\bibnamefont{Sangiovanni}}, \bibinfo{author}{\bibfnamefont{A.}~\bibnamefont{Toschi}}, \bibnamefont{and} \bibinfo{author}{\bibfnamefont{K.}~\bibnamefont{Held}}, \bibinfo{journal}{Phys. Rev. B} \textbf{\bibinfo{volume}{86}}, \bibinfo{pages}{115418} (\bibinfo{year}{2012}), \bibinfo{note}{\href{http://link.aps.org/doi/10.1103/PhysRevB.86.115418}{URL}}.

\bibitem[{\citenamefont{Okamoto}(2007)}]{ok.07}
\bibinfo{author}{\bibfnamefont{S.}~\bibnamefont{Okamoto}}, \bibinfo{journal}{Phys. Rev. B} \textbf{\bibinfo{volume}{76}}, \bibinfo{pages}{035105} (\bibinfo{year}{2007}), \bibinfo{note}{\href{http://link.aps.org/doi/10.1103/PhysRevB.76.035105}{URL}}.

\bibitem[{\citenamefont{Okamoto}(2008)}]{ok.08}
\bibinfo{author}{\bibfnamefont{S.}~\bibnamefont{Okamoto}}, \bibinfo{journal}{Phys. Rev. Lett.} \textbf{\bibinfo{volume}{101}}, \bibinfo{pages}{116807} (\bibinfo{year}{2008}), \bibinfo{note}{\href{http://link.aps.org/doi/10.1103/PhysRevLett.101.116807}{URL}}.

\bibitem[{\citenamefont{Droghetti et~al.}(2022{\natexlab{b}})\citenamefont{Droghetti, Radonji\ifmmode~\acute{c}\else \'{c}\fi{}, Chioncel, and Rungger}}]{andrea_Cu_co}
\bibinfo{author}{\bibfnamefont{A.}~\bibnamefont{Droghetti}}, \bibinfo{author}{\bibfnamefont{M.~c. v.~M.} \bibnamefont{Radonji\ifmmode~\acute{c}\else \'{c}\fi{}}}, \bibinfo{author}{\bibfnamefont{L.}~\bibnamefont{Chioncel}}, \bibnamefont{and} \bibinfo{author}{\bibfnamefont{I.}~\bibnamefont{Rungger}}, \bibinfo{journal}{Phys. Rev. B} \textbf{\bibinfo{volume}{106}}, \bibinfo{pages}{075156} (\bibinfo{year}{2022}{\natexlab{b}}), \bibinfo{note}{\href{https://link.aps.org/doi/10.1103/PhysRevB.106.075156}{URL}}.

\bibitem[{\citenamefont{Chioncel et~al.}(2015)\citenamefont{Chioncel, Morari, \"Ostlin, Appelt, Droghetti, Radonji\ifmmode~\acute{c}\else \'{c}\fi{}, Rungger, Vitos, Eckern, and Postnikov}}]{liviu_Cu_Co_dmft}
\bibinfo{author}{\bibfnamefont{L.}~\bibnamefont{Chioncel}}, \bibinfo{author}{\bibfnamefont{C.}~\bibnamefont{Morari}}, \bibinfo{author}{\bibfnamefont{A.}~\bibnamefont{\"Ostlin}}, \bibinfo{author}{\bibfnamefont{W.~H.} \bibnamefont{Appelt}}, \bibinfo{author}{\bibfnamefont{A.}~\bibnamefont{Droghetti}}, \bibinfo{author}{\bibfnamefont{M.~M.} \bibnamefont{Radonji\ifmmode~\acute{c}\else \'{c}\fi{}}}, \bibinfo{author}{\bibfnamefont{I.}~\bibnamefont{Rungger}}, \bibinfo{author}{\bibfnamefont{L.}~\bibnamefont{Vitos}}, \bibinfo{author}{\bibfnamefont{U.}~\bibnamefont{Eckern}}, \bibnamefont{and} \bibinfo{author}{\bibfnamefont{A.~V.} \bibnamefont{Postnikov}}, \bibinfo{journal}{Phys. Rev. B} \textbf{\bibinfo{volume}{92}}, \bibinfo{pages}{054431} (\bibinfo{year}{2015}), \bibinfo{note}{\href{http://link.aps.org/doi/10.1103/PhysRevB.92.054431}{URL}}.

\bibitem[{\citenamefont{Jacob et~al.}(2009)\citenamefont{Jacob, Haule, and Kotliar}}]{ja.ha.09}
\bibinfo{author}{\bibfnamefont{D.}~\bibnamefont{Jacob}}, \bibinfo{author}{\bibfnamefont{K.}~\bibnamefont{Haule}}, \bibnamefont{and} \bibinfo{author}{\bibfnamefont{G.}~\bibnamefont{Kotliar}}, \bibinfo{journal}{Phys. Rev. Lett.} \textbf{\bibinfo{volume}{103}}, \bibinfo{pages}{016803} (\bibinfo{year}{2009}), \bibinfo{note}{\href{http://link.aps.org/doi/10.1103/PhysRevLett.103.016803}{URL}}.

\bibitem[{\citenamefont{Jacob et~al.}(2010)\citenamefont{Jacob, Haule, and Kotliar}}]{ja.ha.10}
\bibinfo{author}{\bibfnamefont{D.}~\bibnamefont{Jacob}}, \bibinfo{author}{\bibfnamefont{K.}~\bibnamefont{Haule}}, \bibnamefont{and} \bibinfo{author}{\bibfnamefont{G.}~\bibnamefont{Kotliar}}, \bibinfo{journal}{Phys. Rev. B} \textbf{\bibinfo{volume}{82}}, \bibinfo{pages}{195115} (\bibinfo{year}{2010}), \bibinfo{note}{\href{http://link.aps.org/doi/10.1103/PhysRevB.82.195115}{URL}}.

\bibitem[{\citenamefont{Jacob}(2015)}]{Ja.15}
\bibinfo{author}{\bibfnamefont{D.}~\bibnamefont{Jacob}}, \bibinfo{journal}{Journal of Physics: Condensed Matter} \textbf{\bibinfo{volume}{27}}, \bibinfo{pages}{245606} (\bibinfo{year}{2015}), \bibinfo{note}{\href{http://doi.org/10.1088/0953-8984/27/24/245606}{URL}}.

\bibitem[{\citenamefont{Jacob et~al.}(2013)\citenamefont{Jacob, Soriano, and Palacios}}]{ja.so.13}
\bibinfo{author}{\bibfnamefont{D.}~\bibnamefont{Jacob}}, \bibinfo{author}{\bibfnamefont{M.}~\bibnamefont{Soriano}}, \bibnamefont{and} \bibinfo{author}{\bibfnamefont{J.~J.} \bibnamefont{Palacios}}, \bibinfo{journal}{Phys. Rev. B} \textbf{\bibinfo{volume}{88}}, \bibinfo{pages}{134417} (\bibinfo{year}{2013}), \bibinfo{note}{\href{http://link.aps.org/doi/10.1103/PhysRevB.88.134417}{URL}}.

\bibitem[{\citenamefont{Droghetti and Rungger}(2017{\natexlab{a}})}]{andrea_ivan_projection}
\bibinfo{author}{\bibfnamefont{A.}~\bibnamefont{Droghetti}} \bibnamefont{and} \bibinfo{author}{\bibfnamefont{I.}~\bibnamefont{Rungger}}, \bibinfo{journal}{Phys. Rev. B} \textbf{\bibinfo{volume}{95}}, \bibinfo{pages}{085131} (\bibinfo{year}{2017}{\natexlab{a}}), \bibinfo{note}{\href{http://link.aps.org/doi/10.1103/PhysRevB.95.085131}{URL}}.

\bibitem[{\citenamefont{Appelt et~al.}(2018)\citenamefont{Appelt, Droghetti, Chioncel, Radonjić, Muñoz, Kirchner, Vollhardt, and Rungger}}]{ap.dr.18}
\bibinfo{author}{\bibfnamefont{W.~H.} \bibnamefont{Appelt}}, \bibinfo{author}{\bibfnamefont{A.}~\bibnamefont{Droghetti}}, \bibinfo{author}{\bibfnamefont{L.}~\bibnamefont{Chioncel}}, \bibinfo{author}{\bibfnamefont{M.~M.} \bibnamefont{Radonjić}}, \bibinfo{author}{\bibfnamefont{E.}~\bibnamefont{Muñoz}}, \bibinfo{author}{\bibfnamefont{S.}~\bibnamefont{Kirchner}}, \bibinfo{author}{\bibfnamefont{D.}~\bibnamefont{Vollhardt}}, \bibnamefont{and} \bibinfo{author}{\bibfnamefont{I.}~\bibnamefont{Rungger}}, \bibinfo{journal}{Nanoscale} \textbf{\bibinfo{volume}{10}}, \bibinfo{pages}{17738} (\bibinfo{year}{2018}), \bibinfo{note}{\href{http://dx.doi.org/10.1039/C8NR03991G}{URL}}.

\bibitem[{\citenamefont{Rumetshofer et~al.}(2019)\citenamefont{Rumetshofer, Bauernfeind, Arrigoni, and von~der Linden}}]{ru.ba.19}
\bibinfo{author}{\bibfnamefont{M.}~\bibnamefont{Rumetshofer}}, \bibinfo{author}{\bibfnamefont{D.}~\bibnamefont{Bauernfeind}}, \bibinfo{author}{\bibfnamefont{E.}~\bibnamefont{Arrigoni}}, \bibnamefont{and} \bibinfo{author}{\bibfnamefont{W.}~\bibnamefont{von~der Linden}}, \bibinfo{journal}{Phys. Rev. B} \textbf{\bibinfo{volume}{99}}, \bibinfo{pages}{045148} (\bibinfo{year}{2019}), \bibinfo{note}{\href{http://link.aps.org/doi/10.1103/PhysRevB.99.045148}{URL}}.

\bibitem[{\citenamefont{Bhandary et~al.}(2021)\citenamefont{Bhandary, Tomczak, and Valli}}]{bh.to.21}
\bibinfo{author}{\bibfnamefont{S.}~\bibnamefont{Bhandary}}, \bibinfo{author}{\bibfnamefont{J.~M.} \bibnamefont{Tomczak}}, \bibnamefont{and} \bibinfo{author}{\bibfnamefont{A.}~\bibnamefont{Valli}}, \bibinfo{journal}{Nanoscale Adv.} \textbf{\bibinfo{volume}{3}}, \bibinfo{pages}{4990} (\bibinfo{year}{2021}), \bibinfo{note}{\href{http://dx.doi.org/10.1039/D1NA00407G}{URL}}.

\bibitem[{\citenamefont{Gandus et~al.}(2022)\citenamefont{Gandus, Passerone, Stadler, Luisier, and Valli}}]{gandus2022strongly}
\bibinfo{author}{\bibfnamefont{G.}~\bibnamefont{Gandus}}, \bibinfo{author}{\bibfnamefont{D.}~\bibnamefont{Passerone}}, \bibinfo{author}{\bibfnamefont{R.}~\bibnamefont{Stadler}}, \bibinfo{author}{\bibfnamefont{M.}~\bibnamefont{Luisier}}, \bibnamefont{and} \bibinfo{author}{\bibfnamefont{A.}~\bibnamefont{Valli}}, \emph{\bibinfo{title}{Strongly correlated physics in organic open-shell quantum systems}} (\bibinfo{year}{2022}), \bibinfo{note}{\href{ttps://arxiv.org/abs/2301.00282}, {URL}}.

\bibitem[{\citenamefont{Jacob}(2018)}]{ja.18}
\bibinfo{author}{\bibfnamefont{D.}~\bibnamefont{Jacob}}, \bibinfo{journal}{Journal of Physics: Condensed Matter} \textbf{\bibinfo{volume}{30}}, \bibinfo{pages}{354003} (\bibinfo{year}{2018}), \bibinfo{note}{\href{http://doi.org/10.1088/1361-648x/aad523}{URL}}.

\bibitem[{\citenamefont{Morari et~al.}(2017)\citenamefont{Morari, Appelt, \"Ostlin, Prinz-Zwick, Schwingenschl\"ogl, Eckern, and Chioncel}}]{mo.ap.17}
\bibinfo{author}{\bibfnamefont{C.}~\bibnamefont{Morari}}, \bibinfo{author}{\bibfnamefont{W.~H.} \bibnamefont{Appelt}}, \bibinfo{author}{\bibfnamefont{A.}~\bibnamefont{\"Ostlin}}, \bibinfo{author}{\bibfnamefont{A.}~\bibnamefont{Prinz-Zwick}}, \bibinfo{author}{\bibfnamefont{U.}~\bibnamefont{Schwingenschl\"ogl}}, \bibinfo{author}{\bibfnamefont{U.}~\bibnamefont{Eckern}}, \bibnamefont{and} \bibinfo{author}{\bibfnamefont{L.}~\bibnamefont{Chioncel}}, \bibinfo{journal}{Phys. Rev. B} \textbf{\bibinfo{volume}{96}}, \bibinfo{pages}{205137} (\bibinfo{year}{2017}), \bibinfo{note}{\href{http://link.aps.org/doi/10.1103/PhysRevB.96.205137}{URL}}.

\bibitem[{\citenamefont{Halder et~al.}(2024)\citenamefont{Halder, Nell, Sihi, Bajaj, Sanvito, and Droghetti}}]{ha.ne.24}
\bibinfo{author}{\bibfnamefont{A.}~\bibnamefont{Halder}}, \bibinfo{author}{\bibfnamefont{D.}~\bibnamefont{Nell}}, \bibinfo{author}{\bibfnamefont{A.}~\bibnamefont{Sihi}}, \bibinfo{author}{\bibfnamefont{A.}~\bibnamefont{Bajaj}}, \bibinfo{author}{\bibfnamefont{S.}~\bibnamefont{Sanvito}}, \bibnamefont{and} \bibinfo{author}{\bibfnamefont{A.}~\bibnamefont{Droghetti}}, \bibinfo{journal}{Nano Letters} \textbf{\bibinfo{volume}{24}}, \bibinfo{pages}{9221} (\bibinfo{year}{2024}), \bibinfo{note}{\href{https://doi.org/10.1021/acs.nanolett.4c01479}{URL}}.

\bibitem[{\citenamefont{Haug and Jauho}(1996)}]{haug1996quantum}
\bibinfo{author}{\bibfnamefont{H.}~\bibnamefont{Haug}} \bibnamefont{and} \bibinfo{author}{\bibfnamefont{A.}~\bibnamefont{Jauho}}, \emph{\bibinfo{title}{Quantum Kinetics in Transport and Optics of Semiconductors}}, Solid-State Sciences Series (\bibinfo{publisher}{Springer Berlin Heidelberg}, \bibinfo{year}{1996}), ISBN \bibinfo{isbn}{9783540616023}.

\bibitem[{\citenamefont{Meir and Wingreen}(1992)}]{MW_current}
\bibinfo{author}{\bibfnamefont{Y.}~\bibnamefont{Meir}} \bibnamefont{and} \bibinfo{author}{\bibfnamefont{N.~S.} \bibnamefont{Wingreen}}, \bibinfo{journal}{Phys. Rev. Lett.} \textbf{\bibinfo{volume}{68}}, \bibinfo{pages}{2512} (\bibinfo{year}{1992}), \bibinfo{note}{\href{http://link.aps.org/doi/10.1103/PhysRevLett.68.2512}{URL}}.

\bibitem[{\citenamefont{Ferretti et~al.}(2005{\natexlab{a}})\citenamefont{Ferretti, Calzolari, Di~Felice, Manghi, Caldas, Nardelli, and Molinari}}]{fe.ca.05}
\bibinfo{author}{\bibfnamefont{A.}~\bibnamefont{Ferretti}}, \bibinfo{author}{\bibfnamefont{A.}~\bibnamefont{Calzolari}}, \bibinfo{author}{\bibfnamefont{R.}~\bibnamefont{Di~Felice}}, \bibinfo{author}{\bibfnamefont{F.}~\bibnamefont{Manghi}}, \bibinfo{author}{\bibfnamefont{M.~J.} \bibnamefont{Caldas}}, \bibinfo{author}{\bibfnamefont{M.~B.} \bibnamefont{Nardelli}}, \bibnamefont{and} \bibinfo{author}{\bibfnamefont{E.}~\bibnamefont{Molinari}}, \bibinfo{journal}{Phys. Rev. Lett.} \textbf{\bibinfo{volume}{94}}, \bibinfo{pages}{116802} (\bibinfo{year}{2005}{\natexlab{a}}), \bibinfo{note}{\href{http://link.aps.org/doi/10.1103/PhysRevLett.94.116802}{URL}}.

\bibitem[{\citenamefont{Ferretti et~al.}(2005{\natexlab{b}})\citenamefont{Ferretti, Calzolari, Di~Felice, and Manghi}}]{fe.ca.05_2}
\bibinfo{author}{\bibfnamefont{A.}~\bibnamefont{Ferretti}}, \bibinfo{author}{\bibfnamefont{A.}~\bibnamefont{Calzolari}}, \bibinfo{author}{\bibfnamefont{R.}~\bibnamefont{Di~Felice}}, \bibnamefont{and} \bibinfo{author}{\bibfnamefont{F.}~\bibnamefont{Manghi}}, \bibinfo{journal}{Phys. Rev. B} \textbf{\bibinfo{volume}{72}}, \bibinfo{pages}{125114} (\bibinfo{year}{2005}{\natexlab{b}}), \bibinfo{note}{\href{http://link.aps.org/doi/10.1103/PhysRevB.72.125114}{URL}}.

\bibitem[{\citenamefont{Sanvito et~al.}(1999)\citenamefont{Sanvito, Lambert, Jefferson, and Bratkovsky}}]{SS99}
\bibinfo{author}{\bibfnamefont{S.}~\bibnamefont{Sanvito}}, \bibinfo{author}{\bibfnamefont{C.~J.} \bibnamefont{Lambert}}, \bibinfo{author}{\bibfnamefont{J.~H.} \bibnamefont{Jefferson}}, \bibnamefont{and} \bibinfo{author}{\bibfnamefont{A.~M.} \bibnamefont{Bratkovsky}}, \bibinfo{journal}{Phys. Rev. B} \textbf{\bibinfo{volume}{59}}, \bibinfo{pages}{11939} (\bibinfo{year}{1999}), \bibinfo{note}{\href{http://doi.org/10.1103/PhysRevB.59.11936}{URL}}.

\bibitem[{\citenamefont{Rungger and Sanvito}(2008)}]{ru.sa.08}
\bibinfo{author}{\bibfnamefont{I.}~\bibnamefont{Rungger}} \bibnamefont{and} \bibinfo{author}{\bibfnamefont{S.}~\bibnamefont{Sanvito}}, \bibinfo{journal}{Phys. Rev. B} \textbf{\bibinfo{volume}{78}}, \bibinfo{pages}{035407} (\bibinfo{year}{2008}), \bibinfo{note}{\href{http://link.aps.org/doi/10.1103/PhysRevB.78.035407}{URL}}.

\bibitem[{\citenamefont{Sancho et~al.}(1984)\citenamefont{Sancho, Sancho, and Rubio}}]{Sancho_1984}
\bibinfo{author}{\bibfnamefont{M.~P.~L.} \bibnamefont{Sancho}}, \bibinfo{author}{\bibfnamefont{J.~M.~L.} \bibnamefont{Sancho}}, \bibnamefont{and} \bibinfo{author}{\bibfnamefont{J.}~\bibnamefont{Rubio}}, \bibinfo{journal}{Journal of Physics F: Metal Physics} \textbf{\bibinfo{volume}{14}}, \bibinfo{pages}{1205} (\bibinfo{year}{1984}), \bibinfo{note}{\href{http://dx.doi.org/10.1088/0305-4608/14/5/016}{URL}}.

\bibitem[{\citenamefont{Held}(2007)}]{held.07}
\bibinfo{author}{\bibfnamefont{K.}~\bibnamefont{Held}}, \bibinfo{journal}{Adv. Phys.} \textbf{\bibinfo{volume}{56}}, \bibinfo{pages}{829} (\bibinfo{year}{2007}).

\bibitem[{Note1()}]{Note1}
Note1, \bibinfo{note}{in our DFT+DMFT implementation, the upfolded DMFT self-energies acquire a $\protect \bm {k}$-dependence because the transformation matrices are $\protect \bm {k}$-dependent \cite {andrea_ivan_projection, andrea_Cu_co,andrea_sigma_2}.}

\bibitem[{\citenamefont{Zhang et~al.}(2004)\citenamefont{Zhang, Zhang, Krsti\ifmmode~\acute{c}\else \'{c}\fi{}, Cheng, Butler, and MacLaren}}]{zhang12}
\bibinfo{author}{\bibfnamefont{C.}~\bibnamefont{Zhang}}, \bibinfo{author}{\bibfnamefont{X.-G.} \bibnamefont{Zhang}}, \bibinfo{author}{\bibfnamefont{P.~S.} \bibnamefont{Krsti\ifmmode~\acute{c}\else \'{c}\fi{}}}, \bibinfo{author}{\bibfnamefont{H.-p.} \bibnamefont{Cheng}}, \bibinfo{author}{\bibfnamefont{W.~H.} \bibnamefont{Butler}}, \bibnamefont{and} \bibinfo{author}{\bibfnamefont{J.~M.} \bibnamefont{MacLaren}}, \bibinfo{journal}{Phys. Rev. B} \textbf{\bibinfo{volume}{69}}, \bibinfo{pages}{134406} (\bibinfo{year}{2004}), \bibinfo{note}{\href{http://link.aps.org/doi/10.1103/PhysRevB.69.134406}{URL}}.

\bibitem[{\citenamefont{Xie et~al.}(2016)\citenamefont{Xie, Rungger, Munira, Stamenova, Sanvito, and Ghosh}}]{xie2016spin}
\bibinfo{author}{\bibfnamefont{Y.}~\bibnamefont{Xie}}, \bibinfo{author}{\bibfnamefont{I.}~\bibnamefont{Rungger}}, \bibinfo{author}{\bibfnamefont{K.}~\bibnamefont{Munira}}, \bibinfo{author}{\bibfnamefont{M.}~\bibnamefont{Stamenova}}, \bibinfo{author}{\bibfnamefont{S.}~\bibnamefont{Sanvito}}, \bibnamefont{and} \bibinfo{author}{\bibfnamefont{A.~W.} \bibnamefont{Ghosh}}, in \emph{\bibinfo{booktitle}{Nanomagnetic and spintronic devices for energy-efficient memory and computing}} (\bibinfo{publisher}{John Wiley \& Sons}, \bibinfo{year}{2016}), p.~\bibinfo{pages}{91}.

\bibitem[{\citenamefont{Heiliger et~al.}(2005)\citenamefont{Heiliger, Zahn, Yavorsky, and Mertig}}]{PhysRevB.72.180406}
\bibinfo{author}{\bibfnamefont{C.}~\bibnamefont{Heiliger}}, \bibinfo{author}{\bibfnamefont{P.}~\bibnamefont{Zahn}}, \bibinfo{author}{\bibfnamefont{B.~Y.} \bibnamefont{Yavorsky}}, \bibnamefont{and} \bibinfo{author}{\bibfnamefont{I.}~\bibnamefont{Mertig}}, \bibinfo{journal}{Phys. Rev. B} \textbf{\bibinfo{volume}{72}}, \bibinfo{pages}{180406} (\bibinfo{year}{2005}), \bibinfo{note}{\href{http://link.aps.org/doi/10.1103/PhysRevB.72.180406}{URL}}.

\bibitem[{\citenamefont{Rudnev et~al.}(2017)\citenamefont{Rudnev, Kaliginedi, Droghetti, Ozawa, Kuzume, aki Haga, Broekmann, and Rungger}}]{Rudnev_Sci_Adv2017}
\bibinfo{author}{\bibfnamefont{A.~V.} \bibnamefont{Rudnev}}, \bibinfo{author}{\bibfnamefont{V.}~\bibnamefont{Kaliginedi}}, \bibinfo{author}{\bibfnamefont{A.}~\bibnamefont{Droghetti}}, \bibinfo{author}{\bibfnamefont{H.}~\bibnamefont{Ozawa}}, \bibinfo{author}{\bibfnamefont{A.}~\bibnamefont{Kuzume}}, \bibinfo{author}{\bibfnamefont{M.}~\bibnamefont{aki Haga}}, \bibinfo{author}{\bibfnamefont{P.}~\bibnamefont{Broekmann}}, \bibnamefont{and} \bibinfo{author}{\bibfnamefont{I.}~\bibnamefont{Rungger}}, \bibinfo{journal}{Science Advances} \textbf{\bibinfo{volume}{3}}, \bibinfo{pages}{e1602297} (\bibinfo{year}{2017}), \bibinfo{note}{\href{https://www.science.org/doi/abs/10.1126/sciadv.1602297}{URL}}.

\bibitem[{\citenamefont{Jacob and Kurth}(2018)}]{ja.ku.18}
\bibinfo{author}{\bibfnamefont{D.}~\bibnamefont{Jacob}} \bibnamefont{and} \bibinfo{author}{\bibfnamefont{S.}~\bibnamefont{Kurth}}, \bibinfo{journal}{Nano Letters} \textbf{\bibinfo{volume}{18}}, \bibinfo{pages}{2086} (\bibinfo{year}{2018}), \bibinfo{note}{\href{https://doi.org/10.1021/acs.nanolett.8b00255}{URL}}.

\bibitem[{\citenamefont{Rungger}(2009)}]{rungger_2009}
\bibinfo{author}{\bibfnamefont{I.}~\bibnamefont{Rungger}}, Ph.D. thesis, \bibinfo{school}{Trinity College (Dublin, Ireland). School of Physics,} (\bibinfo{year}{2009}).

\bibitem[{\citenamefont{Ness et~al.}(2010)\citenamefont{Ness, Dash, and Godby}}]{ne.da.10}
\bibinfo{author}{\bibfnamefont{H.}~\bibnamefont{Ness}}, \bibinfo{author}{\bibfnamefont{L.~K.} \bibnamefont{Dash}}, \bibnamefont{and} \bibinfo{author}{\bibfnamefont{R.~W.} \bibnamefont{Godby}}, \bibinfo{journal}{Phys. Rev. B} \textbf{\bibinfo{volume}{82}}, \bibinfo{pages}{085426} (\bibinfo{year}{2010}), \bibinfo{note}{\href{http://link.aps.org/doi/10.1103/PhysRevB.82.085426}{URL}}.

\bibitem[{\citenamefont{Gull et~al.}(2011)\citenamefont{Gull, Millis, Lichtenstein, Rubtsov, Troyer, and Werner}}]{gu.mi.11}
\bibinfo{author}{\bibfnamefont{E.}~\bibnamefont{Gull}}, \bibinfo{author}{\bibfnamefont{A.~J.} \bibnamefont{Millis}}, \bibinfo{author}{\bibfnamefont{A.~I.} \bibnamefont{Lichtenstein}}, \bibinfo{author}{\bibfnamefont{A.~N.} \bibnamefont{Rubtsov}}, \bibinfo{author}{\bibfnamefont{M.}~\bibnamefont{Troyer}}, \bibnamefont{and} \bibinfo{author}{\bibfnamefont{P.}~\bibnamefont{Werner}}, \bibinfo{journal}{Rev. Mod. Phys.} \textbf{\bibinfo{volume}{83}}, \bibinfo{pages}{349} (\bibinfo{year}{2011}), \bibinfo{note}{\href{http://link.aps.org/doi/10.1103/RevModPhys.83.349}{URL}}.

\bibitem[{\citenamefont{Katsnelson and Lichtenstein}(1999)}]{ka.li.99}
\bibinfo{author}{\bibfnamefont{M.~I.} \bibnamefont{Katsnelson}} \bibnamefont{and} \bibinfo{author}{\bibfnamefont{A.~I.} \bibnamefont{Lichtenstein}}, \bibinfo{journal}{J. Phys.: Condens. Matter} \textbf{\bibinfo{volume}{11}}, \bibinfo{pages}{1037} (\bibinfo{year}{1999}), \bibinfo{note}{\href{}{UR}}.

\bibitem[{\citenamefont{Katsnelson and Lichtenstein}(2002)}]{ka.li.02}
\bibinfo{author}{\bibfnamefont{M.~I.} \bibnamefont{Katsnelson}} \bibnamefont{and} \bibinfo{author}{\bibfnamefont{A.~I.} \bibnamefont{Lichtenstein}}, \bibinfo{journal}{Eur. Phys. J. B} \textbf{\bibinfo{volume}{30}}, \bibinfo{pages}{9} (\bibinfo{year}{2002}).

\bibitem[{\citenamefont{Pourovskii et~al.}(2006)\citenamefont{Pourovskii, Katsnelson, and Lichtenstein}}]{po.ka.06}
\bibinfo{author}{\bibfnamefont{L.~V.} \bibnamefont{Pourovskii}}, \bibinfo{author}{\bibfnamefont{M.~I.} \bibnamefont{Katsnelson}}, \bibnamefont{and} \bibinfo{author}{\bibfnamefont{A.~I.} \bibnamefont{Lichtenstein}}, \bibinfo{journal}{Phys. Rev. B} \textbf{\bibinfo{volume}{73}}, \bibinfo{pages}{060506} (\bibinfo{year}{2006}), \bibinfo{note}{\href{http://link.aps.org/doi/10.1103/PhysRevB.73.060506}{URL}}.

\bibitem[{\citenamefont{Jarrell and Gubernatis}(1996)}]{ja.gu.96}
\bibinfo{author}{\bibfnamefont{M.}~\bibnamefont{Jarrell}} \bibnamefont{and} \bibinfo{author}{\bibfnamefont{J.}~\bibnamefont{Gubernatis}}, \bibinfo{journal}{Physics Reports} \textbf{\bibinfo{volume}{269}}, \bibinfo{pages}{133} (\bibinfo{year}{1996}), ISSN \bibinfo{issn}{0370-1573}, \bibinfo{note}{\href{http://www.sciencedirect.com/science/article/pii/0370157395000747}{URL}}.

\bibitem[{\citenamefont{Sandvik}(1998)}]{sa.98}
\bibinfo{author}{\bibfnamefont{A.~W.} \bibnamefont{Sandvik}}, \bibinfo{journal}{Phys. Rev. B} \textbf{\bibinfo{volume}{57}}, \bibinfo{pages}{10287} (\bibinfo{year}{1998}), \bibinfo{note}{\href{http://link.aps.org/doi/10.1103/PhysRevB.57.10287}{URL}}.

\bibitem[{\citenamefont{Mishchenko et~al.}(2000)\citenamefont{Mishchenko, Prokof'ev, Sakamoto, and Svistunov}}]{mi.pr.20}
\bibinfo{author}{\bibfnamefont{A.~S.} \bibnamefont{Mishchenko}}, \bibinfo{author}{\bibfnamefont{N.~V.} \bibnamefont{Prokof'ev}}, \bibinfo{author}{\bibfnamefont{A.}~\bibnamefont{Sakamoto}}, \bibnamefont{and} \bibinfo{author}{\bibfnamefont{B.~V.} \bibnamefont{Svistunov}}, \bibinfo{journal}{Phys. Rev. B} \textbf{\bibinfo{volume}{62}}, \bibinfo{pages}{6317} (\bibinfo{year}{2000}), \bibinfo{note}{\href{http://link.aps.org/doi/10.1103/PhysRevB.62.6317}{URL}}.

\bibitem[{\citenamefont{Fuchs et~al.}(2010)\citenamefont{Fuchs, Pruschke, and Jarrell}}]{fu.pr.10}
\bibinfo{author}{\bibfnamefont{S.}~\bibnamefont{Fuchs}}, \bibinfo{author}{\bibfnamefont{T.}~\bibnamefont{Pruschke}}, \bibnamefont{and} \bibinfo{author}{\bibfnamefont{M.}~\bibnamefont{Jarrell}}, \bibinfo{journal}{Phys. Rev. E} \textbf{\bibinfo{volume}{81}}, \bibinfo{pages}{056701} (\bibinfo{year}{2010}), \bibinfo{note}{\href{http://link.aps.org/doi/10.1103/PhysRevE.81.056701}{URL}}.

\bibitem[{\citenamefont{Janas}(2023)}]{andrea_FeO}
\bibinfo{author}{\bibfnamefont{D.~M.} \bibnamefont{Janas}}, \bibinfo{journal}{Adv Mater} \textbf{\bibinfo{volume}{35}} (\bibinfo{year}{2023}), \bibinfo{note}{\href{http://doi.org/10.1002/adma.202205698}{URL}}.

\bibitem[{\citenamefont{Droghetti and Rungger}(2017{\natexlab{b}})}]{dr.ru.17}
\bibinfo{author}{\bibfnamefont{A.}~\bibnamefont{Droghetti}} \bibnamefont{and} \bibinfo{author}{\bibfnamefont{I.}~\bibnamefont{Rungger}}, \bibinfo{journal}{Phys. Rev. B} \textbf{\bibinfo{volume}{95}}, \bibinfo{pages}{085131} (\bibinfo{year}{2017}{\natexlab{b}}), \bibinfo{note}{\href{http://link.aps.org/doi/10.1103/PhysRevB.95.085131}{URL}}.

\bibitem[{\citenamefont{Soler et~al.}(2002)\citenamefont{Soler, Artacho, Gale, Garc{\'{\i}}a, Junquera, Ordej{\'{o}}n, and S{\'{a}}nchez-Portal}}]{so.ar.02}
\bibinfo{author}{\bibfnamefont{J.~M.} \bibnamefont{Soler}}, \bibinfo{author}{\bibfnamefont{E.}~\bibnamefont{Artacho}}, \bibinfo{author}{\bibfnamefont{J.~D.} \bibnamefont{Gale}}, \bibinfo{author}{\bibfnamefont{A.}~\bibnamefont{Garc{\'{\i}}a}}, \bibinfo{author}{\bibfnamefont{J.}~\bibnamefont{Junquera}}, \bibinfo{author}{\bibfnamefont{P.}~\bibnamefont{Ordej{\'{o}}n}}, \bibnamefont{and} \bibinfo{author}{\bibfnamefont{D.}~\bibnamefont{S{\'{a}}nchez-Portal}}, \bibinfo{journal}{Journal of Physics: Condensed Matter} \textbf{\bibinfo{volume}{14}}, \bibinfo{pages}{2745} (\bibinfo{year}{2002}), \bibinfo{note}{\href{http://doi.org/10.1088/0953-8984/14/11/302}{URL}}.

\bibitem[{\citenamefont{Troullier and Martins}(1991)}]{Tr.Ma.91}
\bibinfo{author}{\bibfnamefont{N.}~\bibnamefont{Troullier}} \bibnamefont{and} \bibinfo{author}{\bibfnamefont{J.~L.} \bibnamefont{Martins}}, \bibinfo{journal}{Phys. Rev. B} \textbf{\bibinfo{volume}{43}}, \bibinfo{pages}{1993} (\bibinfo{year}{1991}), \bibinfo{note}{\href{http://link.aps.org/doi/10.1103/PhysRevB.43.1993}{URL}}.

\bibitem[{\citenamefont{Imada et~al.}(1998)\citenamefont{Imada, Fujimori, and Tokura}}]{im.fu.98}
\bibinfo{author}{\bibfnamefont{M.}~\bibnamefont{Imada}}, \bibinfo{author}{\bibfnamefont{A.}~\bibnamefont{Fujimori}}, \bibnamefont{and} \bibinfo{author}{\bibfnamefont{Y.}~\bibnamefont{Tokura}}, \bibinfo{journal}{Rev. Mod. Phys.} \textbf{\bibinfo{volume}{70}}, \bibinfo{pages}{1039} (\bibinfo{year}{1998}).

\bibitem[{\citenamefont{Anisimov and Gunnarsson}(1991)}]{an.gu.91}
\bibinfo{author}{\bibfnamefont{V.~I.} \bibnamefont{Anisimov}} \bibnamefont{and} \bibinfo{author}{\bibfnamefont{O.}~\bibnamefont{Gunnarsson}}, \bibinfo{journal}{Phys. Rev. B} \textbf{\bibinfo{volume}{43}}, \bibinfo{pages}{7570} (\bibinfo{year}{1991}), \bibinfo{note}{\href{http://link.aps.org/doi/10.1103/PhysRevB.43.7570}{URL}}.

\bibitem[{\citenamefont{Dudarev et~al.}(1998{\natexlab{a}})\citenamefont{Dudarev, Botton, Savrasov, Humphreys, and Sutton}}]{dudarev}
\bibinfo{author}{\bibfnamefont{S.~L.} \bibnamefont{Dudarev}}, \bibinfo{author}{\bibfnamefont{G.~A.} \bibnamefont{Botton}}, \bibinfo{author}{\bibfnamefont{S.~Y.} \bibnamefont{Savrasov}}, \bibinfo{author}{\bibfnamefont{C.~J.} \bibnamefont{Humphreys}}, \bibnamefont{and} \bibinfo{author}{\bibfnamefont{A.~P.} \bibnamefont{Sutton}}, \bibinfo{journal}{Phys. Rev. B} \textbf{\bibinfo{volume}{57}}, \bibinfo{pages}{1505} (\bibinfo{year}{1998}{\natexlab{a}}), \bibinfo{note}{\href{http://link.aps.org/doi/10.1103/PhysRevB.57.1505}{URL}}.

\bibitem[{\citenamefont{Butler et~al.}(2001{\natexlab{b}})\citenamefont{Butler, Zhang, Schulthess, and MacLaren}}]{Bu_Zh+2001}
\bibinfo{author}{\bibfnamefont{W.~H.} \bibnamefont{Butler}}, \bibinfo{author}{\bibfnamefont{X.-G.} \bibnamefont{Zhang}}, \bibinfo{author}{\bibfnamefont{T.~C.} \bibnamefont{Schulthess}}, \bibnamefont{and} \bibinfo{author}{\bibfnamefont{J.~M.} \bibnamefont{MacLaren}}, \bibinfo{journal}{Phys. Rev. B} \textbf{\bibinfo{volume}{63}}, \bibinfo{pages}{054416} (\bibinfo{year}{2001}{\natexlab{b}}), \bibinfo{note}{\href{http://link.aps.org/doi/10.1103/PhysRevB.63.054416}{URL}}.

\bibitem[{\citenamefont{Faleev et~al.}(2012)\citenamefont{Faleev, Mryasov, and van Schilfgaarde}}]{FeMgO_GW}
\bibinfo{author}{\bibfnamefont{S.~V.} \bibnamefont{Faleev}}, \bibinfo{author}{\bibfnamefont{O.~N.} \bibnamefont{Mryasov}}, \bibnamefont{and} \bibinfo{author}{\bibfnamefont{M.}~\bibnamefont{van Schilfgaarde}}, \bibinfo{journal}{Phys. Rev. B} \textbf{\bibinfo{volume}{85}}, \bibinfo{pages}{174433} (\bibinfo{year}{2012}), \bibinfo{note}{\href{http://link.aps.org/doi/10.1103/PhysRevB.85.174433}{URL}}.

\bibitem[{\citenamefont{Anisimov et~al.}(1991)\citenamefont{Anisimov, Zaanen, and Andersen}}]{an.za.91}
\bibinfo{author}{\bibfnamefont{V.~I.} \bibnamefont{Anisimov}}, \bibinfo{author}{\bibfnamefont{J.}~\bibnamefont{Zaanen}}, \bibnamefont{and} \bibinfo{author}{\bibfnamefont{O.~K.} \bibnamefont{Andersen}}, \bibinfo{journal}{Phys. Rev. B} \textbf{\bibinfo{volume}{44}}, \bibinfo{pages}{943} (\bibinfo{year}{1991}), \bibinfo{note}{\href{http://link.aps.org/doi/10.1103/PhysRevB.44.943}{URL}}.

\bibitem[{\citenamefont{Liechtenstein et~al.}(1995)\citenamefont{Liechtenstein, Anisimov, and Zaanen}}]{li.an.95}
\bibinfo{author}{\bibfnamefont{A.~I.} \bibnamefont{Liechtenstein}}, \bibinfo{author}{\bibfnamefont{V.~I.} \bibnamefont{Anisimov}}, \bibnamefont{and} \bibinfo{author}{\bibfnamefont{J.}~\bibnamefont{Zaanen}}, \bibinfo{journal}{Phys. Rev. B} \textbf{\bibinfo{volume}{52}}, \bibinfo{pages}{R5467} (\bibinfo{year}{1995}), \bibinfo{note}{\href{http://link.aps.org/doi/10.1103/PhysRevB.52.R5467}{URL}}.

\bibitem[{\citenamefont{Dudarev et~al.}(1998{\natexlab{b}})\citenamefont{Dudarev, Botton, Savrasov, Humphreys, and Sutton}}]{du.bo.98}
\bibinfo{author}{\bibfnamefont{S.~L.} \bibnamefont{Dudarev}}, \bibinfo{author}{\bibfnamefont{G.~A.} \bibnamefont{Botton}}, \bibinfo{author}{\bibfnamefont{S.~Y.} \bibnamefont{Savrasov}}, \bibinfo{author}{\bibfnamefont{C.~J.} \bibnamefont{Humphreys}}, \bibnamefont{and} \bibinfo{author}{\bibfnamefont{A.~P.} \bibnamefont{Sutton}}, \bibinfo{journal}{Phys. Rev. B} \textbf{\bibinfo{volume}{57}}, \bibinfo{pages}{1505} (\bibinfo{year}{1998}{\natexlab{b}}), \bibinfo{note}{\href{http://link.aps.org/doi/10.1103/PhysRevB.57.1505}{URL}}.

\bibitem[{\citenamefont{Cococcioni and de~Gironcoli}(2005)}]{co.gi.05}
\bibinfo{author}{\bibfnamefont{M.}~\bibnamefont{Cococcioni}} \bibnamefont{and} \bibinfo{author}{\bibfnamefont{S.}~\bibnamefont{de~Gironcoli}}, \bibinfo{journal}{Phys. Rev. B} \textbf{\bibinfo{volume}{71}}, \bibinfo{pages}{035105} (\bibinfo{year}{2005}), \bibinfo{note}{\href{http://link.aps.org/doi/10.1103/PhysRevB.71.035105}{URL}}.

\bibitem[{\citenamefont{Butler et~al.}(1995)\citenamefont{Butler, Zhang, Nicholson, and MacLaren}}]{bu.zh.95}
\bibinfo{author}{\bibfnamefont{W.~H.} \bibnamefont{Butler}}, \bibinfo{author}{\bibfnamefont{X.-G.} \bibnamefont{Zhang}}, \bibinfo{author}{\bibfnamefont{D.~M.~C.} \bibnamefont{Nicholson}}, \bibnamefont{and} \bibinfo{author}{\bibfnamefont{J.~M.} \bibnamefont{MacLaren}}, \bibinfo{journal}{Phys. Rev. B} \textbf{\bibinfo{volume}{52}}, \bibinfo{pages}{13399} (\bibinfo{year}{1995}), \bibinfo{note}{\href{http://link.aps.org/doi/10.1103/PhysRevB.52.13399}{URL}}.

\bibitem[{\citenamefont{Peralta-Ramos et~al.}(2008{\natexlab{b}})\citenamefont{Peralta-Ramos, Llois, Rungger, and Sanvito}}]{PhysRevB.78.024430}
\bibinfo{author}{\bibfnamefont{J.}~\bibnamefont{Peralta-Ramos}}, \bibinfo{author}{\bibfnamefont{A.~M.} \bibnamefont{Llois}}, \bibinfo{author}{\bibfnamefont{I.}~\bibnamefont{Rungger}}, \bibnamefont{and} \bibinfo{author}{\bibfnamefont{S.}~\bibnamefont{Sanvito}}, \bibinfo{journal}{Phys. Rev. B} \textbf{\bibinfo{volume}{78}}, \bibinfo{pages}{024430} (\bibinfo{year}{2008}{\natexlab{b}}), \bibinfo{note}{\href{http://link.aps.org/doi/10.1103/PhysRevB.78.024430}{URL}}.

\bibitem[{\citenamefont{Hertz and Edwards}(1973)}]{Hertz_1973}
\bibinfo{author}{\bibfnamefont{J.~A.} \bibnamefont{Hertz}} \bibnamefont{and} \bibinfo{author}{\bibfnamefont{D.~M.} \bibnamefont{Edwards}}, \bibinfo{journal}{Journal of Physics F: Metal Physics} \textbf{\bibinfo{volume}{3}}, \bibinfo{pages}{2174} (\bibinfo{year}{1973}), \bibinfo{note}{\href{http://dx.doi.org/10.1088/0305-4608/3/12/018}{URL}}.

\bibitem[{\citenamefont{Edwards and Hertz}(1973)}]{Edwards_1973}
\bibinfo{author}{\bibfnamefont{D.~M.} \bibnamefont{Edwards}} \bibnamefont{and} \bibinfo{author}{\bibfnamefont{J.~A.} \bibnamefont{Hertz}}, \bibinfo{journal}{Journal of Physics F: Metal Physics} \textbf{\bibinfo{volume}{3}}, \bibinfo{pages}{2191} (\bibinfo{year}{1973}), \bibinfo{note}{\href{http://dx.doi.org/10.1088/0305-4608/3/12/019}{URL}}.

\bibitem[{\citenamefont{Jamneala et~al.}(2000)\citenamefont{Jamneala, Madhavan, Chen, and Crommie}}]{ja.ma.2000}
\bibinfo{author}{\bibfnamefont{T.}~\bibnamefont{Jamneala}}, \bibinfo{author}{\bibfnamefont{V.}~\bibnamefont{Madhavan}}, \bibinfo{author}{\bibfnamefont{W.}~\bibnamefont{Chen}}, \bibnamefont{and} \bibinfo{author}{\bibfnamefont{M.~F.} \bibnamefont{Crommie}}, \bibinfo{journal}{Phys. Rev. B} \textbf{\bibinfo{volume}{61}}, \bibinfo{pages}{9990} (\bibinfo{year}{2000}), \bibinfo{note}{\href{http://link.aps.org/doi/10.1103/PhysRevB.61.9990}{URL}}.

\bibitem[{\citenamefont{Heinrich et~al.}(2004)\citenamefont{Heinrich, Gupta, Lutz, and Eigler}}]{he.gu.04}
\bibinfo{author}{\bibfnamefont{A.~J.} \bibnamefont{Heinrich}}, \bibinfo{author}{\bibfnamefont{J.~A.} \bibnamefont{Gupta}}, \bibinfo{author}{\bibfnamefont{C.~P.} \bibnamefont{Lutz}}, \bibnamefont{and} \bibinfo{author}{\bibfnamefont{D.~M.} \bibnamefont{Eigler}}, \bibinfo{journal}{Science} \textbf{\bibinfo{volume}{306}}, \bibinfo{pages}{466} (\bibinfo{year}{2004}), \bibinfo{note}{\href{https://www.science.org/doi/abs/10.1126/science.1101077}{URL}}.

\bibitem[{\citenamefont{Hirjibehedin et~al.}(2006)\citenamefont{Hirjibehedin, Lutz, and Heinrich}}]{hi.lu.06}
\bibinfo{author}{\bibfnamefont{C.~F.} \bibnamefont{Hirjibehedin}}, \bibinfo{author}{\bibfnamefont{C.~P.} \bibnamefont{Lutz}}, \bibnamefont{and} \bibinfo{author}{\bibfnamefont{A.~J.} \bibnamefont{Heinrich}}, \bibinfo{journal}{Science} \textbf{\bibinfo{volume}{312}}, \bibinfo{pages}{1021} (\bibinfo{year}{2006}), \bibinfo{note}{\href{https://www.science.org/doi/abs/10.1126/science.1125398}{URL}}.

\bibitem[{\citenamefont{Balashov et~al.}(2009)\citenamefont{Balashov, Schuh, Tak\'acs, Ernst, Ostanin, Henk, Mertig, Bruno, Miyamachi, Suga et~al.}}]{ba.sc.09}
\bibinfo{author}{\bibfnamefont{T.}~\bibnamefont{Balashov}}, \bibinfo{author}{\bibfnamefont{T.}~\bibnamefont{Schuh}}, \bibinfo{author}{\bibfnamefont{A.~F.} \bibnamefont{Tak\'acs}}, \bibinfo{author}{\bibfnamefont{A.}~\bibnamefont{Ernst}}, \bibinfo{author}{\bibfnamefont{S.}~\bibnamefont{Ostanin}}, \bibinfo{author}{\bibfnamefont{J.}~\bibnamefont{Henk}}, \bibinfo{author}{\bibfnamefont{I.}~\bibnamefont{Mertig}}, \bibinfo{author}{\bibfnamefont{P.}~\bibnamefont{Bruno}}, \bibinfo{author}{\bibfnamefont{T.}~\bibnamefont{Miyamachi}}, \bibinfo{author}{\bibfnamefont{S.}~\bibnamefont{Suga}}, \bibnamefont{et~al.}, \bibinfo{journal}{Phys. Rev. Lett.} \textbf{\bibinfo{volume}{102}}, \bibinfo{pages}{257203} (\bibinfo{year}{2009}), \bibinfo{note}{\href{http://link.aps.org/doi/10.1103/PhysRevLett.102.257203}{URL}}.

\bibitem[{\citenamefont{Lounis et~al.}(2010)\citenamefont{Lounis, Costa, Muniz, and Mills}}]{lo.co.10}
\bibinfo{author}{\bibfnamefont{S.}~\bibnamefont{Lounis}}, \bibinfo{author}{\bibfnamefont{A.~T.} \bibnamefont{Costa}}, \bibinfo{author}{\bibfnamefont{R.~B.} \bibnamefont{Muniz}}, \bibnamefont{and} \bibinfo{author}{\bibfnamefont{D.~L.} \bibnamefont{Mills}}, \bibinfo{journal}{Phys. Rev. Lett.} \textbf{\bibinfo{volume}{105}}, \bibinfo{pages}{187205} (\bibinfo{year}{2010}), \bibinfo{note}{\href{http://link.aps.org/doi/10.1103/PhysRevLett.105.187205}{URL}}.

\bibitem[{\citenamefont{Khajetoorians et~al.}(2011)\citenamefont{Khajetoorians, Lounis, Chilian, Costa, Zhou, Mills, Wiebe, and Wiesendanger}}]{kh.lo.11}
\bibinfo{author}{\bibfnamefont{A.~A.} \bibnamefont{Khajetoorians}}, \bibinfo{author}{\bibfnamefont{S.}~\bibnamefont{Lounis}}, \bibinfo{author}{\bibfnamefont{B.}~\bibnamefont{Chilian}}, \bibinfo{author}{\bibfnamefont{A.~T.} \bibnamefont{Costa}}, \bibinfo{author}{\bibfnamefont{L.}~\bibnamefont{Zhou}}, \bibinfo{author}{\bibfnamefont{D.~L.} \bibnamefont{Mills}}, \bibinfo{author}{\bibfnamefont{J.}~\bibnamefont{Wiebe}}, \bibnamefont{and} \bibinfo{author}{\bibfnamefont{R.}~\bibnamefont{Wiesendanger}}, \bibinfo{journal}{Phys. Rev. Lett.} \textbf{\bibinfo{volume}{106}}, \bibinfo{pages}{037205} (\bibinfo{year}{2011}), \bibinfo{note}{\href{http://link.aps.org/doi/10.1103/PhysRevLett.106.037205}{URL}}.

\bibitem[{\citenamefont{Fern\'andez-Rossier}(2009)}]{Rossier2009}
\bibinfo{author}{\bibfnamefont{J.}~\bibnamefont{Fern\'andez-Rossier}}, \bibinfo{journal}{Phys. Rev. Lett.} \textbf{\bibinfo{volume}{102}}, \bibinfo{pages}{256802} (\bibinfo{year}{2009}), \bibinfo{note}{\href{http://link.aps.org/doi/10.1103/PhysRevLett.102.256802}{URL}}.

\bibitem[{\citenamefont{Hnoteey}(2011)}]{hu.ba.11}
\bibinfo{author}{\bibfnamefont{h.}~\bibnamefont{Hnoteey}}, \bibinfo{journal}{Phys. Rev. B} \textbf{\bibinfo{volume}{84}}, \bibinfo{pages}{035427} (\bibinfo{year}{2011}), \bibinfo{note}{\href{http://link.aps.org/doi/10.1103/PhysRevB.84.035427}{URL}}.

\bibitem[{\citenamefont{Ternes}(2015)}]{Ternes_2015}
\bibinfo{author}{\bibfnamefont{M.}~\bibnamefont{Ternes}}, \bibinfo{journal}{New Journal of Physics} \textbf{\bibinfo{volume}{17}}, \bibinfo{pages}{063016} (\bibinfo{year}{2015}), \bibinfo{note}{\href{http://dx.doi.org/10.1088/1367-2630/17/6/063016}{URL}}.

\bibitem[{\citenamefont{Jacob et~al.}(2021)\citenamefont{Jacob, Ortiz, and Fern\'andez-Rossier}}]{ja.or.21}
\bibinfo{author}{\bibfnamefont{D.}~\bibnamefont{Jacob}}, \bibinfo{author}{\bibfnamefont{R.}~\bibnamefont{Ortiz}}, \bibnamefont{and} \bibinfo{author}{\bibfnamefont{J.}~\bibnamefont{Fern\'andez-Rossier}}, \bibinfo{journal}{Phys. Rev. B} \textbf{\bibinfo{volume}{104}}, \bibinfo{pages}{075404} (\bibinfo{year}{2021}), \bibinfo{note}{\href{http://link.aps.org/doi/10.1103/PhysRevB.104.075404}{URL}}.

\bibitem[{\citenamefont{Gaudenzi et~al.}(2017)\citenamefont{Gaudenzi, de~Bruijckere, Reta, Moreira, Rovira, Veciana, van~der Zant, and Burzurí}}]{ga.de.17}
\bibinfo{author}{\bibfnamefont{R.}~\bibnamefont{Gaudenzi}}, \bibinfo{author}{\bibfnamefont{J.}~\bibnamefont{de~Bruijckere}}, \bibinfo{author}{\bibfnamefont{D.}~\bibnamefont{Reta}}, \bibinfo{author}{\bibfnamefont{I.~d. P.~R.} \bibnamefont{Moreira}}, \bibinfo{author}{\bibfnamefont{C.}~\bibnamefont{Rovira}}, \bibinfo{author}{\bibfnamefont{J.}~\bibnamefont{Veciana}}, \bibinfo{author}{\bibfnamefont{H.~S.~J.} \bibnamefont{van~der Zant}}, \bibnamefont{and} \bibinfo{author}{\bibfnamefont{E.}~\bibnamefont{Burzurí}}, \bibinfo{journal}{ACS Nano} \textbf{\bibinfo{volume}{11}}, \bibinfo{pages}{5879} (\bibinfo{year}{2017}), \bibinfo{note}{\href{ https://doi.org/10.1021/acsnano.7b01578}{URL}}.

\bibitem[{\citenamefont{Scheike et~al.}(2021)\citenamefont{Scheike, Xiang, Wen, Sukegawa, Ohkubo, Hono, and Mitani}}]{Scheike2021}
\bibinfo{author}{\bibfnamefont{T.}~\bibnamefont{Scheike}}, \bibinfo{author}{\bibfnamefont{Q.}~\bibnamefont{Xiang}}, \bibinfo{author}{\bibfnamefont{Z.}~\bibnamefont{Wen}}, \bibinfo{author}{\bibfnamefont{H.}~\bibnamefont{Sukegawa}}, \bibinfo{author}{\bibfnamefont{T.}~\bibnamefont{Ohkubo}}, \bibinfo{author}{\bibfnamefont{K.}~\bibnamefont{Hono}}, \bibnamefont{and} \bibinfo{author}{\bibfnamefont{S.}~\bibnamefont{Mitani}}, \bibinfo{journal}{Applied Physics Letters} \textbf{\bibinfo{volume}{118}}, \bibinfo{pages}{042411} (\bibinfo{year}{2021}), ISSN \bibinfo{issn}{0003-6951}, \bibinfo{note}{\href{http://doi.org/10.1063/5.0037972}{URL}}.

\bibitem[{\citenamefont{Scheike et~al.}(2023)\citenamefont{Scheike, Wen, Sukegawa, and Mitani}}]{Scheike2023}
\bibinfo{author}{\bibfnamefont{T.}~\bibnamefont{Scheike}}, \bibinfo{author}{\bibfnamefont{Z.}~\bibnamefont{Wen}}, \bibinfo{author}{\bibfnamefont{H.}~\bibnamefont{Sukegawa}}, \bibnamefont{and} \bibinfo{author}{\bibfnamefont{S.}~\bibnamefont{Mitani}}, \bibinfo{journal}{Applied Physics Letters} \textbf{\bibinfo{volume}{122}}, \bibinfo{pages}{112404} (\bibinfo{year}{2023}), ISSN \bibinfo{issn}{0003-6951}, \bibinfo{note}{\href{http://doi.org/10.1063/5.0145873}{URL}}.

\bibitem[{\citenamefont{Itoh}(2007)}]{disorder_1}
\bibinfo{author}{\bibfnamefont{H.}~\bibnamefont{Itoh}}, \bibinfo{journal}{Journal of Physics D: Applied Physics} \textbf{\bibinfo{volume}{40}}, \bibinfo{pages}{1228} (\bibinfo{year}{2007}), \bibinfo{note}{\href{http://dx.doi.org/10.1088/0022-3727/40/5/S03}{URL}}.

\bibitem[{\citenamefont{Xu et~al.}(2006)\citenamefont{Xu, Karpan, Xia, Zwierzycki, Marushchenko, and Kelly}}]{disorder_2}
\bibinfo{author}{\bibfnamefont{P.~X.} \bibnamefont{Xu}}, \bibinfo{author}{\bibfnamefont{V.~M.} \bibnamefont{Karpan}}, \bibinfo{author}{\bibfnamefont{K.}~\bibnamefont{Xia}}, \bibinfo{author}{\bibfnamefont{M.}~\bibnamefont{Zwierzycki}}, \bibinfo{author}{\bibfnamefont{I.}~\bibnamefont{Marushchenko}}, \bibnamefont{and} \bibinfo{author}{\bibfnamefont{P.~J.} \bibnamefont{Kelly}}, \bibinfo{journal}{Phys. Rev. B} \textbf{\bibinfo{volume}{73}}, \bibinfo{pages}{180402} (\bibinfo{year}{2006}), \bibinfo{note}{\href{http://link.aps.org/doi/10.1103/PhysRevB.73.180402}{URL}}.

\bibitem[{\citenamefont{Bose et~al.}(2008)\citenamefont{Bose, Ernst, Mertig, and Henk}}]{FeO_oxidation}
\bibinfo{author}{\bibfnamefont{P.}~\bibnamefont{Bose}}, \bibinfo{author}{\bibfnamefont{A.}~\bibnamefont{Ernst}}, \bibinfo{author}{\bibfnamefont{I.}~\bibnamefont{Mertig}}, \bibnamefont{and} \bibinfo{author}{\bibfnamefont{J.}~\bibnamefont{Henk}}, \bibinfo{journal}{Phys. Rev. B} \textbf{\bibinfo{volume}{78}}, \bibinfo{pages}{092403} (\bibinfo{year}{2008}), \bibinfo{note}{\href{http://link.aps.org/doi/10.1103/PhysRevB.78.092403}{URL}}.

\bibitem[{\citenamefont{Tiusan et~al.}(2006)\citenamefont{Tiusan, Sicot, Hehn, Belouard, Andrieu, Montaigne, and Schuhl}}]{cm.sm_06}
\bibinfo{author}{\bibfnamefont{C.}~\bibnamefont{Tiusan}}, \bibinfo{author}{\bibfnamefont{M.}~\bibnamefont{Sicot}}, \bibinfo{author}{\bibfnamefont{M.}~\bibnamefont{Hehn}}, \bibinfo{author}{\bibfnamefont{C.}~\bibnamefont{Belouard}}, \bibinfo{author}{\bibfnamefont{S.}~\bibnamefont{Andrieu}}, \bibinfo{author}{\bibfnamefont{F.}~\bibnamefont{Montaigne}}, \bibnamefont{and} \bibinfo{author}{\bibfnamefont{A.}~\bibnamefont{Schuhl}}, \bibinfo{journal}{Applied Physics Letters} \textbf{\bibinfo{volume}{88}}, \bibinfo{pages}{062512} (\bibinfo{year}{2006}), ISSN \bibinfo{issn}{0003-6951}, \bibinfo{note}{\href{http://doi.org/10.1063/1.2172717}{URL}}.

\bibitem[{\citenamefont{Bowen et~al.}(2001{\natexlab{b}})\citenamefont{Bowen, Cros, Petroff, Fert, Martinez~Boubeta, Costa-Krämer, Anguita, Cebollada, Briones, de~Teresa et~al.}}]{M.Bowen_A.Fert_01}
\bibinfo{author}{\bibfnamefont{M.}~\bibnamefont{Bowen}}, \bibinfo{author}{\bibfnamefont{V.}~\bibnamefont{Cros}}, \bibinfo{author}{\bibfnamefont{F.}~\bibnamefont{Petroff}}, \bibinfo{author}{\bibfnamefont{A.}~\bibnamefont{Fert}}, \bibinfo{author}{\bibfnamefont{C.}~\bibnamefont{Martinez~Boubeta}}, \bibinfo{author}{\bibfnamefont{J.~L.} \bibnamefont{Costa-Krämer}}, \bibinfo{author}{\bibfnamefont{J.~V.} \bibnamefont{Anguita}}, \bibinfo{author}{\bibfnamefont{A.}~\bibnamefont{Cebollada}}, \bibinfo{author}{\bibfnamefont{F.}~\bibnamefont{Briones}}, \bibinfo{author}{\bibfnamefont{J.~M.} \bibnamefont{de~Teresa}}, \bibnamefont{et~al.}, \bibinfo{journal}{Applied Physics Letters} \textbf{\bibinfo{volume}{79}}, \bibinfo{pages}{1655} (\bibinfo{year}{2001}{\natexlab{b}}), ISSN \bibinfo{issn}{0003-6951}, \bibinfo{note}{\href{http://doi.org/10.1063/1.1404125}{URL}}.

\end{thebibliography}

\end{document}